\documentclass{aa} 
\usepackage{color} 
\usepackage[squaren, Gray, cdot]{SIunits}
\usepackage{graphicx}
\usepackage[varg]{txfonts}
\usepackage{natbib}
\bibpunct{(}{)}{;}{a}{}{,} % to follow the A&A style
\usepackage{amsmath}

\newcounter{eqn}

\makeatletter
\newcommand{\putindeepbox}[2][0.7\baselineskip]{{%
    \setbox0=\hbox{#2}%
    \setbox0=\vbox{\noindent\hsize=\wd0\unhbox0}
    \@tempdima=\dp0
    \advance\@tempdima by \ht0
    \advance\@tempdima by -#1\relax
    \dp0=\@tempdima
    \ht0=#1\relax
    \box0
}}
\makeatother
\begin{document}

   \title{H$_2$ distribution during the formation of multiphase  molecular clouds}

   \author{Valeska Valdivia \inst{\ref{inst1},\ref{inst2},\ref{inst3}}
          \and
          Patrick Hennebelle\inst{\ref{inst2},\ref{inst1}}
          \and
          Maryvonne G\'erin \inst{\ref{inst1},\ref{inst3}}
          \and
          Pierre Lesaffre \inst{\ref{inst1},\ref{inst3}}
        }

        \institute{%Laboratoire de radioastronomie, LERMA, Observatoire de Paris, \'Ecole Normale Sup\'erieure (UMR 8112 CNRS), 24 rue Lhomond, 75231 Paris Cedex 05, France\\
Laboratoire de radioastronomie, LERMA, Observatoire de Paris, \'Ecole Normale Sup\'erieure, PSL Research University, CNRS, UMR 8112, 75005, Paris, France \\
        \email{valeska.valdivia@lra.ens.fr}\label{inst1}
        %Universit\'e Pierre et Marie Curie, 4 place Jussieu 75005 Paris, France \label{inst3}
        \and
        Laboratoire AIM, Paris-Saclay, CEA/IRFU/SAp - CNRS - Universit\'e Paris Diderot, 91191 Gif-sur-Yvette Cedex, France\\
        \email{patrick.hennebelle@lra.ens.fr} \label{inst2}
        %\and Sorbonne Universités, UPMC Univ Paris06, IFD, 4 Place Jussieu, 75252 PARIS cedex05
        %Sorbonne Universit\'es, UPMC Univ Paris06, IFD, 4 place Jussieu, 75252 Paris Cedex 05, France
        \and Sorbonne Universit\'es, UPMC Univ. Paris 06, UMR 8112, LERMA, Paris, France, F-75005
\label{inst3}
             }

   \date{Received Month dd, yyyy; accepted Month dd, yyyy}

% \abstract{}{}{}{}{} 
% 5 {} token are mandatory
 
  \abstract
  % context heading (optional)
  % {} leave it empty if necessary  
   {H$_2$ is the simplest and the most abundant molecule in the interstellar medium (ISM), and its formation precedes the formation of other molecules. }
%do not take place in a It is well know that it is but several aspects remain unclear.  }
  % aims heading (mandatory)
   {Understanding the dynamical influence of the environment and the interplay between the thermal processes related to the formation and 
destruction of H$_2$ and the structure of the cloud is mandatory  to understand correctly the observations of H$_2$.}
  % methods heading (mandatory)
   {We performed high-resolution
 magnetohydrodynamical colliding-flow simulations with the adaptive mesh refinement code RAMSES in which 
the physics of H$_2$ has been included. We compared the simulation results with various observations of the 
H$_2$ molecule, including  the column densities of excited rotational levels.}
  % results heading (mandatory)
   {As a result of a combination of thermal pressure, ram pressure, and gravity, the clouds produced at the converging point 
of HI streams are highly inhomogeneous. H$_2$ molecules quickly form in relatively dense clumps and spread into the diffuse 
interclump gas. This in particular leads to the existence of significant abundances of H$_2$ in the diffuse and warm gas that lies in between clumps. Simulations and observations show similar trends, especially for the HI-to-H$_2$ transition (H$_2$ fraction vs total hydrogen column density).
%
%Comparison of the observed H$_2$ fraction vs total hydrogen column density reveals very similar trends and numbers.
Moreover, the abundances of excited rotational levels, calculated at equilibrium in the simulations, turn out to be 
very similar to the observed abundances inferred from \emph{FUSE} results. This is a direct consequence of the presence of the 
H$_2$ enriched diffuse and warm gas.}
  % conclusions heading (optional), leave it empty if necessary 
   {Our simulations, which self-consistently form molecular clouds out of the diffuse atomic gas, show that 
H$_2$ rapidly forms  in the dense clumps and, due to the complex structure of molecular clouds, 
quickly spreads at lower densities. Consequently, a significant fraction of warm H$_2$ exists 
in the low-density gas. This warm H$_2$ leads to column densities of excited rotational levels close to the observed ones
and probably reveals the complex intermix between the warm and cold gas in molecular clouds.
This  suggests that the two-phase structure of molecular clouds is an essential ingredient  for fully understanding 
molecular hydrogen in these objects. }


\titlerunning{H$_2$ distribution during formation of molecular clouds} 
\authorrunning{Valdivia et al.}
   \maketitle

%
%________________________________________________________________

\section{Introduction}

It is well known that stars form in dense and self-gravitating molecular clouds and that the star formation rate per unit area ($\Sigma_{\mathrm{SFR}}$) is on average relatively well
correlated with the H$_2$ surface density ($\Sigma_{\mathrm{H_2}}$) \citep{leroy2013, lada2012, wong2002, bigiel2008}.

Although H$_2$ is the simplest and most abundant molecule in the interstellar medium (ISM),  it is very hard to observe directly. 
Because of its homonuclear nature, H$_2$ lacks a permament dipole and only weak quadrupolar transitions are allowed. Moreover, excitation energies are very high and require very high temperatures or strong ultraviolet (UV) fields to excite its rotational levels \citep{kennicutt2012}. H$_2$ can be 
observed in emission by infrared (IR) rovibrational transitions \citep{burton1992, santangelo2014, habart2011}, or in absorption at far-ultraviolet 
(far-UV) wavelengths from the Lyman and Werner bands. These bands were first observed by \citet{spitzer1975} using the \emph{Copernicus} satellite. 
The \emph{Far Ultraviolet Spectroscopic Explorer (\emph{FUSE})} \citep{moos2000} offered new perspectives on the study of H$_2$ in the 
ISM through its sensitivity, which is $10^5$ times higher than \emph{Copernicus} in the far-UV part of the spectrum, providing measurements of H$_2$ 
column densities (of the total column density of H$_2$ and for several rotational levels $J$) along translucent lines of sight \citep{wakker2006, sheffer2008}. 
 
While it seems to be clear that molecular gas is well correlated with star formation at different scales,  it is still an open
question how the gas becomes molecular. Since understanding the atomic-to-molecular hydrogen (HI-to-H$_2$) transition is of the highest importance for understanding the star-forming process,
 numerous models have been developed \citep[e.g.][]{krumholz+2008, sternberg2014} and seem able to reproduce many of the 
observed constraints. While extremely useful, these models leave aside the dynamical aspects of H$_2$ formation. 
In particular, the question of the relatively long timescale that is needed to form the H$_2$ molecules has for many years 
been difficult to reconcile with short-lived and quickly forming molecular clouds. The typical timescale for H$_2$ formation is of the order 
of $10^9 {\rm yr} / n$ \citep{hollen1971}, which would lead to ages
older than 10 Myr for
 molecular clouds of mean densities of the order of 
10-100 cm$^{-3}$ \citep{blitz-shu1980}. 
However, \citet{glover2007a} and \citet{glover2007b}, who have simulated the formation of molecular hydrogen in  supersonic clouds, showed that 
H$_2$ can form relatively quickly in the dense clumps induced by the shocks, leading to a significantly shorter timescale (typically of about 3 Myr). 
Recently, \citet{micic+2012} explored the influence of the turbulent forcing and showed that it has a significant influence on the timescale 
of H$_2$ formation. This is because compressible forcing leads to denser clumps than solenoidal forcing, for which the motions are less 
compressible. This clearly shows that dynamics is playing an important role in the process of HI-to-H$_2$ transition, at least at the scale of 
a molecular cloud.

In the Milky Way most of the molecular hydrogen is in the form of low-temperature gas, but
different observations have brought to light the existence of large amounts of fairly warm H$_2$ in the ISM 
\citep{valentijn1999, verstraete1999}. \citet{gry2002} and \citet{lacour2005} have shown that the H$_2$ excitation resulting from UV 
pumping and from H$_2$ formation cannot account for the observed population of H$_2$ excited levels  ($ J > 2$). Excited H$_2$ is 
likely explained by the presence of a warm and turbulent layer associated with the molecular cloud. But in such warm gas, which
is characterised by very 
low densities ($n_\mathrm{H} \approx 1-10~ \mathrm{cm^{-3}}$) and very high temperatures ($T \approx 10^{3}-10^{4}~ \mathrm{K}$), H$_2$ formation 
on grain surfaces becomes negligible, which accordingly leads to an apparent contradiction. Recently, \citet{godard+2009} have investigated the possibility that 
warm H$_2$ could form during intermittent high-energy dissipation events such as vortices and shocks. They showed in particular that under plausible 
assumptions for the distributions of these events, the observed abundance of excited H$_2$ molecules could be reproduced. 
Another possibility to explain the abundance of these  excited H$_2$ molecules is that they could be formed in dense gas and then 
be transported in more diffuse and warmer medium. This possibility has been investigated by \citet{lesaf+2007} using a 1D model and 
a prescription to take into account the turbulent diffusion between the phases. In particular, they found that the abundance of H$_2$ molecules
 at low density and high temperature increases with the turbulent diffusion efficiency.

While different in nature, previous works \citep{glover2007a,glover2007b,lesaf+2007,godard+2009} therefore consistently found that 
the formation of H$_2$ is significantly influenced by the dynamics of the flow in which it forms. Although the exact 
mechanism that leads to the formation of molecular clouds is still under investigation, it seems unavoidable to consider 
converging streams of diffuse gas \citep[e.g.][]{hennefalg2012,dobbs2014}. What exactly triggers these flows is not fully 
elucidated yet but is most likely due a combination of gravity, large-scale turbulence, and large-scale shell expansion.
Another important issue is the thermodynamical state of the gas. The diffuse gas out of which molecular clouds form is 
most likely a mixture of phases made of warm and cold atomic hydrogen (HI) and even possibly somewhat diffuse molecular gas. 
Such a multiphase medium presents strong temperature variations (typically from 8000 K to lower than 50 K). This 
causes the dynamics of the flow to be significantly different from isothermal or nearly isothermal flow \citep[see][for a comparison between barotropic  
and two-phase flows]{audithenne2010}. Moreover, the two-phase nature of the diffuse atomic medium has been found to persist 
in molecular clouds \citep{hennebelle+2008,heitsch+2008,banerjee+2009,vazquez2010,inoue+2012}. This in particular 
produces a medium in which supersonic clumps of cold gas are embedded in a diffuse phase of subsonic warm gas. 
Consequently, the question about the consequences of the multiphase nature of a
molecular cloud are for the formation of H$_2$. Clearly, various processes may occur. 
First of all, the dense clumps are continuously forming and accreting out of the diffuse gas because 
molecular clouds continue to accrete.
Second, 
some of the H$_2$ formed at high density is likely to be spread in low-density high-temperature gas because of the 
phase exchanges. 

We here present numerical simulations using a simple model for H$_2$ formation within a dynamically evolving turbulent molecular cloud that is formed 
through colliding streams of atomic gas. Such flows, sometimes called colliding flows, have been widely used to study 
the formation of molecular clouds because they represent a good compromise between the need to assemble the diffuse material 
from the large scales and the need to  accurately describe the small scales within the molecular clouds \citep{hp99,bal99,heitsch+06,vazquez+06, clark2012}. 
The content of this paper is as follows: in Sect. \ref{process} we present  the governing equations and the physical processes, such as  H$_2$ formation. In Sect. 
\ref{setup} we describe the numerical setup that we use to perform the simulations. In Sect.
\ref{results} we present the results of our simulations, first describing the general structure of the cloud and then focussing 
on the H$_2$ distribution. We perform a few complementary calculations that aim at better understanding the physical mechanisms at play in the simulations.
Then in Sect. \ref{observation} we compare our results with various observations that have quantified the abundance of H$_2$ and find 
 very reasonable agreements.  
Finally, we summarise and discuss the implications of our work
in Sect. 6.

\begin{figure*}
\centering
\includegraphics[width=0.95\textwidth]{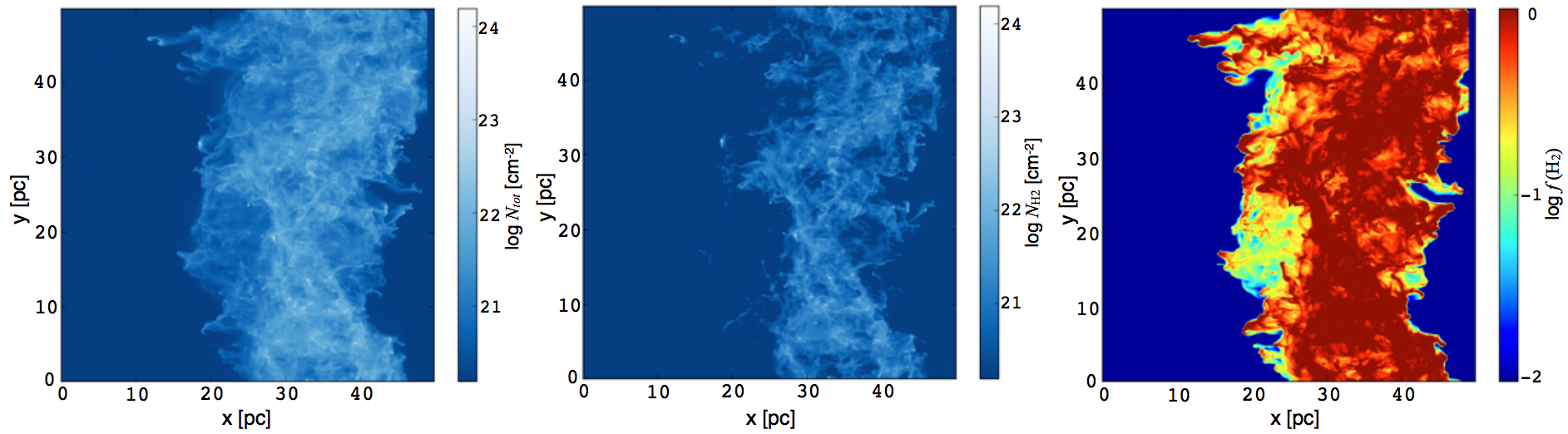}
\caption{Column density maps. Total hydrogen column density map (left), H$_2$ column density map (centre), and molecular fraction (right). The gas enters the box from both sides ($x$-axis), and the conditions for the other sides are periodic. The gas is compressed along the $x$-direction and the cloud forms toward the middle of the box.}
\label{NcolH2fH2}
\end{figure*}

\begin{figure*}[htb]
\centering 
\begin{tabular}{@{}ccc@{}}
%    \includegraphics[width=.32\textwidth]{Figures/scPaper_slice_bin_8_10_rho_cm_3_164.png} &
%     \includegraphics[width=.32\textwidth]{Figures/scPaper_slice_bin_8_10_nH2_cm_3_164.png} &
%   \includegraphics[width=.32\textwidth]{Figures/Paper_slice_bin_8_10_kphot_cm_3_164.png} 
%\hspace{-0.4cm}
\includegraphics[width=.32\textwidth]{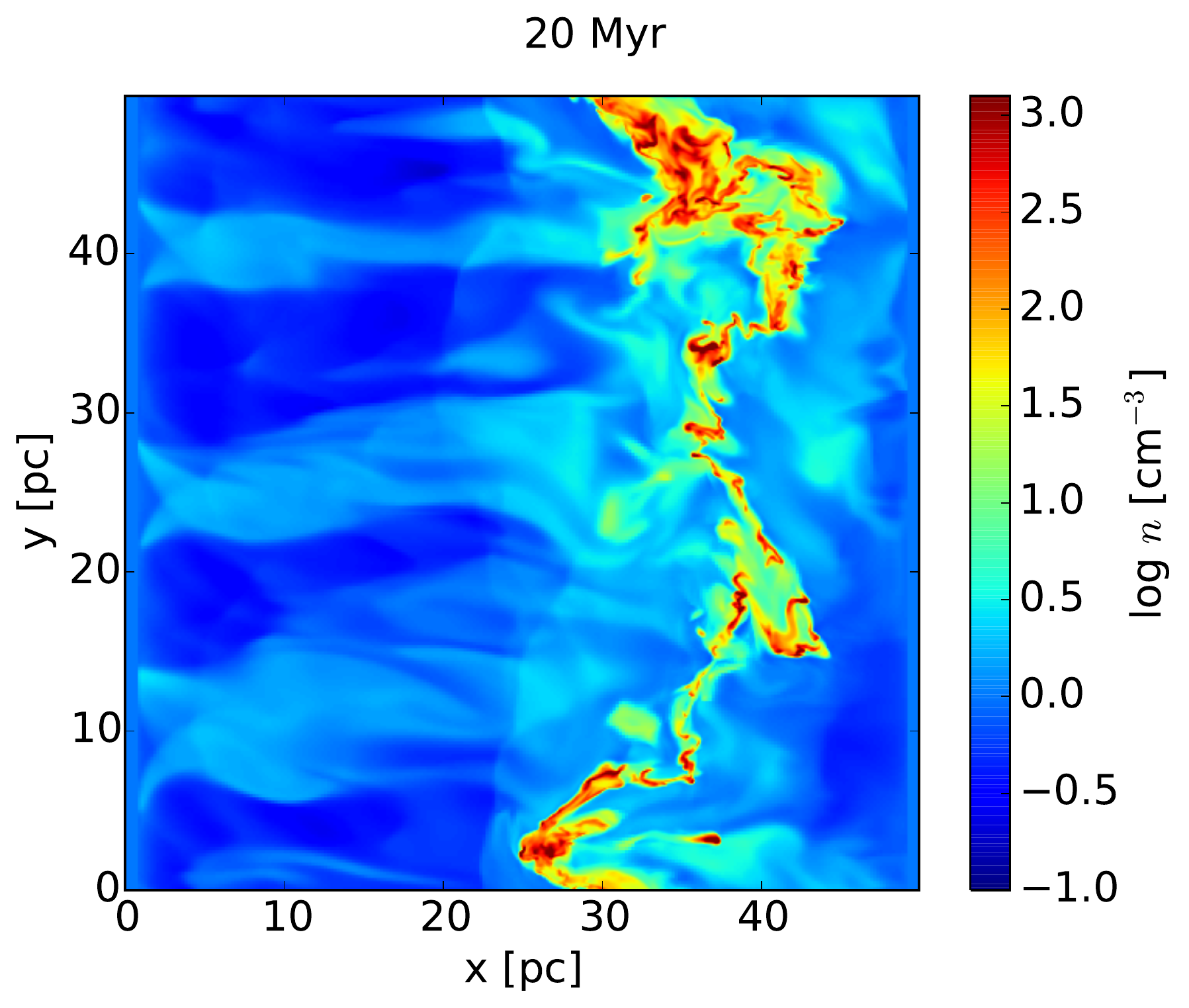} &
\includegraphics[width=.32\textwidth]{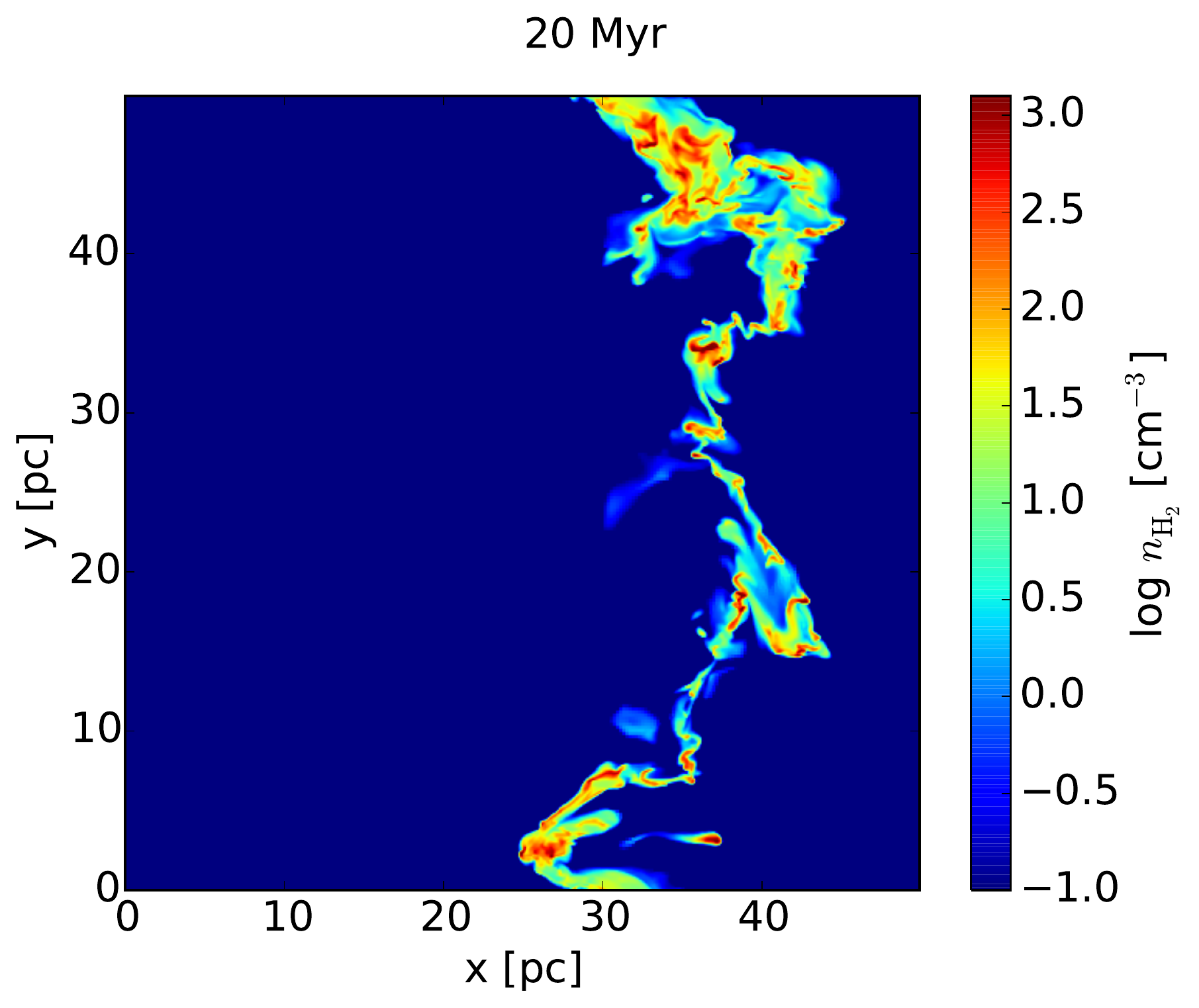} &
\includegraphics[width=.32\textwidth]{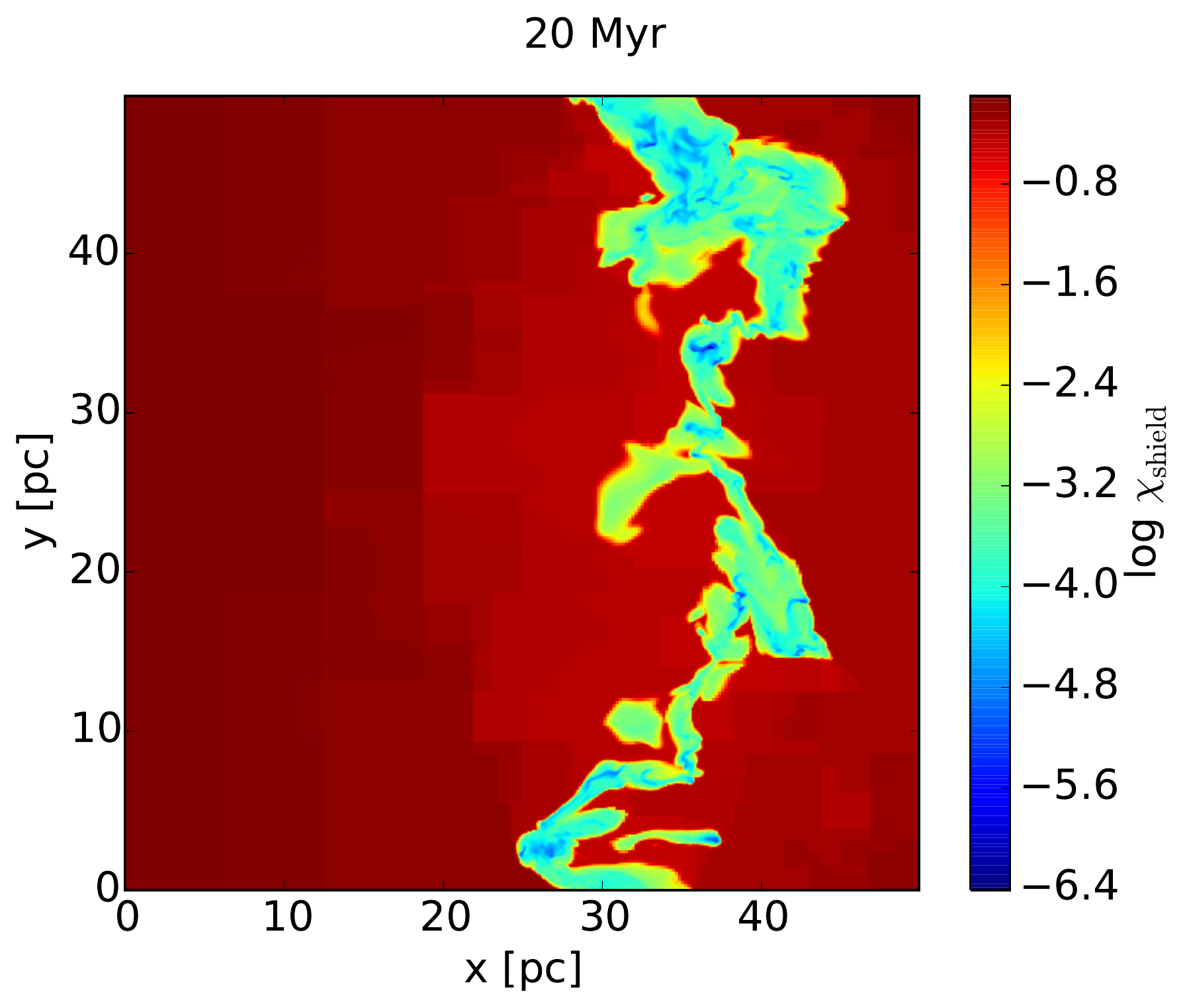} 
\end{tabular}
\caption{Slices cut through the mid-plane. The panel on the left shows the total number density of hydrogen nucleons, the panel in the centre shows the number density of molecular hydrogen (on the same scale), and the panel on the right shows the mean shielding factor for the H$_2$ photodissociation rate coefficient, calculated as $\langle \exp(-\tau_{d, 1000}) f_\mathrm{shield}\rangle$.}
\label{shield_f}
\end{figure*}

\section{Physical processes}\label{process}

\subsection{Governing fluid equations}
We consider  the usual compressive magnetohydrodynamical equations that govern the behaviour of the gas. These equations written in their conservative form are

\begin{eqnarray}
\frac{\partial \rho}{\partial t} + \nabla\cdot(\rho \mathbf{v})&=& 0, \label{mhd1}\\
\frac{\partial \rho \mathbf{v}}{\partial t} + \nabla\cdot(\rho\mathbf{v}\mathbf{v} -\mathbf{B}\mathbf{B}) +\nabla P &=& -\rho \nabla \phi, \label{mhd2}    \\
\frac{\partial E}{\partial t} + \nabla\cdot[(E+P)\mathbf{v} - \mathbf{B}(\mathbf{B}\mathbf{v})] &=& -\rho {\cal L}, \label{mhd3}\\
\frac{\partial \mathbf{B}}{\partial t} +\nabla\cdot(\mathbf{v}\mathbf{B} - \mathbf{B}\mathbf{v}) &=& 0, \label{mhd4}\\
\nabla^2 \phi &=& 4\pi G \rho \label{mhd5} 
,\end{eqnarray} 

\noindent where $\rho$, $\mathbf{v}$, $\mathbf{B}$, $P$, $E$,
and $\phi$ are the mass density, velocity field, magnetic field, total energy, and the gravitational potential of the gas, respectively,
and $\cal L$ is the net loss function that describes gas cooling and heating.

\subsection{Chemistry of H$_2$}
The fluid equations (Eqs. \ref{mhd1}-\ref{mhd5}) are complemented by an equation that describes the formation of the H$_2$
molecules,

\begin{eqnarray}
\frac{\partial n_\mathrm{H_2}}{\partial t} + \nabla\cdot(n_\mathrm{H_2} \mathbf{v})&=& k_\mathrm{form} n (n - 2 n_\mathrm{H_2})- k_\mathrm{ph} n_\mathrm{H_2} , \label{eqH2}
\end{eqnarray} 
where $n$ is the total hydrogen density,  $n_\mathrm{H_2}$ represents the density of H$_2$, and $k_\mathrm{form}$ and $k_\mathrm{ph}$ represent the formation and
destruction rates of H$_2$.

\subsubsection{H$_2$ formation}

When two hydrogen atoms encounter each other in the gas phase, they cannot radiate the excess of energy because of  the lack of 
an electric dipole moment, and therefore H$_2$ formation in the
gas phase is negligible. It is widely accepted today that H$_2$ 
is formed through grain catalysis \citep{hollenbach1971, gould1963}. Hydrogen atoms can adsorb onto the grain surfaces and 
encounter another hydrogen atom to form a H$_2$ molecule through two mechanisms: the Langmuir-Hinshelwood mechanism, where atoms are 
physisorbed (efficient in shielded gas), and the Eley-Rideal
mechanism, or chemisorption (efficient on warm grains) \citep{lebourlot2012, bron2014}. 
The detailed physical description of these two mechanisms is complex and the numerical treatment is computationally expensive. For this reason we adopted a simpler description for H$_2$ formation on grain surfaces, given by the following mean formation rate:
\begin{equation}
k_\mathrm{form, 0} = 3 \times 10^{-17}~\mathrm{cm^{3} s^{-1}}
.\end{equation}
\noindent This rate was first derived by \citet{jura1974}, using \emph{Copernicus} observations, and was confirmed by \citet{gry2002} using \emph{FUSE} observations. The formation rate depends  on the local gas temperature and on the adsorption properties of the grain, therefore this value is corrected for by the dependence on temperature and by a sticking coefficient: 
\begin{equation}
k_\mathrm{form} =  k_\mathrm{form, 0}\sqrt{\frac{T}{100 \ \mathrm{K}}} \times S(T).\end{equation}
\noindent $S(T)$ is the empirical expression for the sticking coefficient of hydrogen atoms on the grain surface as described in \cite{lebourlot2012},
\begin{equation}
S(T) = \frac{1}{ 1+\left(\frac{T}{T_2} \right)^{\beta}},
\end{equation}
\noindent where we use the same fitting values as \cite{bron2014}, namely $T_2 = 464~\mathrm{K}$, and $\beta = 1.5$.

%T2 = 464 K and β = 1.5, which gives results close to the expression of Sternberg & Dalgarno (1995).

\subsubsection{H$_2$ destruction}
The main mechanism that destroys the H$_2$ molecule is photodissociation by absorption of UV photons in the $912$ - $1100$ $\angstrom$ range (Lyman and Werner transitions). In a UV field of strength $G_0$, H$_2$ is photodissociated in optically thin gas at a rate \citep{draine1996} of 
\begin{equation}
k_{\mathrm{ph,0}} = 3.3 \times 10^{-11} G_0 \ \mathrm{s^{-1}},   \label{kph0}
\end{equation} 

\noindent but the H$_2$ gas can protect itself against photodissociation by two shielding effects. The first shielding effect is the \emph{\textup{continuous dust absorption}}, while the second effect is the line absorption due to other H$_2$ molecules, called \textup{\emph{\textup{self-shielding}}}. \cite{draine1996} showed that the photodissociation rate can be written as
\begin{equation}
k_{\mathrm{ph}} =  e^{-\tau_{d,1000}}  f_\mathrm{shield} (\mathcal{N}_{\mathrm{H_2}} )  k_{\mathrm{ph,0}}.   \label{kdest}
\end{equation} 

\noindent The first term is the effect of the dust. Here $\tau_{d,1000} = \sigma_{d,1000} \mathcal{N}_{tot}$ is the optical depth along a line of sight ($\sigma_{d,1000} = 2\times 10^{-21}\mathrm{cm^{-2}}$ is the effective attenuation cross section for dust grains at $\lambda = 1000 \ \angstrom$ and $\mathcal{N}_{tot}$ is the total column density of hydrogen). The second term is the self-shielding factor, and we use the same approximation,
\begin{equation}
f_{\mathrm{shield}} = \frac{0.965}{(1 + x/b_5)^2} + \frac{0.035}{(1+x)^{1/2}}\mathrm{exp}(-8.5\times10^{-4}(1+x)^{1/2}) , \label{fshield}
\end{equation} 

\noindent where $x= \mathcal{N}_\mathrm{H_2}/5\times 10^{14}~\mathrm{cm^{-2}}$, $b_5 = b/10^5~\mathrm{cm \ s^{-1}}$, where $b$ is the Doppler-broadening parameter, which is typically $2~\mathrm{km\ s^{-1}}$, but it can reach values as high as $10~\mathrm{km\ s^{-1}}$ \citep{shull2000}. We assume that the turbulent contribution dominates, then we use $b = 2~\mathrm{km \ s^{-1}}$ (see Appendix \ref{bdop_infl} for the influence of this choice on the shielding coefficient).

\subsection{Thermal processes}

For the heating and cooling of the gas, we used the same treatment as we did previously \citep{valdivia2014} \citep[see also][]{audit2005}, which we describe in this section and call the standard heating and cooling, to which we have added the thermal feedback from H$_2$, described in Sect. \ref{thermfeedbH2}.
%\textbf{The standard heating and cooling processes are the same as our previous work \citep{valdivia2014} \citep[see also][]{audit2005}, to which we have added the thermal feedback (cooling and heating processes) from H$_2$, described in Sect. \ref{thermfeedbH2}}.
%In this section we describe the standard treatment
%heating and cooling processes included in our simulations and in 
%
%For the heating and cooling of the gas, we perform the same treatment as our previous work \citep{valdivia2014} \citep[see also][]{audit2005}\textbf{. This simple treatment is what we call the "standard" heating and cooling, described in this section, to which we have added the thermal feedback (cooling and heating processes) from H$_2$, described in Sect. \ref{thermfeedbH2}}. 

The dominant heating process in the gas is due to the ultraviolet flux from the interstellar radiation field (ISRF) through the photoelectric effect on grains \citep{bakestielens1994, wolfire1995}, where we use the effective UV field strength, calculated using our attenuation factor $\chi$ for the UV field. We also include the heating by cosmic rays \citep{goldsmith2001}, which is important in well shielded regions.

The primary coolant at low temperature is the $^2P_{3/2} \rightarrow \ ^2P_{1/2}$ [CII] fine-structure transition at $158~\mathrm{\micro m}$ \citep{launay1977, hayes1984, wolfire1995}. At higher temperature, fine-structure levels of OI can be excited and thus contribute to the cooling of the gas. We include the [OI] $63~\mathrm{\micro m}$ and $146~\mathrm{\micro m}$ line emission \citep{flower1986, tielens2005book, wolfire2003}. For the Lyman $\alpha$ emission, which becomes dominant at high temperatures, we use the classical expression of \citet{spitzer1978book}. We also include the cooling by electron recombination onto dust grains using the prescription of \citet{wolfire1995, wolfire2003} and \citet{bakestielens1994}.
We note that strictly speaking, this cooling function does not include molecular cooling in spite of the high densities reached in the simulation. However, it leads to temperatures that are entirely reasonable even at high densities and therefore appears to be sufficiently accurate, in particular 
because we are not explicitly solving for the formation of molecules in addition to H$_2$ \citep[see][for a quantitative estimate]{levrier+2012}. 

\subsubsection{Thermal feedback from H$_2$}
\label{thermfeedbH2}
To the atomic cooling, described before, we added the molecular cooling by H$_2$ lines and the heating by H$_2$ formation and destruction.\\
%  chaud_2 = 1.5_dp*eVtoerg*kform * n*(1.0_dp - fH2) !Heating by H2 formation: 1/3*4.5eV in erg s-1 cm-3                                                                
%  chaudH2ph = 6.4d-13*kdest*fH2/2.0_dp              !Heating by photodiss of H2: O.4eV (Eq 44 Glover&MacLow 2006, Black&Dalgarno 1977)   
%
During the H$_2$ formation process, about $4.5~\mathrm{eV}$ are released. The distribution of this energy into translational energy, H$_2$ internal energy (rotational and vibrational excitation), and into the grain heating is not well known, and the derived
fraction of this energy that actually heats the gas varies from one author to another \citep{lebourlot2012, glover2007a}. We consider an equipartition of the energy, which me$\text{ans that
one-third}$ of the energy released goes into heating the gas,
\begin{equation}
\Gamma_{\mathrm{form}} = 2.4\times 10^{-12} k_{\mathrm{form}} n_{\mathrm{H}} n\ \mathrm{erg \ s^{-1} cm^{-3}} \label{heat_H2form}
.\end{equation}
H$_2$ photodissociation provides an additional heating source. Following \citet{black1977} and \citet{glover2007a}, we assumed $0.4\ \mathrm{eV}$ released into the gas per photodissociation,
\begin{equation}
\Gamma_{\mathrm{ph}} = 6.4\times 10^{-13} k_{\mathrm{ph}} n_{\mathrm{H_2}} \ \mathrm{erg \ s^{-1} cm^{-3}}  \label{heat_H2dest}
.\end{equation}

H$_2$ contributes to the cooling of the gas through line emission and can become strong for the diffuse medium under certain conditions \citep{glover2014, gnedin2011}. As H$_2$ is a homonuclear molecule, only quadrupolar transitions are allowed, and thus, the ortho and para states can be treated independently. In our simulations we adopted the cooling function from \citet{lebourlot1999}. This is a function of temperature, total density, $n(\mathrm{HI})/n(\mathrm{H_2})$ relative abundance, and of the ortho-to-para-H$_2$ ratio (OPR). It considers transitions between the first $51$ rovibrational energy levels. As the cooling function is quite insensitive to the OPR, we fixed it to 3 for simplicity.

\begin{figure}
\centering
\includegraphics[width=7.8cm]{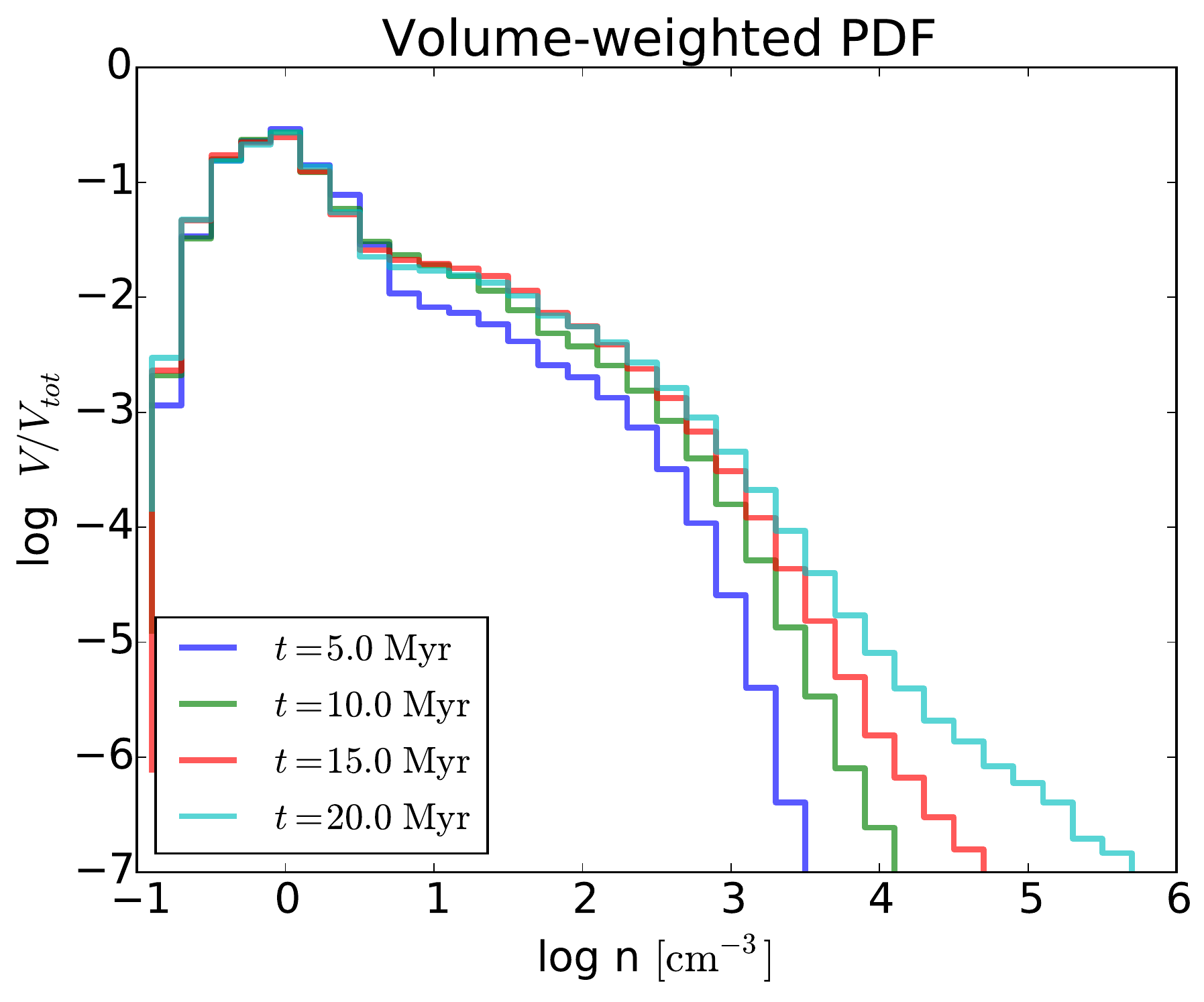}
\caption{Density PDF evolution. Colour lines show the density PDF at $t=5$ (blue), $10$ (green), $15$ (red), and $20$ Myr (light blue).}
\label{gen_PDF}
\end{figure}

\begin{figure}
\centering
\includegraphics[width=7.8cm]{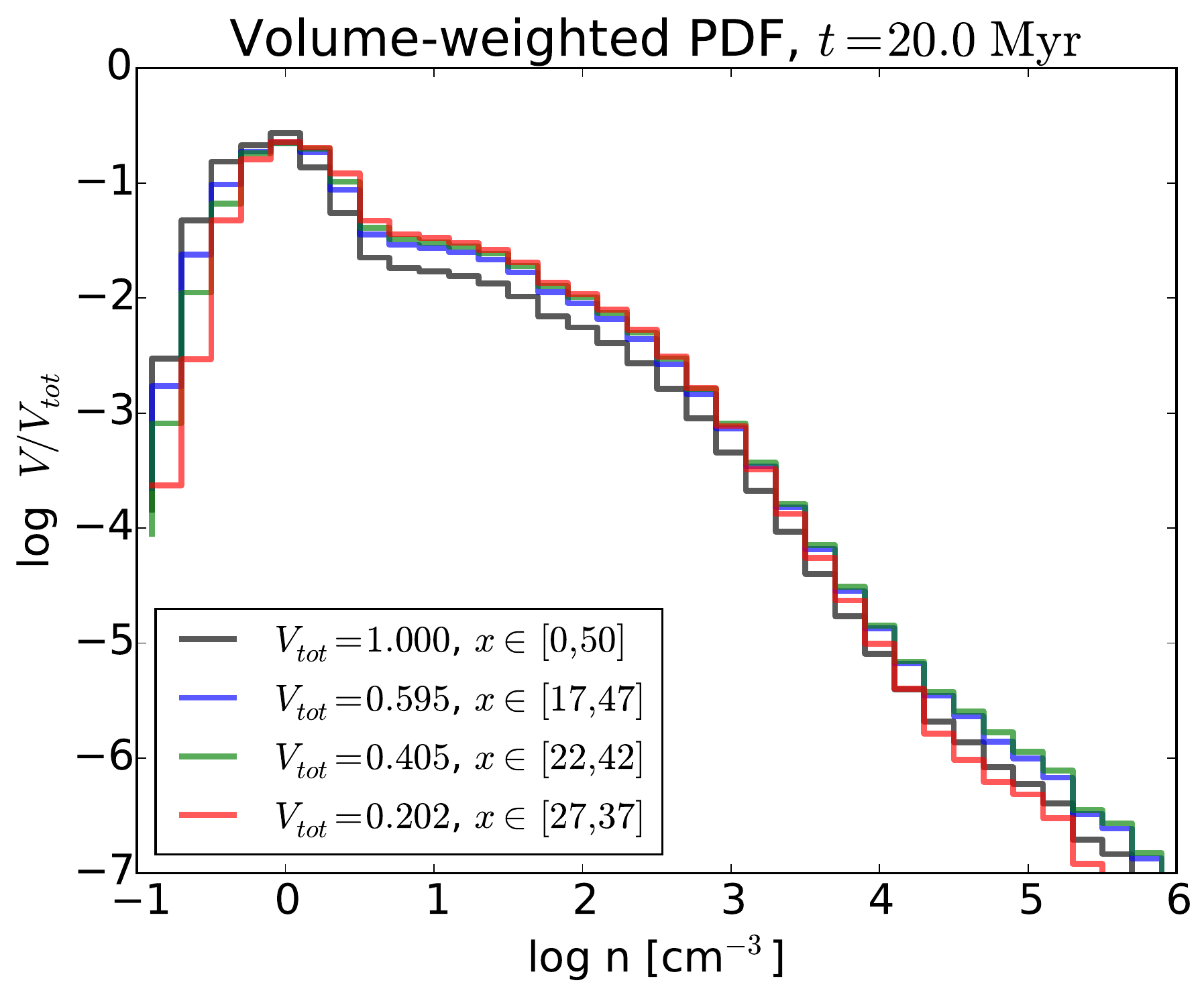}
\caption{Volume filling factor as a function of density in different regions. $V_{tot}$ is the volume fraction of each region. }
\label{vol_fill}
\end{figure}

\begin{figure}
\centering
{\includegraphics[width=7cm]{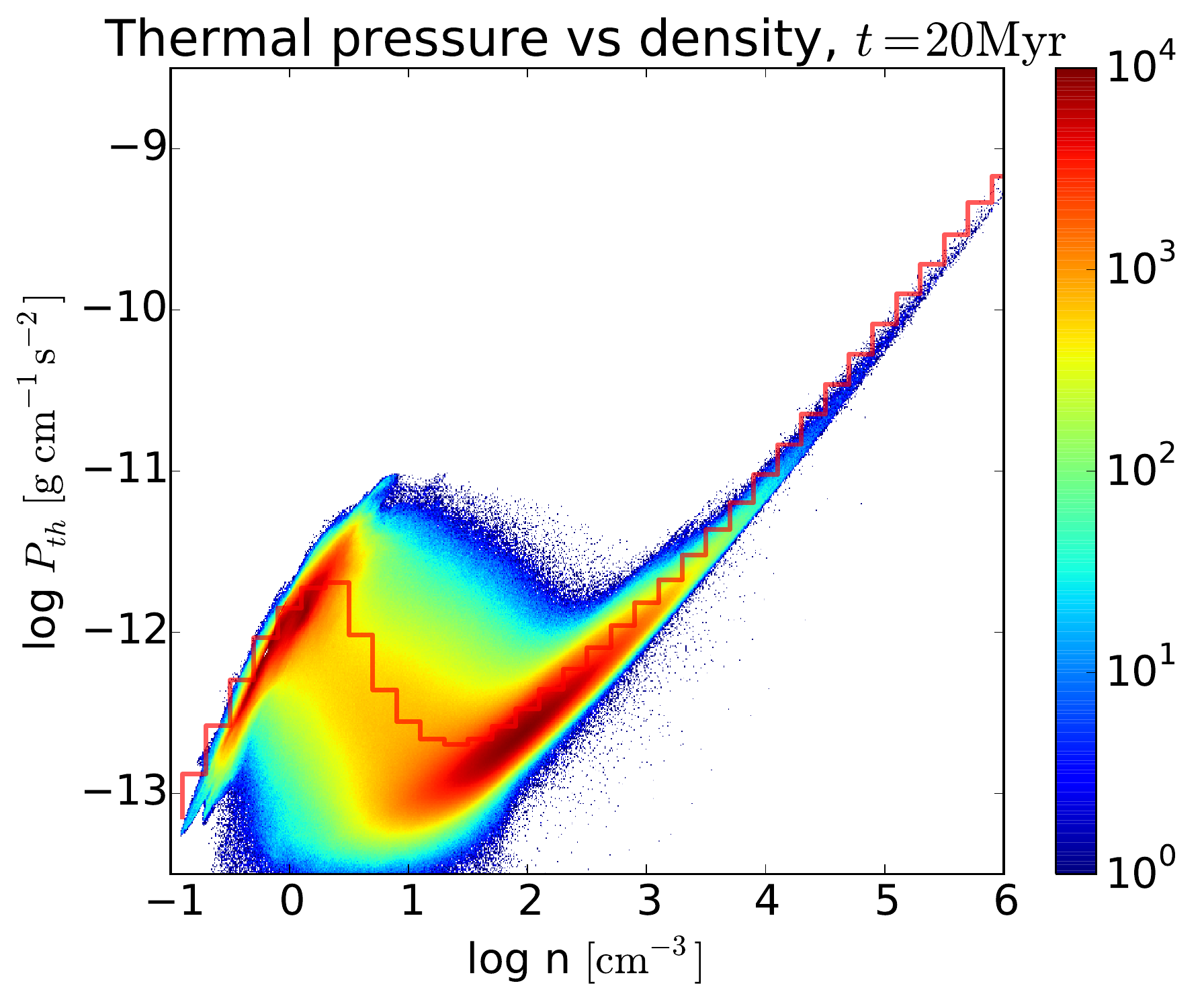}}\\
{\includegraphics[width=7cm]{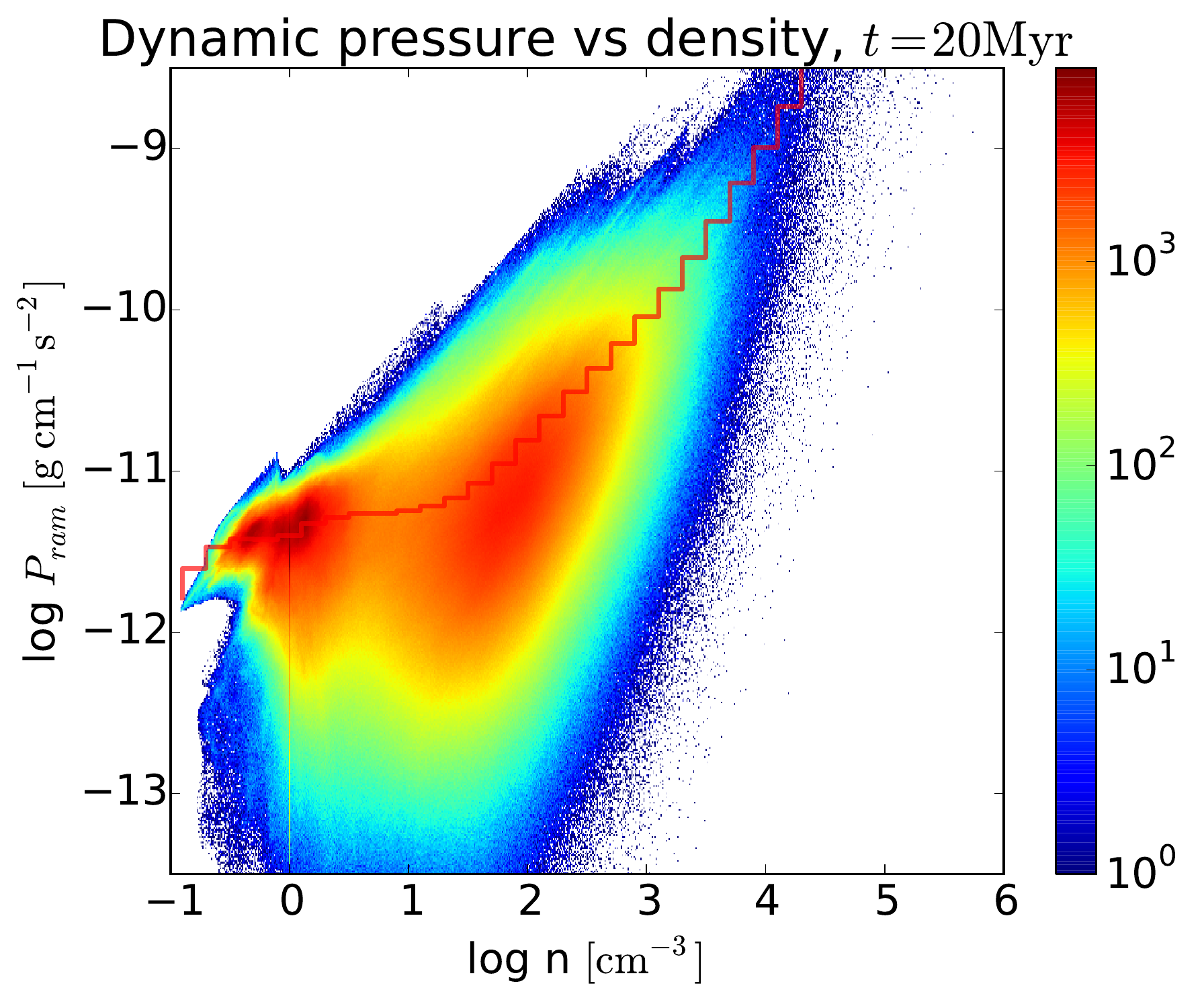}}\\
{\includegraphics[width=7cm]{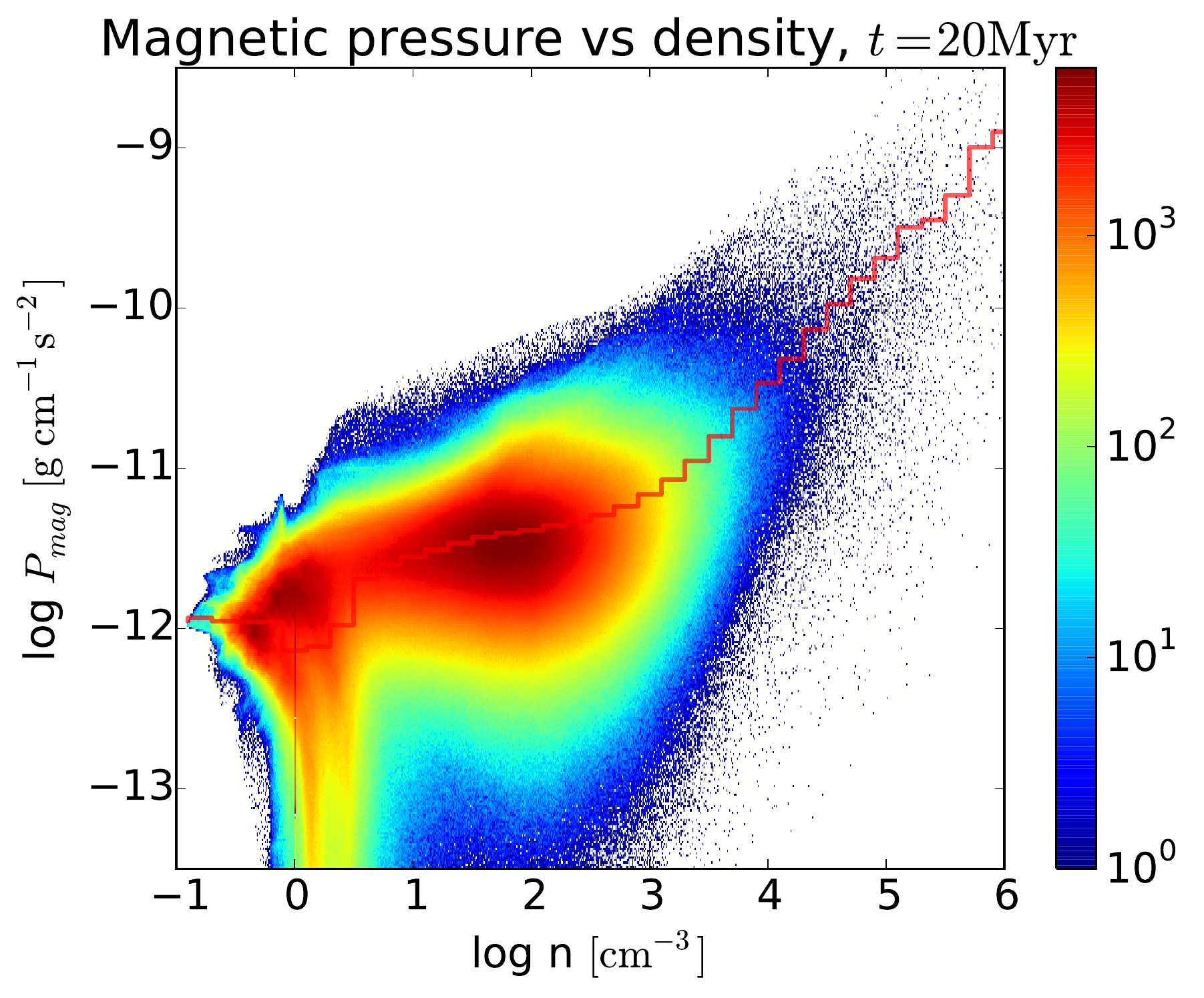}}
\caption{Different pressures in the ISM vs density. The top panel shows the thermal pressure, $P_{th}=nkT$, the middle panel the ram pressure, $P_{ram} = \rho V^2$, and the
bottom panel the magnetic pressure, $P_{mag} = B^2/8\pi$ in the simulation box. The red line shows the mean value per density bin. 
The colour scale indicates the number of points. }
\label{rapport_pressions}
\end{figure}

\begin{figure}
\centering
\includegraphics[width=7.7cm]{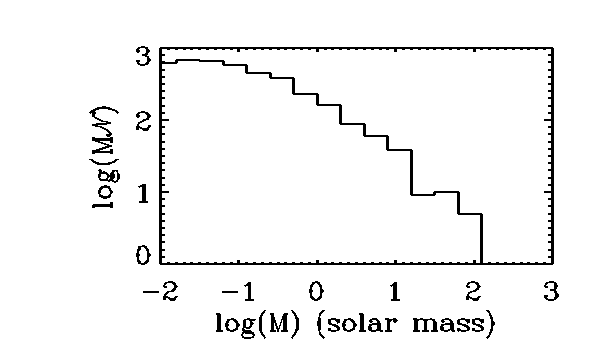}\\
\includegraphics[width=7.7cm]{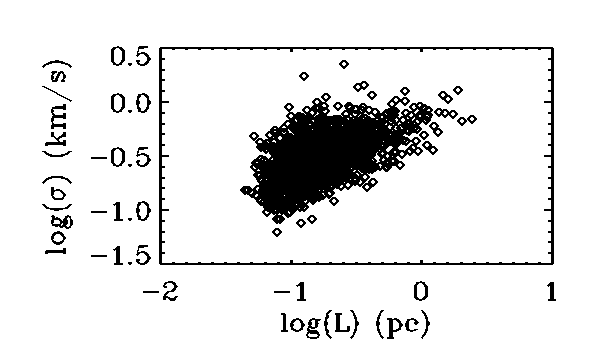}\\
\includegraphics[width=7.7cm]{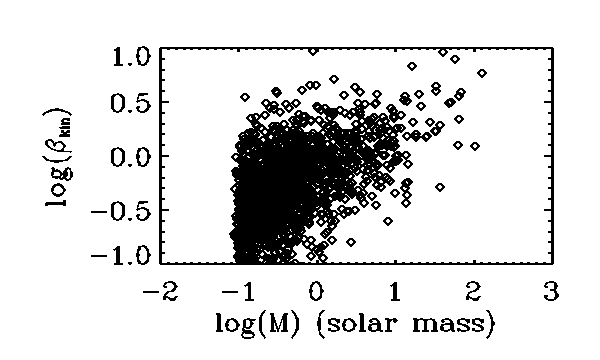} \\
\includegraphics[width=7.7cm]{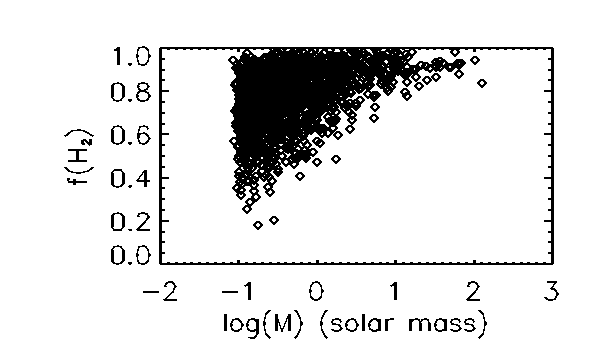} 
\caption{Statistics of clumps. The top panel shows the mass spectrum, the second panel their 
internal velocity dispersion as a function of size, the third panel the kinetic $\beta = P_{ram}/P_{mag}$ , and 
the bottom panel displays the H$_2$ fraction as a function of mass.}
\label{fig_clump}
\end{figure}

\section{Numerical setup and initial conditions}
\label{setup}

As already discussed, colliding flows are a practical ansatz to gather matter to form a molecular cloud 
\citep{inoue+2012, audit2005, vazquezsemadeni2007, heitsch_et_al_2005, heitsch+06, heitsch_et_al_2008, micic2013} to mimic large-scale converging flows, as in Galactic spiral arms or in super-bubble collisions. 

\subsection{General setup}

The setup is very similar to the one that \citet{valdivia2014} used. The size of the simulation box is $L=50~\mathrm{pc}$ and it is initially uniformly filled with atomic gas, with initial number density $n_\mathrm{tot} = 1~\mathrm{cm^{-3}}$ and initial temperature $T=8000~\mathrm{K}$. Atomic gas, with the same number density and temperature, is injected from the left and right sides of the box with an average velocity $V_0 = 15~\mathrm{km\ s^{-1}}$, aligned with the $x$-axis, which is modulated by a function of amplitude $\epsilon= 0.5$, producing a slightly turbulent profile, as in \citet{audit2005}. We use periodic boundary conditions for the remaining sides. The gas is 
initially uniformly magnetised, with a moderate magnetic field ($\sim 2.5~\mathrm{\micro G}$) aligned with the inflowing gas.
This configuration avoids boundary issues and facilitates the building of the cloud. Introducing an angle between the magnetic and the velocity 
fields would be more realistic, but this very simplified framework requires relatively small angles 
\citep{henne2000,kortgen2015, inoue+2012}. 

\subsection{Radiative transfer and dust shielding}
Molecular clouds are embedded in the interstellar radiation field (ISRF), which is assumed to be isotropic and constant throughout the space \citep{habing1968}. The whole simulation is embedded in an external and isotropic UV field that is assumed to be monochromatic, of strength $G_0 = 1.7$ in Habing units, and which heats the gas. UV photons are assumed to enter the computational box from all directions and undergo interactions with dust grains in their pathways before reaching any cell. The intensity of the UV field varies from one point to another according to the dust shielding factor $\chi$, which is calculated at each point of the simulation box for each timestep using our \textup{\emph{\textup{tree-based method}}}, which is fully described in \citet{valdivia2014}. Hence, the strength of the local UV field, after dust shielding, can be written $\overline{G}_0~=~\chi G_0$, where $\chi = \langle e ^{-2.5 \mathrm{A_V}} \rangle = \langle e ^{-1.3 \times 10^{-21} \mathcal{N}_\mathrm{tot}} \rangle $, is the mean shielding factor due to dust. This mean value is calculated using a fixed number of directions. In our previous work \citep{valdivia2014} we have shown that the number of directions is not crucial for the dynamical evolution of the cloud. For simplicity and numerical efficiency, 
we therefore used 12 directions ($M = 3$ intervals for the polar angle and $N = 4$ for the azimuthal angle), which has turned out to be an excellent compromise
between accuracy and efficiency.

We adapted the code to also compute the self-shielding (stated by Eq.~\ref{fshield}). More precisely, along each direction we 
computed not only the total gas density, but also the H$_2$ column density from which we computed the shielding coefficient, $f_\mathrm{shield}$, and the 
extinction due to dust, $\tau_{d,1000}$. Since the UV flux is assumed to be isotropic, the final photodissociation rate is obtained 
by taking the mean value over all directions. We define $\chi _{\rm shield}$ as
\begin{equation}
\chi _{\rm shield} = { k_{\mathrm{ph}} \over k_{\mathrm{ph,0}} }= \langle e^{-\tau_{d,1000}}  f_\mathrm{shield} (\mathcal{N}_{\mathrm{H_2}} )   \rangle.   \label{kdest_mean}
\end{equation}

\subsection{Numerical resolution and runs performed}
To perform our simulations we employed the adaptive mesh refinement code RAMSES \citep{teyssier2002,fromang2006}. RAMSES solves the 
MHD equations using the HLLD Riemann solvers. It preserves the nullity of the divergence of the magnetic field thanks to 
the use of a staggered mesh. To solve Eq.~(\ref{eqH2}), we used operator splitting, solving first for the advection 
(which is identical to the conservation equation) and then subcycling to solve for the right-hand side. 
In AMR codes the refinement is done on a cell-by-cell basis. In our simulations, the refinement criterion is the density. 
When a cell reaches a given density threshold, it is split into eight smaller cells, each one having the same mass and volume. 
The process is repeated recursively until the maximum resolution is reached. 
In our fiducial run, we
 allowed two AMR levels $\ell_{min} = 8$ and $\ell_{max} = 10$, leading to an equivalent numerical resolution of $1024^3$ cells and an 
effective spatial resolution of about $0.05~\mathrm{pc}$. For the first refinement level (from level $\ell=8$ to $\ell=9$) 
the density threshold is $n_\mathrm{thresh} = 50~\mathrm{cm^{-3}}$ and for the next refinement, the density threshold is $n_\mathrm{thresh} = 100~\mathrm{cm^{-3}}$. 
To investigate numerical convergence, a particularly crucial issue when chemistry is considered, we also performed a high-resolution run
for which the resolution was doubled (that is to say, using levels from 9 to 11 with the same refinement criterion).

We ran our simulations for about 20 Myr. We note that these calculations are somewhat demanding because of the short timesteps 
induced by high temperatures, therefore it was not possible to run the high-resolution simulations up to this point. This run
reaches up to 15 Myr. 
To check further for convergence, we also performed low-resolution runs (128$^3$ and 256$^3$ , see Appendix) that we compared with our highest 
resolution simulations.
We also performed complementary runs in which we modified the physics of H$_2$ formation to better understand how 
H$_2$ is formed. More precisely, in these runs we suppressed
the H$_2$ formation above a certain density threshold,

\begin{figure}
\centering
\includegraphics[width=7.8cm]{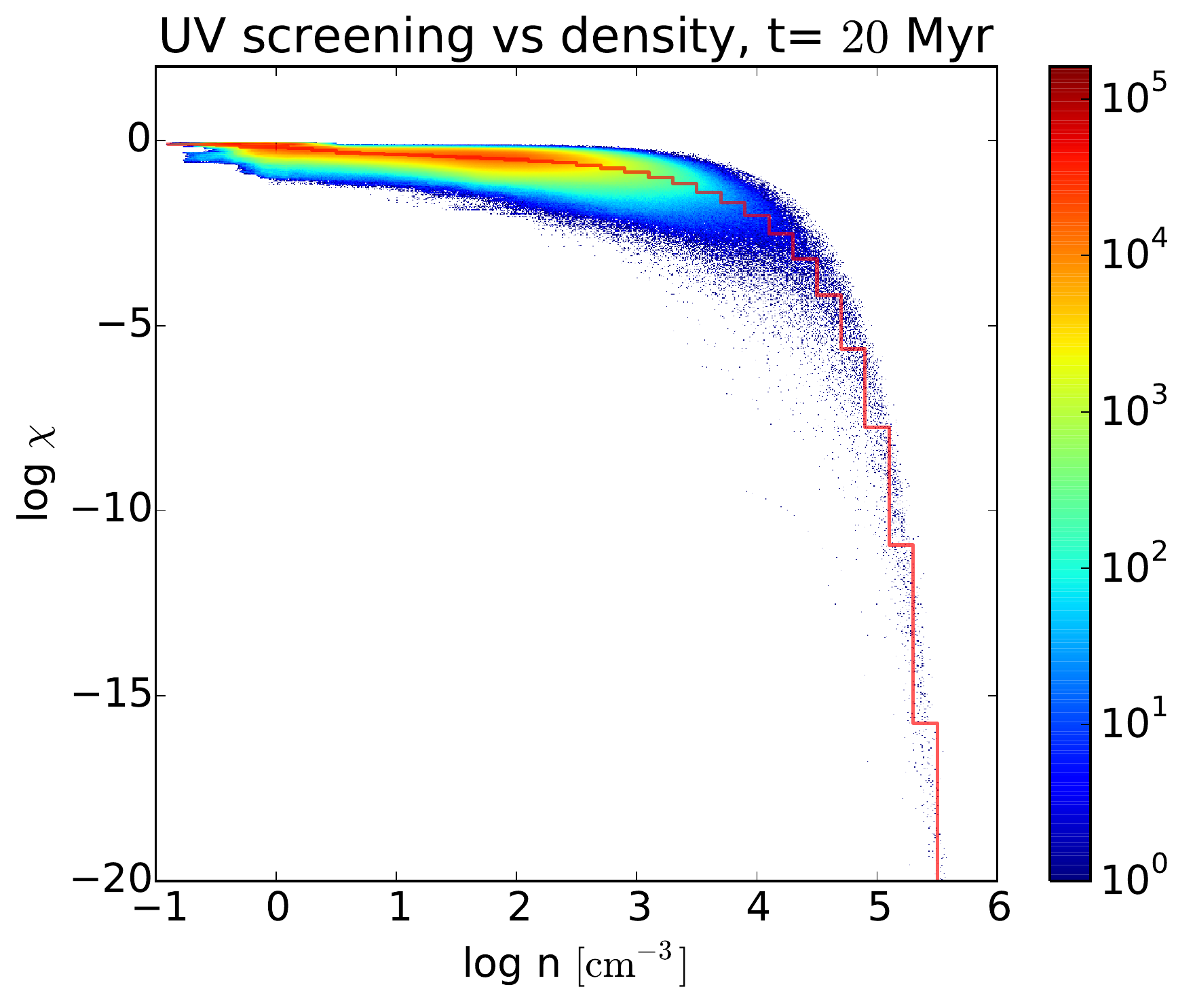} \\
\includegraphics[width=7.8cm]{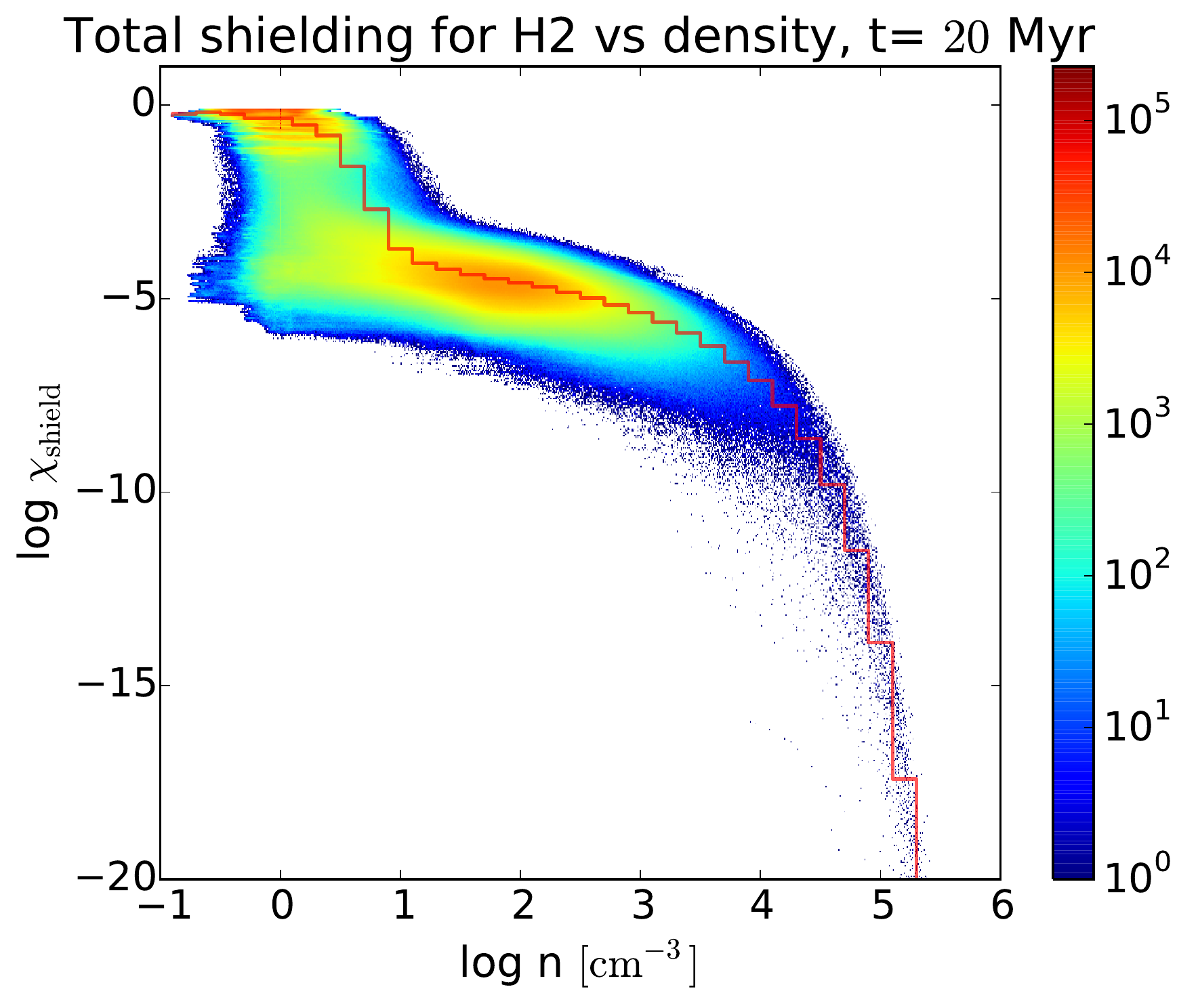}
\caption{UV screening factor vs density (top panel) and total shielding coefficient for H$_2$ vs density
(bottom panel) at $t=20~\mathrm{Myr}$. The red line shows the mean value per density bin, and the colour scale indicates the number of points.}
\label{UV_scatt}
\end{figure}

\begin{figure}
\centering
\includegraphics[width=7.8cm]{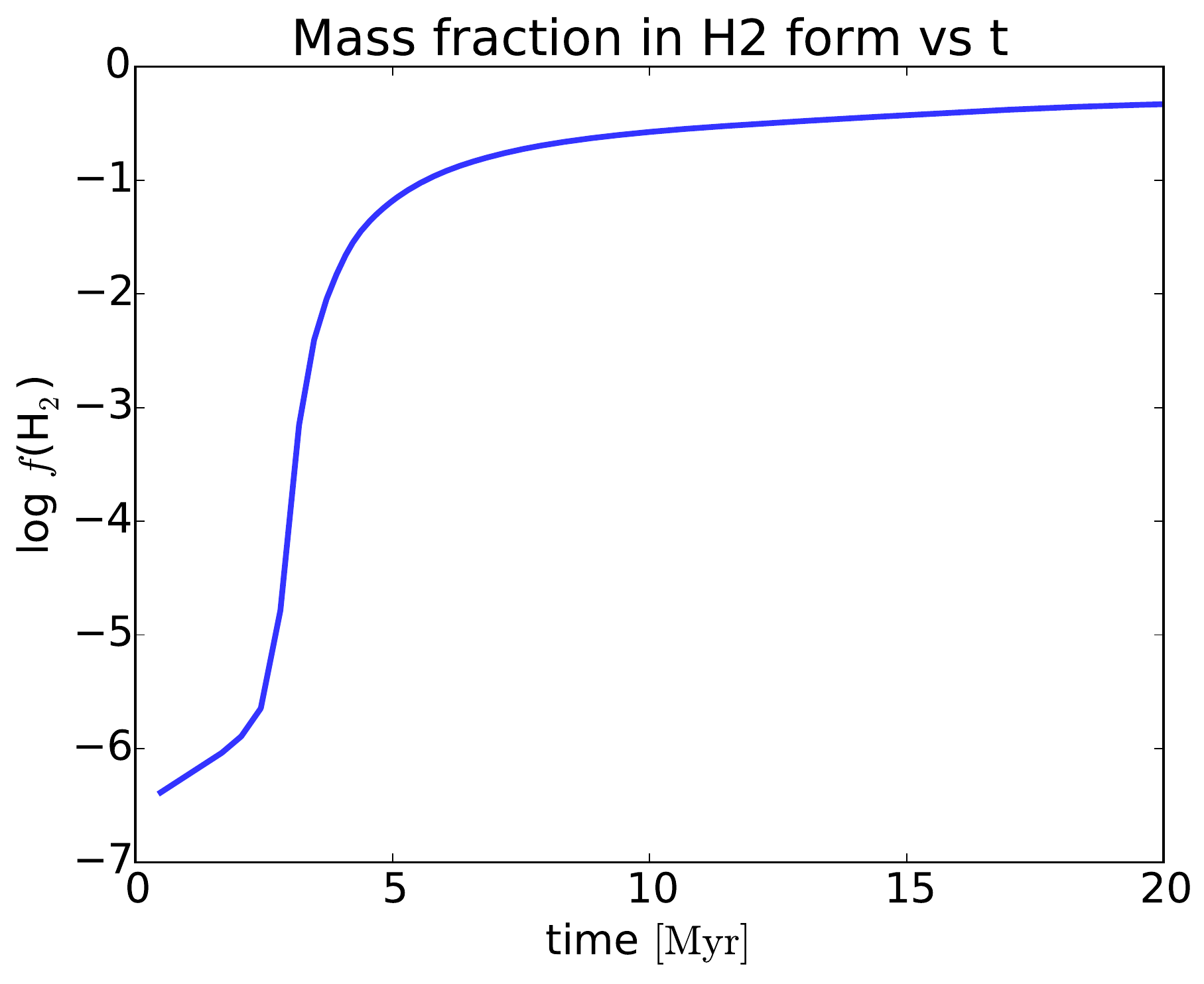}
\caption{Evolution of the molecular fraction in the simulation box as a function of time.}
\label{fH2_evo}
\end{figure}

\begin{figure}
\centering
\includegraphics[width=7.8cm]{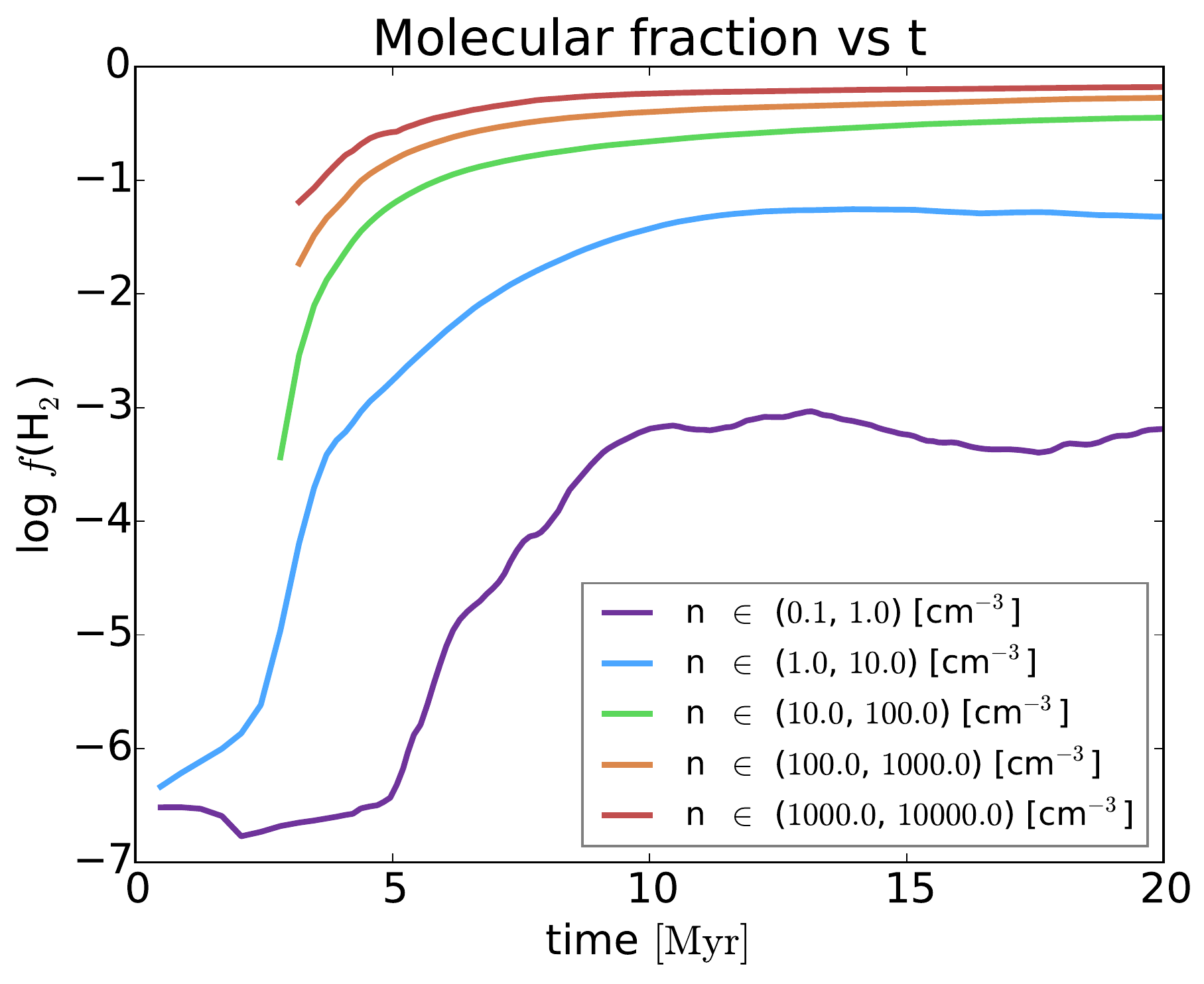}\\
\includegraphics[width=7.8cm]{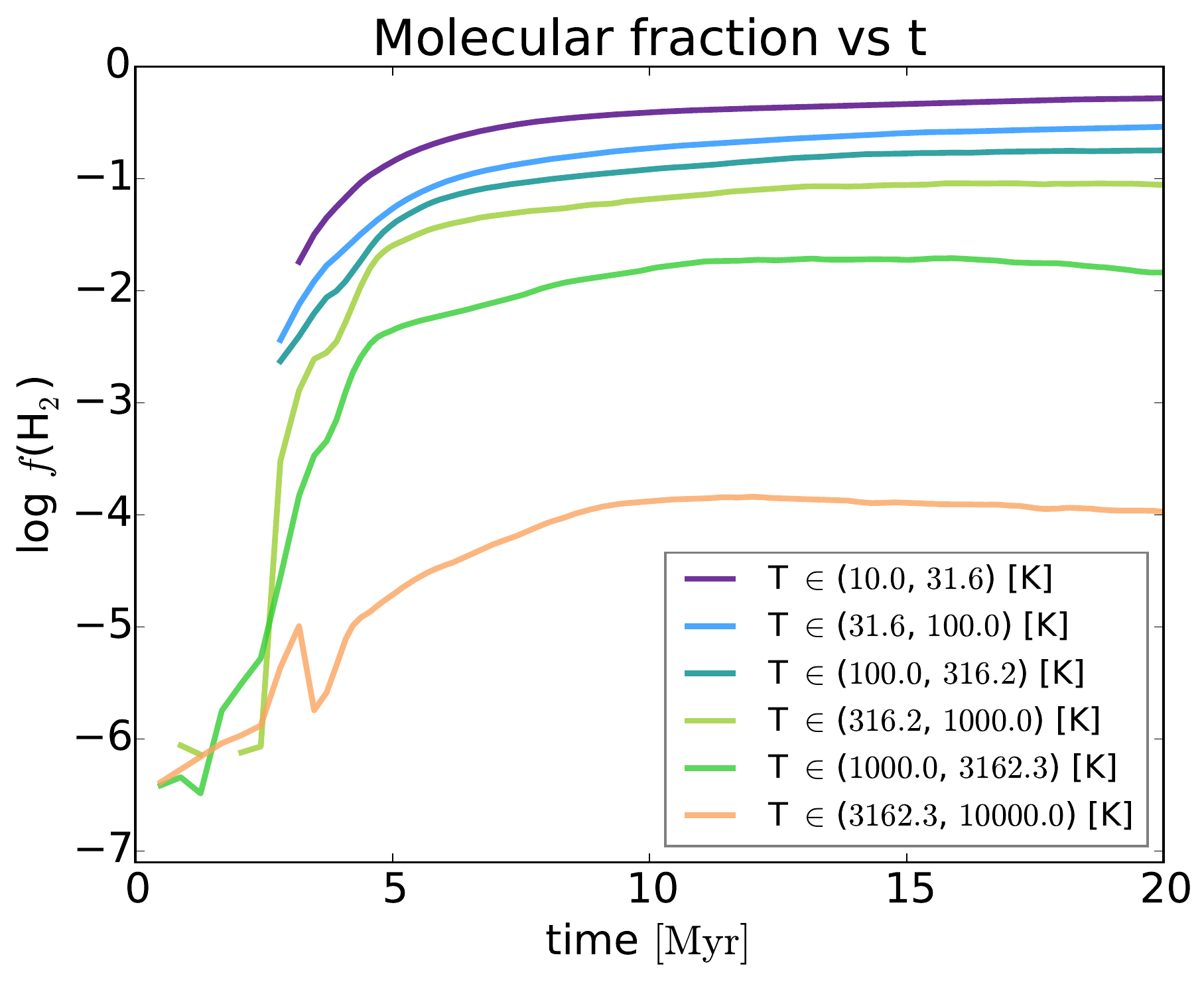}
\caption{Molecular fraction evolution. In the top panel we show
the evolution per density bin (the density increases from purple to red), in the bottom panel the evolution per temperature bin (the temperature increases from purple to salmon). }
\label{evof_tbin}
\end{figure}

\begin{figure*}[htb]
\centering
  \begin{tabular}{@{}cccc@{}}
    $t = 5~\mathrm{Myr}$ & $t = 10~\mathrm{Myr}$ & $t = 15~\mathrm{Myr}$ & $t = 20~\mathrm{Myr.}$\\
    \includegraphics[width=.23\textwidth]{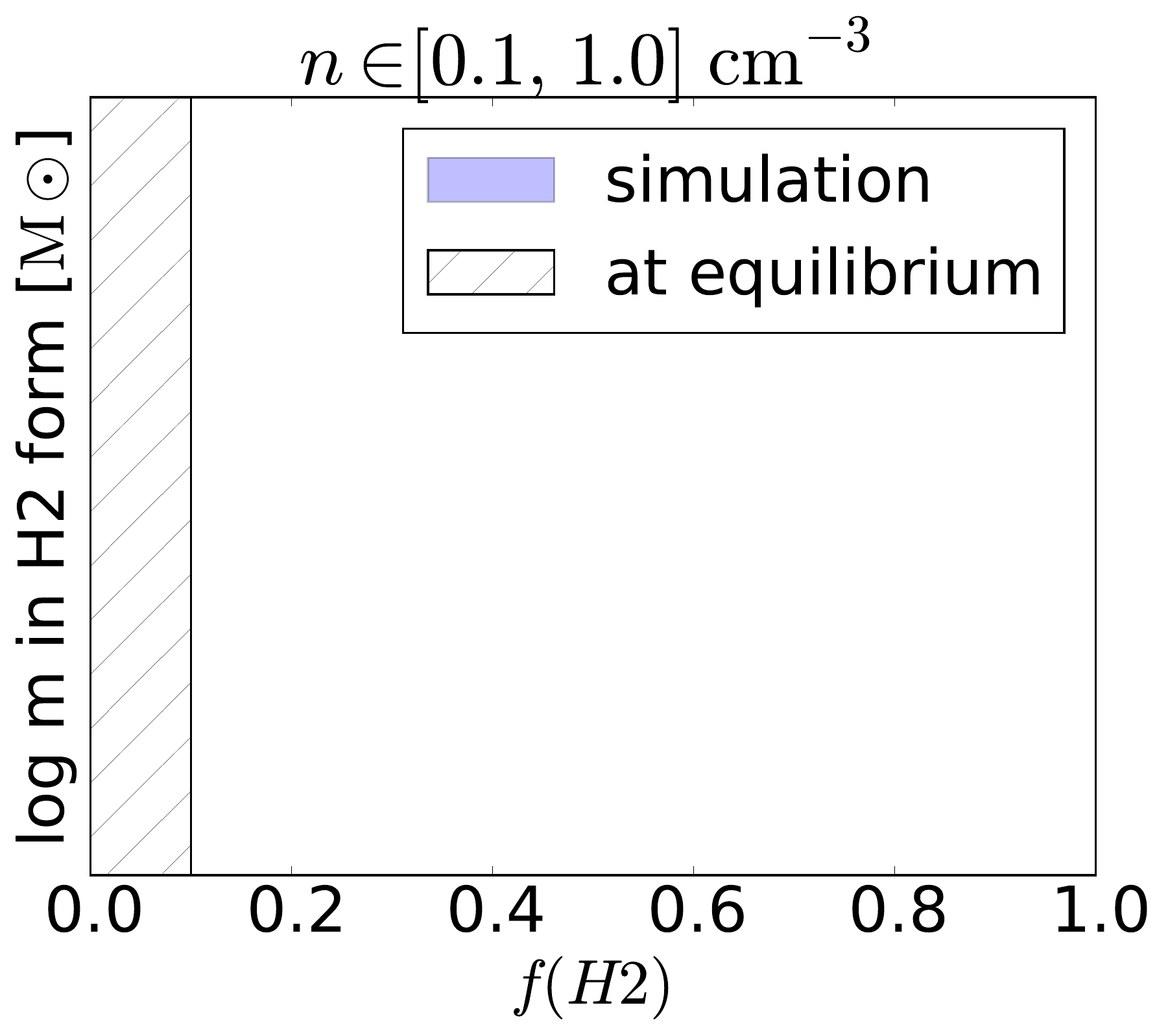} &
    \includegraphics[width=.23\textwidth]{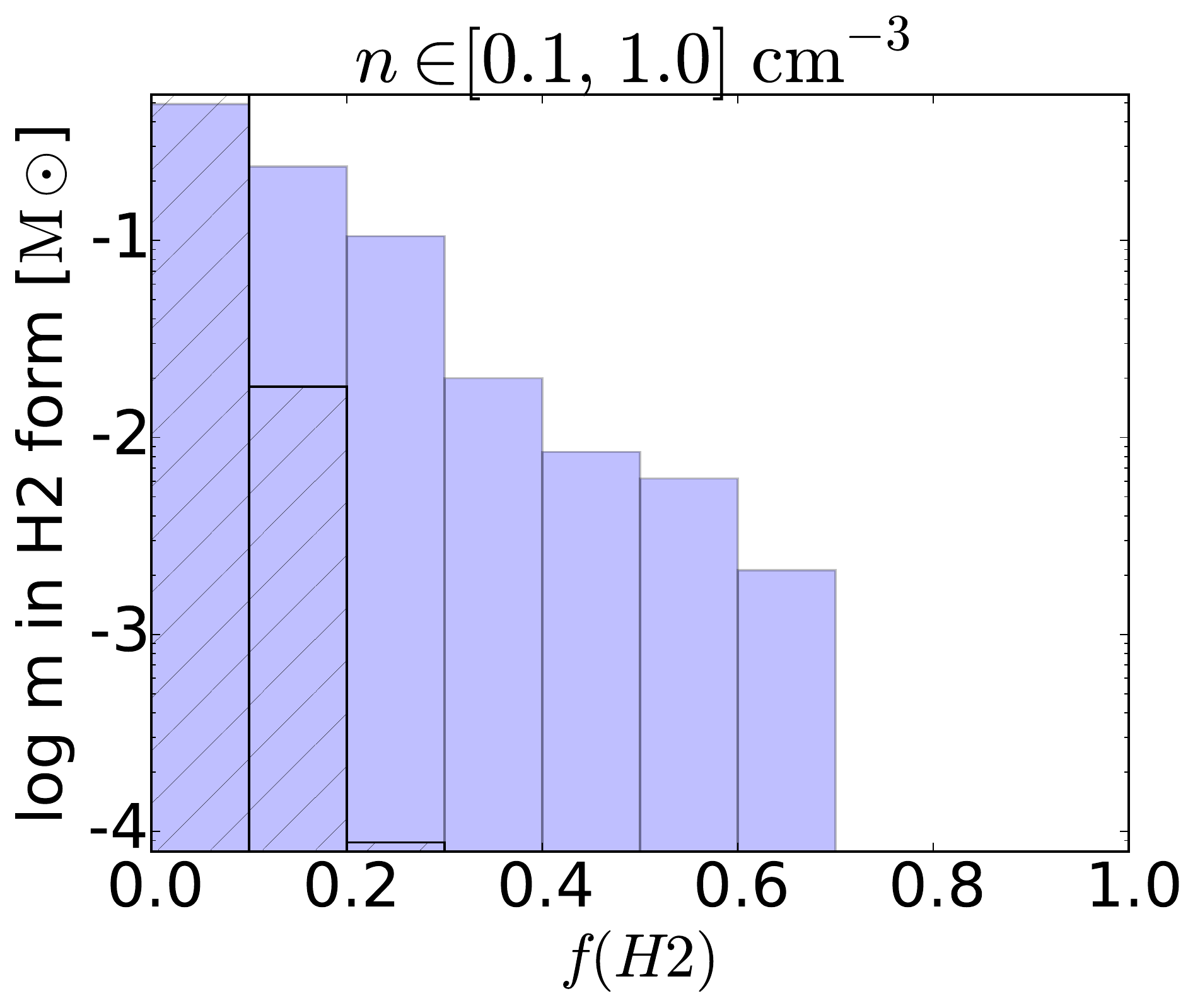} &
    \includegraphics[width=.23\textwidth]{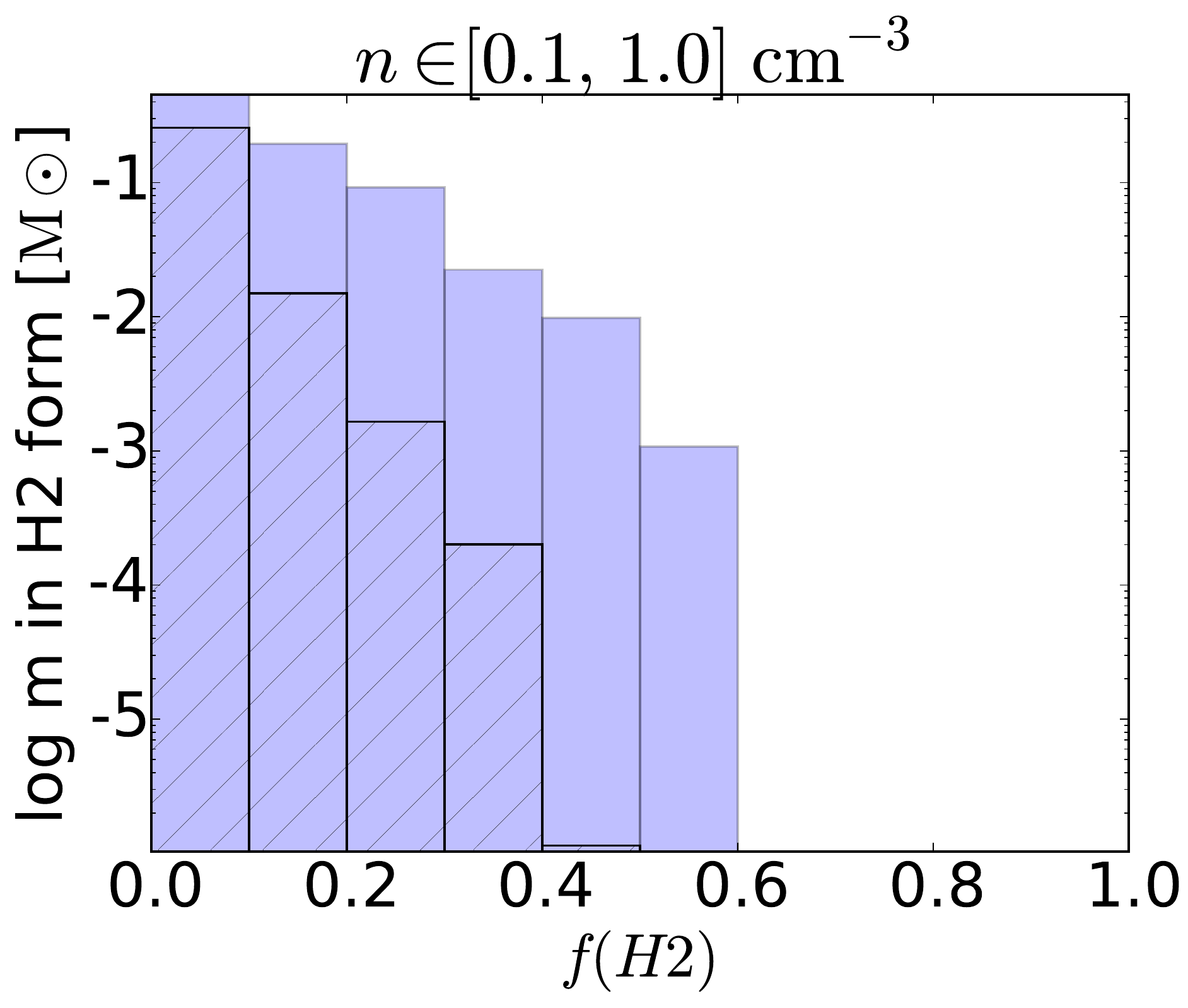} &
    \includegraphics[width=.23\textwidth]{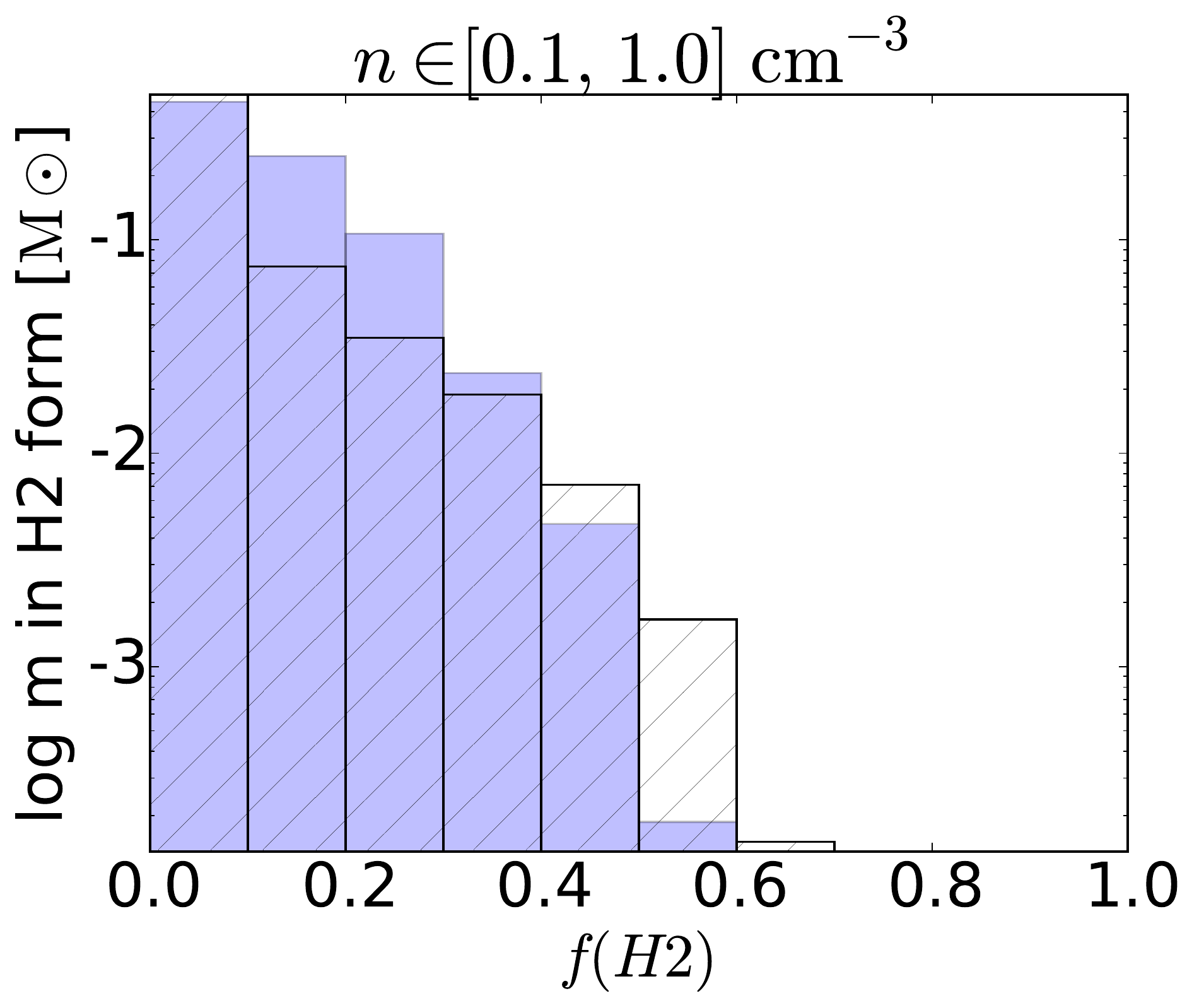}   \\
    \includegraphics[width=.23\textwidth]{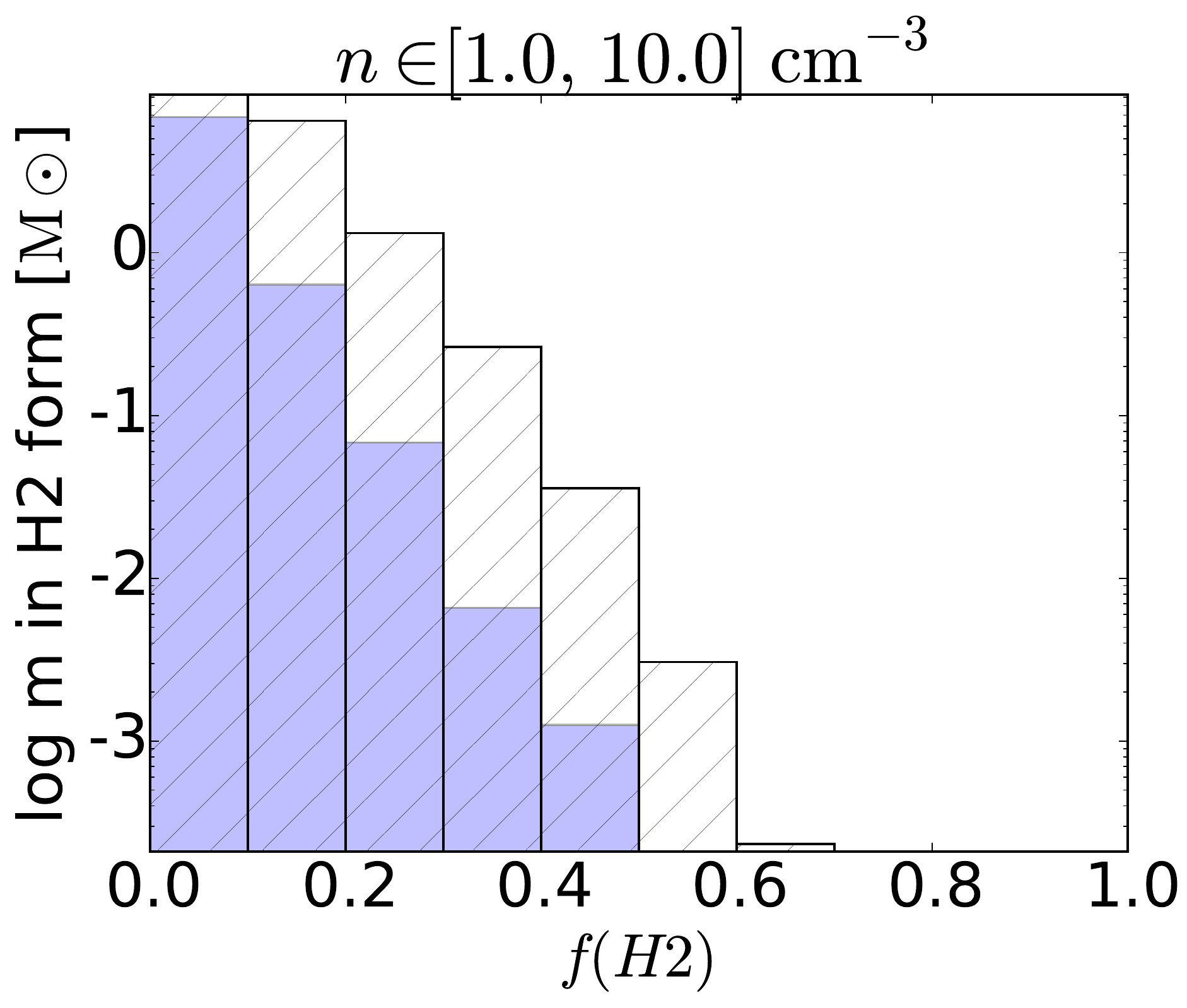} &
    \includegraphics[width=.23\textwidth]{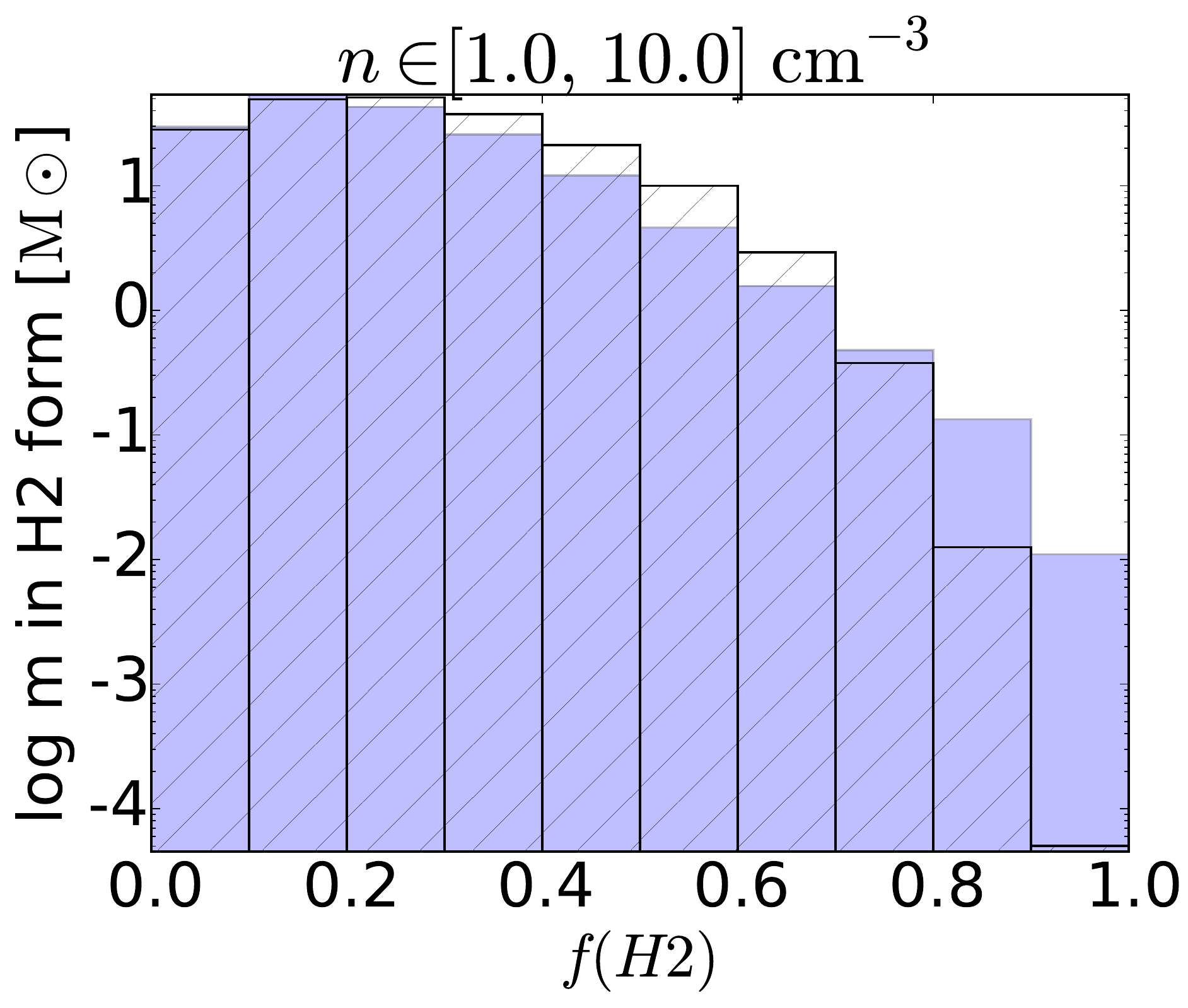} &
    \includegraphics[width=.23\textwidth]{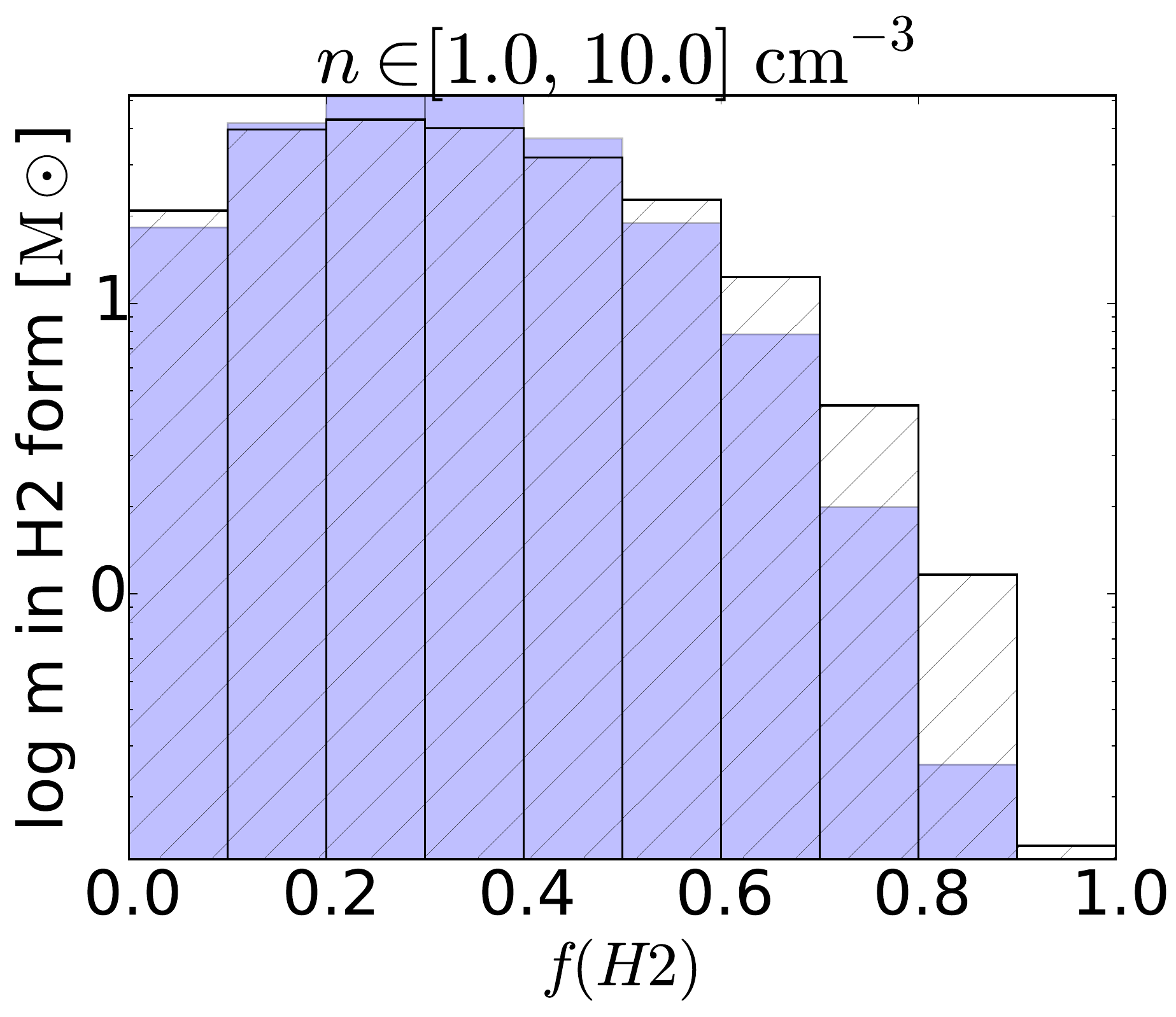} &
    \includegraphics[width=.23\textwidth]{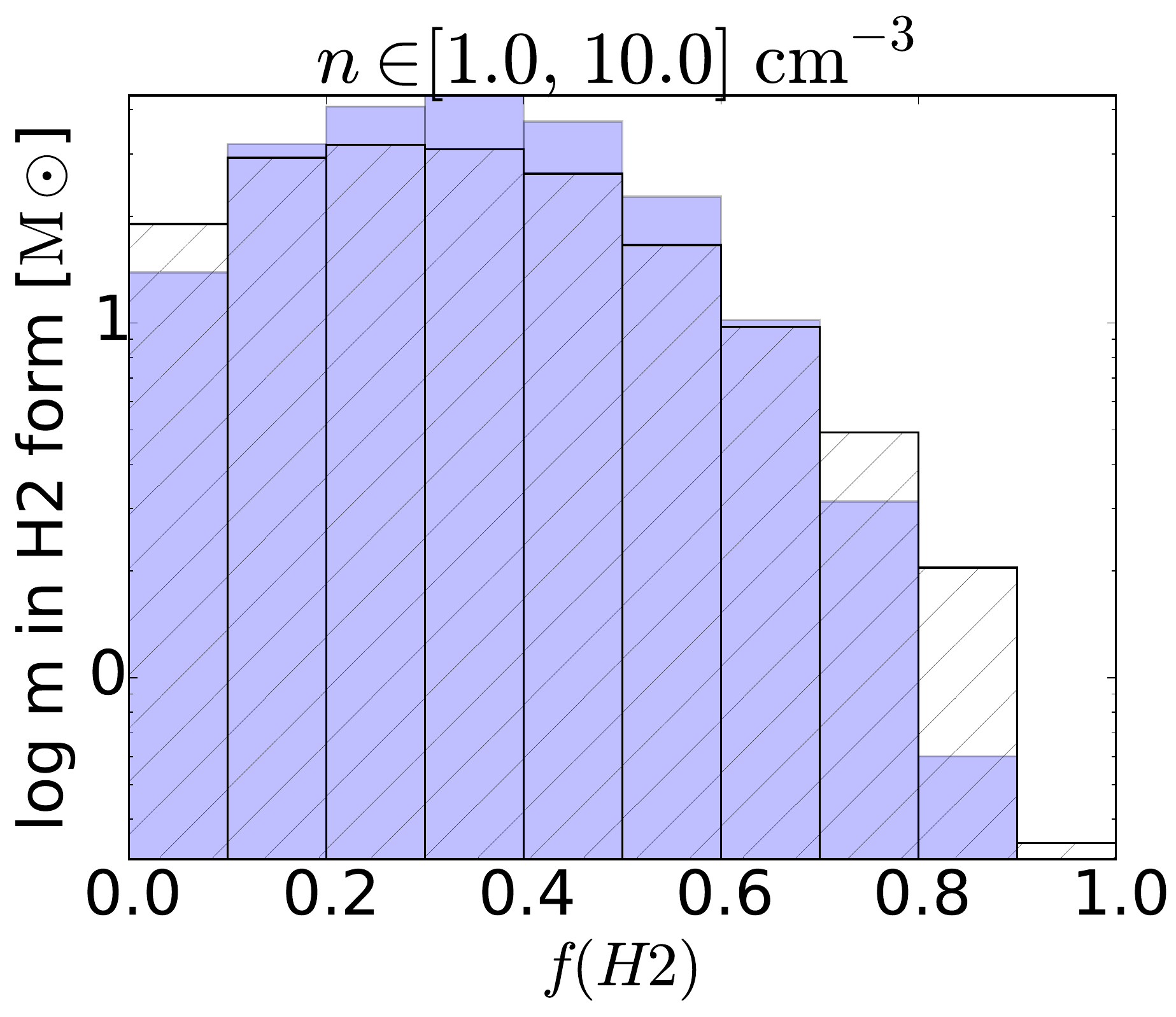}   \\
    \includegraphics[width=.23\textwidth]{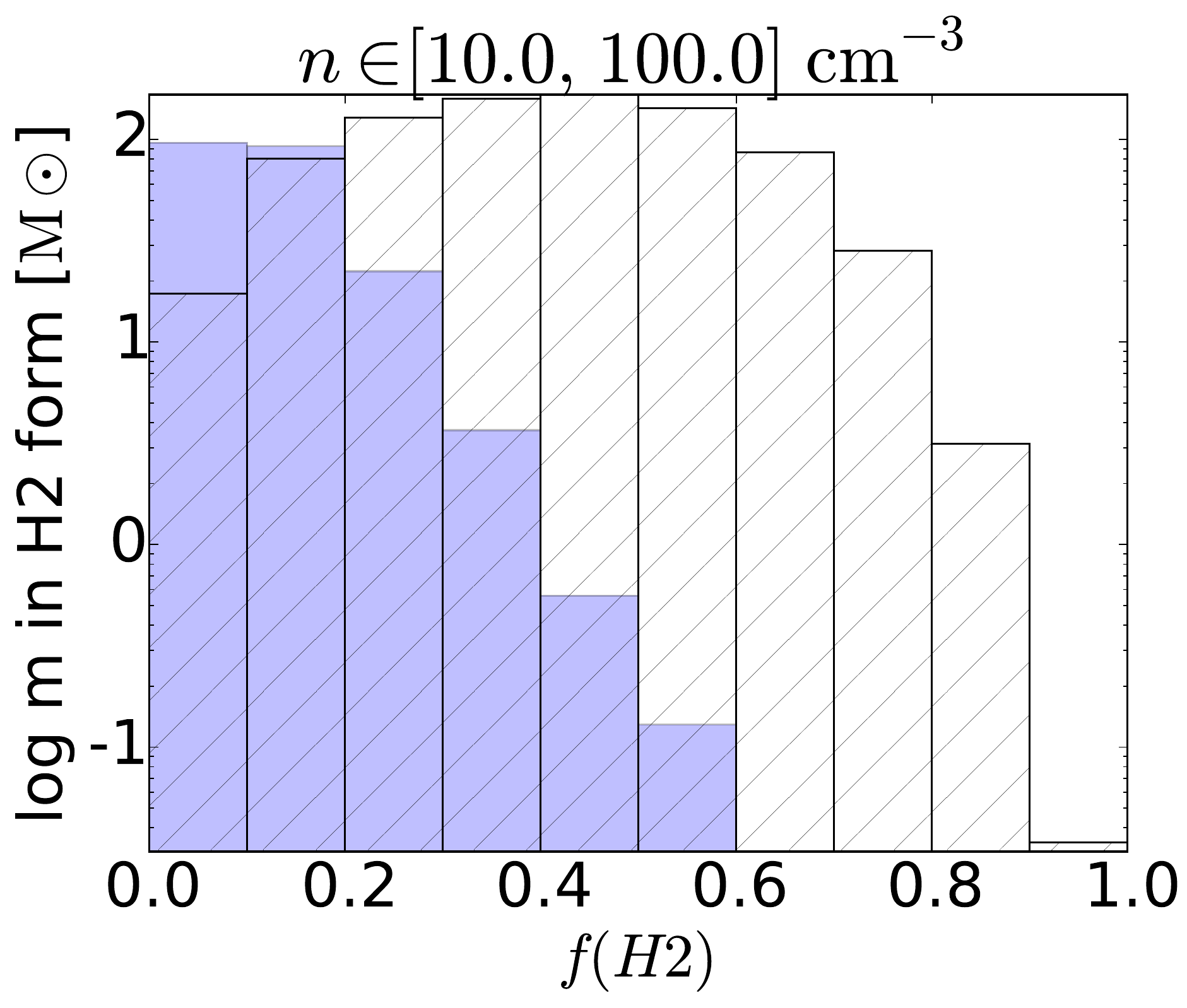} &
    \includegraphics[width=.23\textwidth]{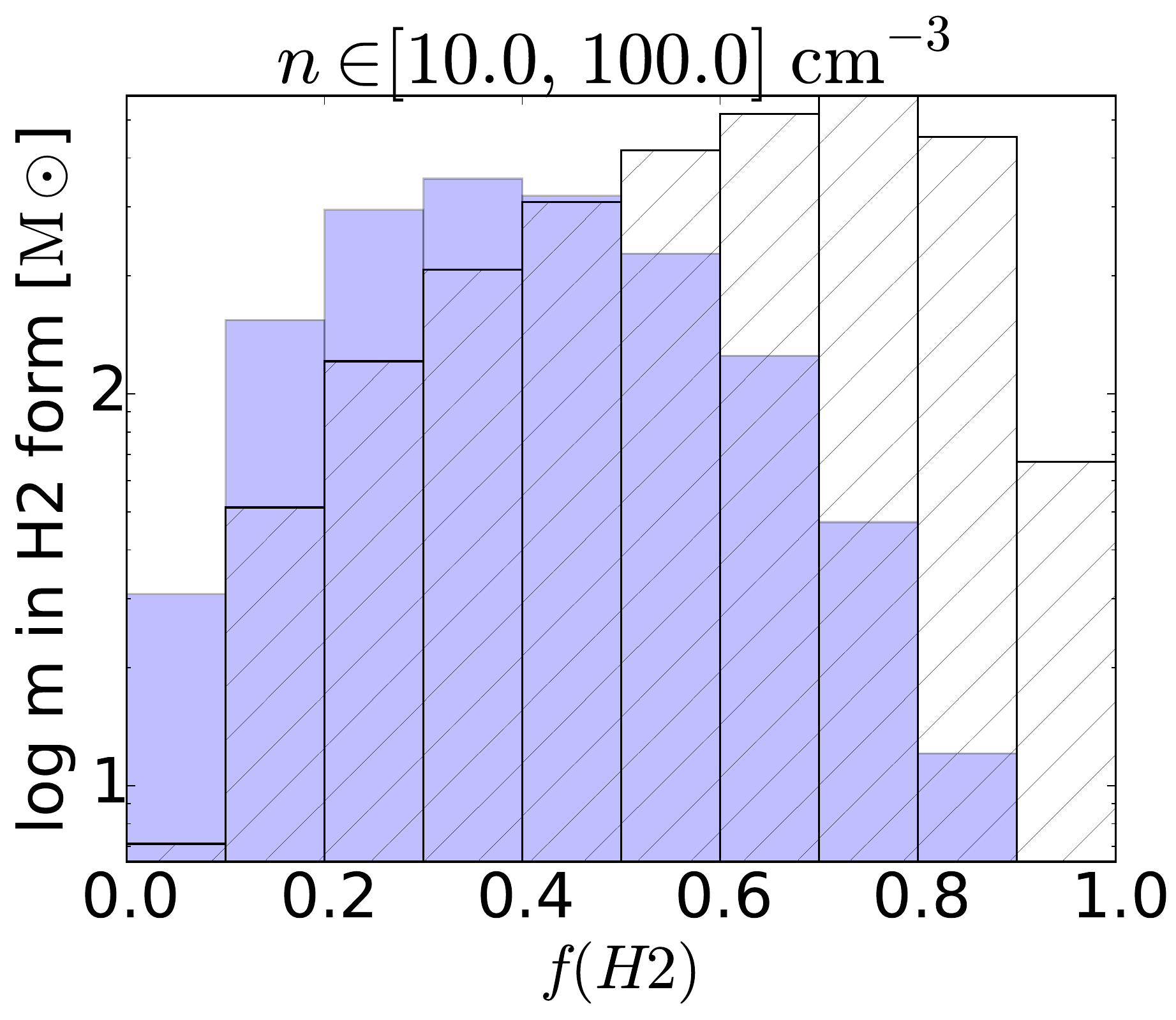} &
    \includegraphics[width=.23\textwidth]{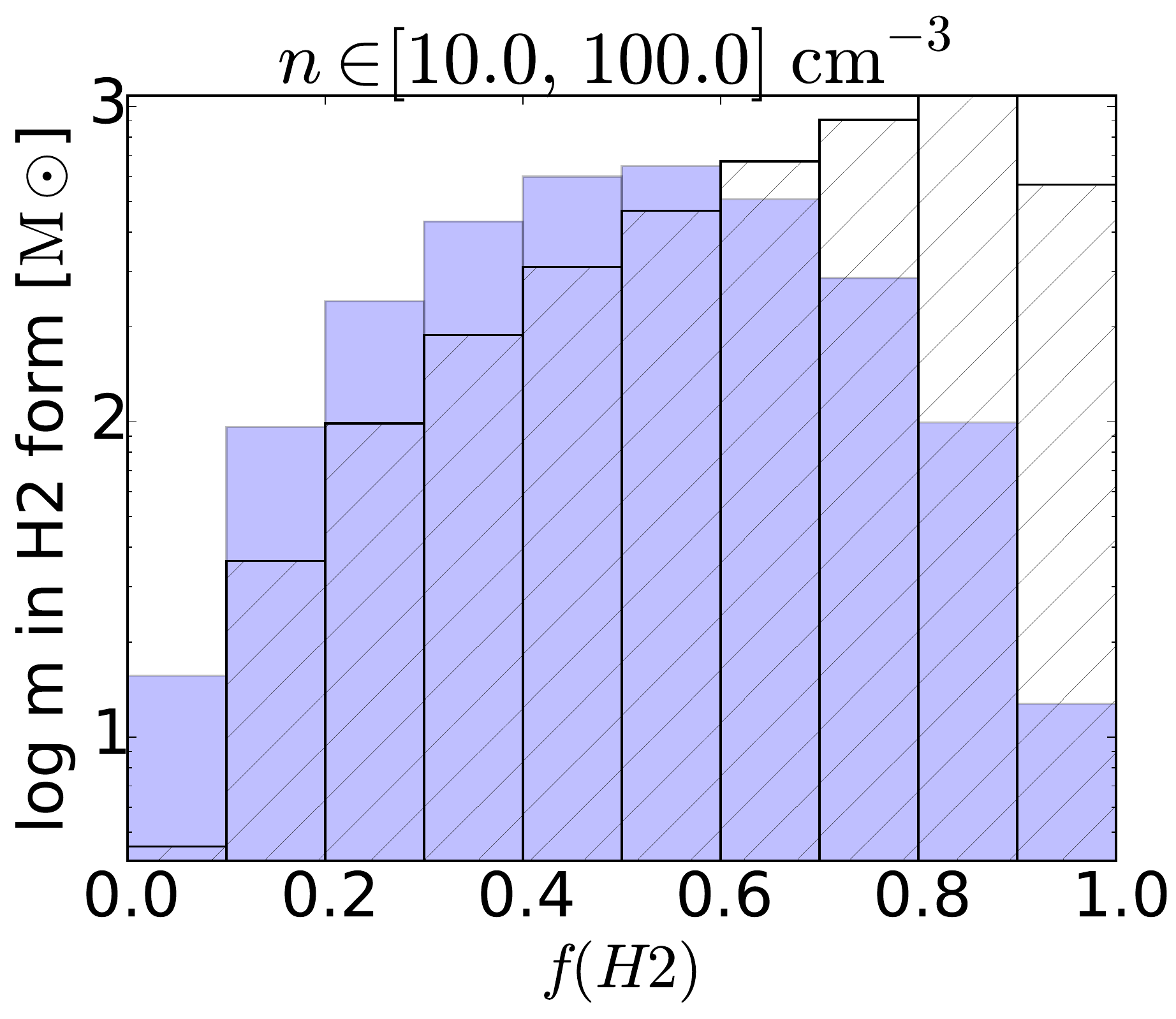} &
    \includegraphics[width=.23\textwidth]{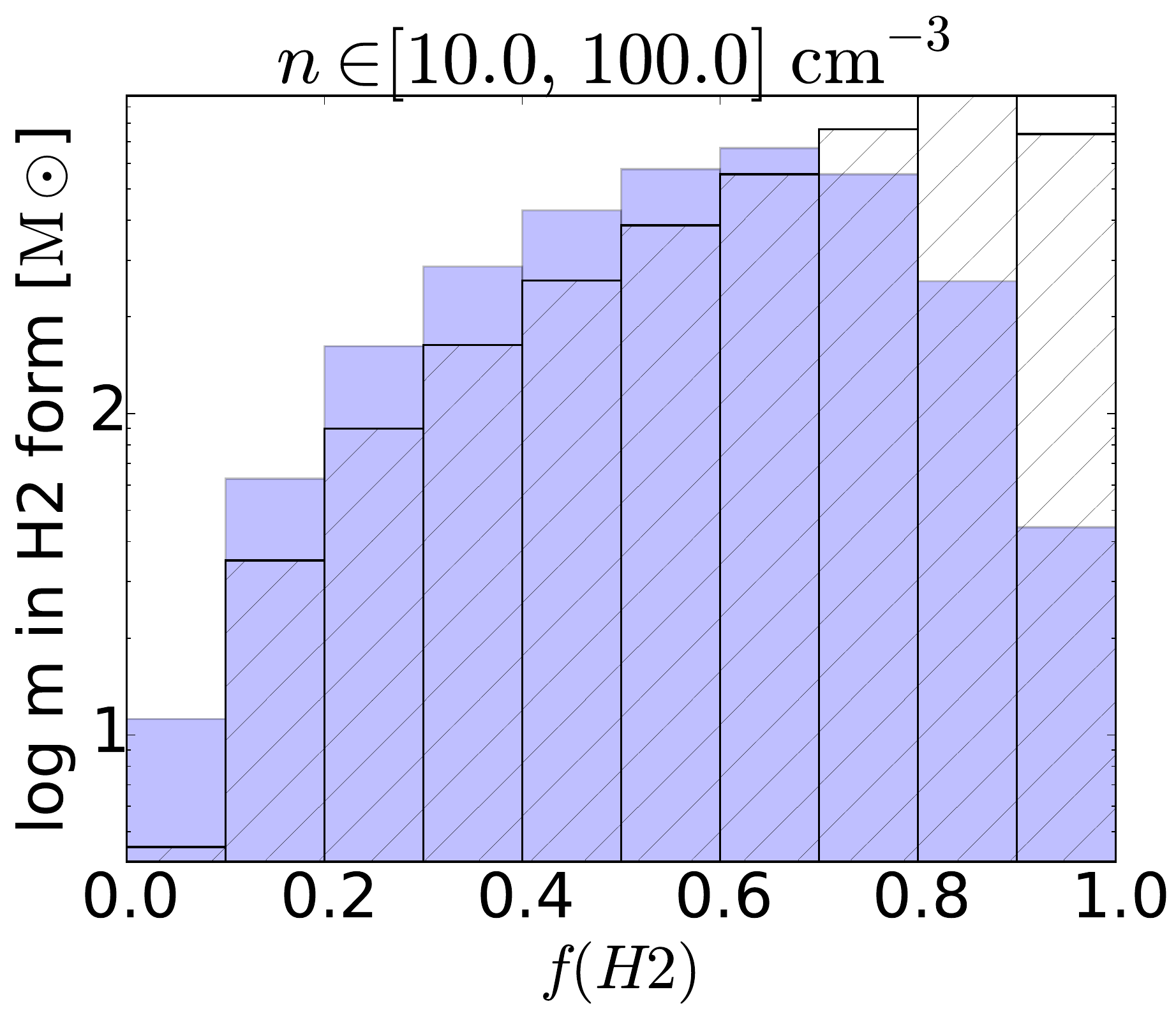}   \\
    \includegraphics[width=.23\textwidth]{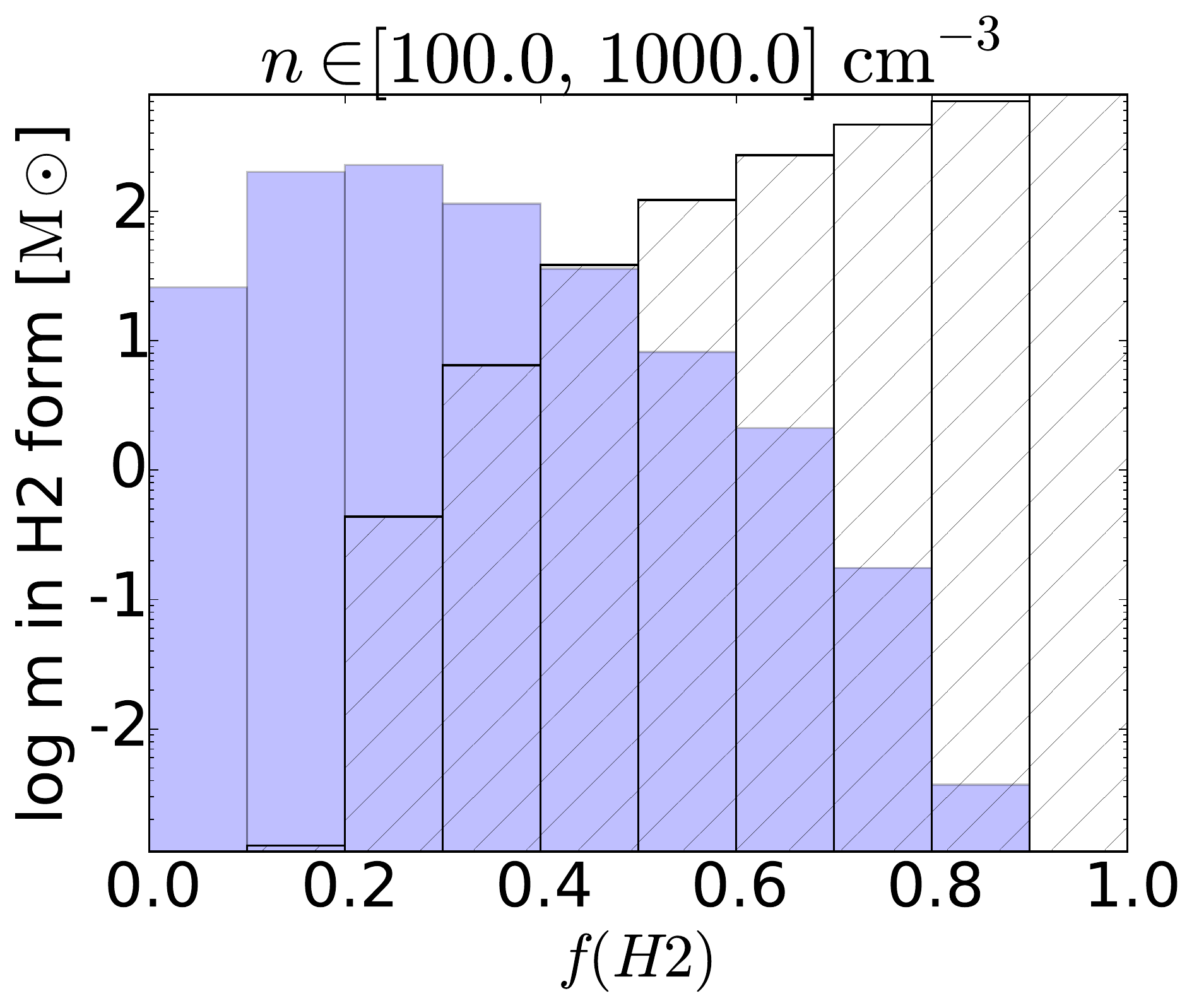} &
    \includegraphics[width=.23\textwidth]{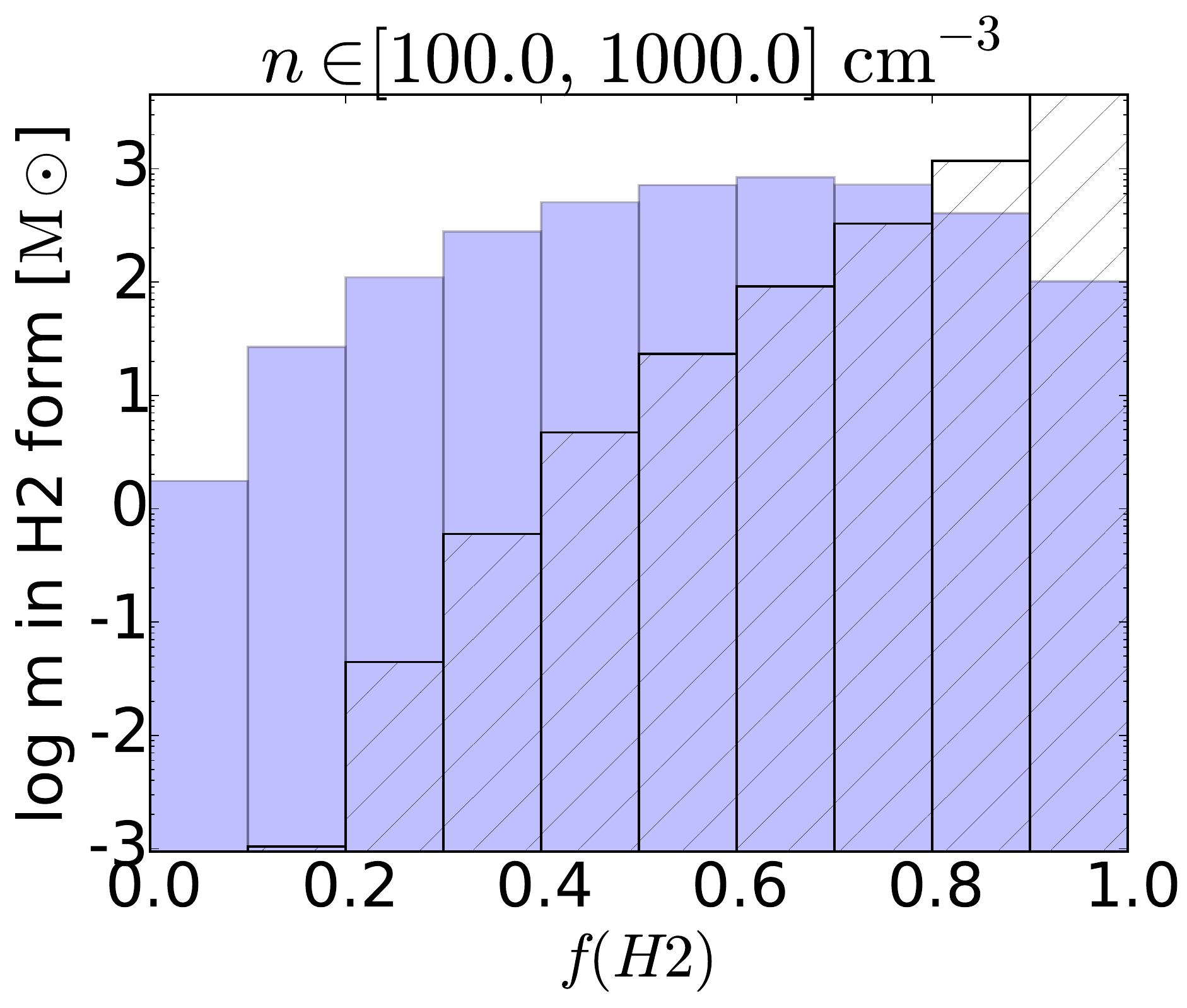} &
    \includegraphics[width=.23\textwidth]{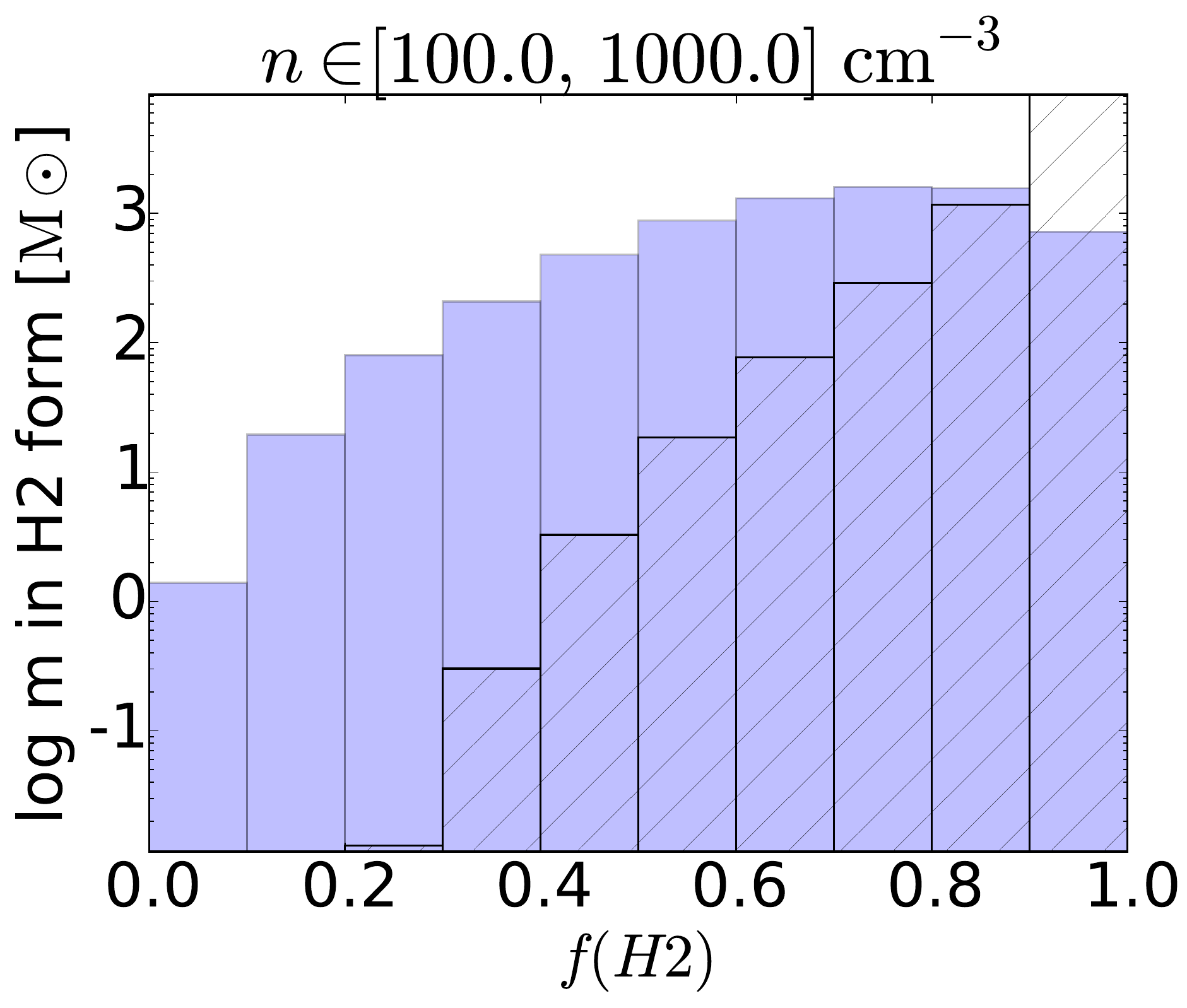} &
    \includegraphics[width=.23\textwidth]{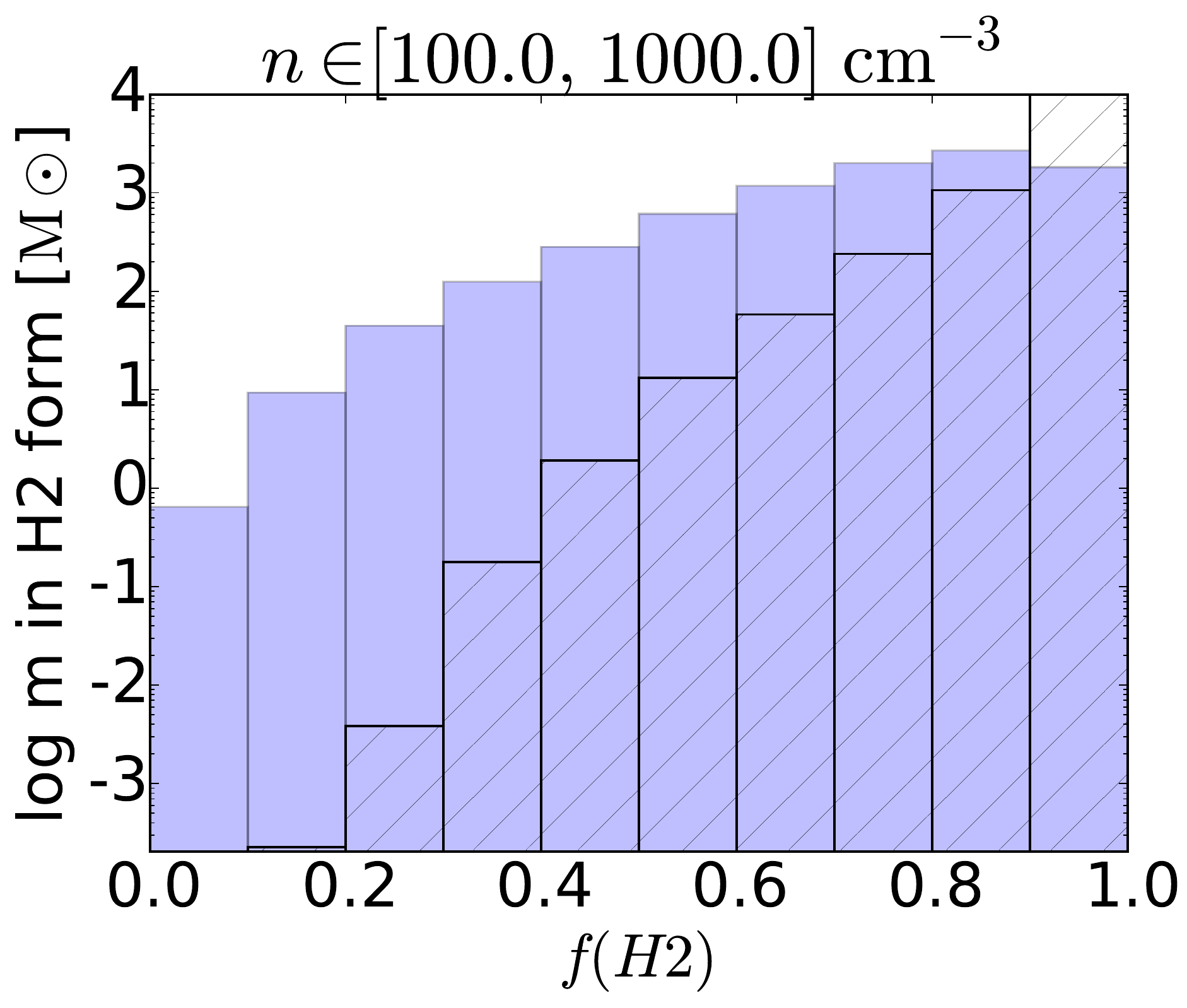}   \\
    \includegraphics[width=.23\textwidth]{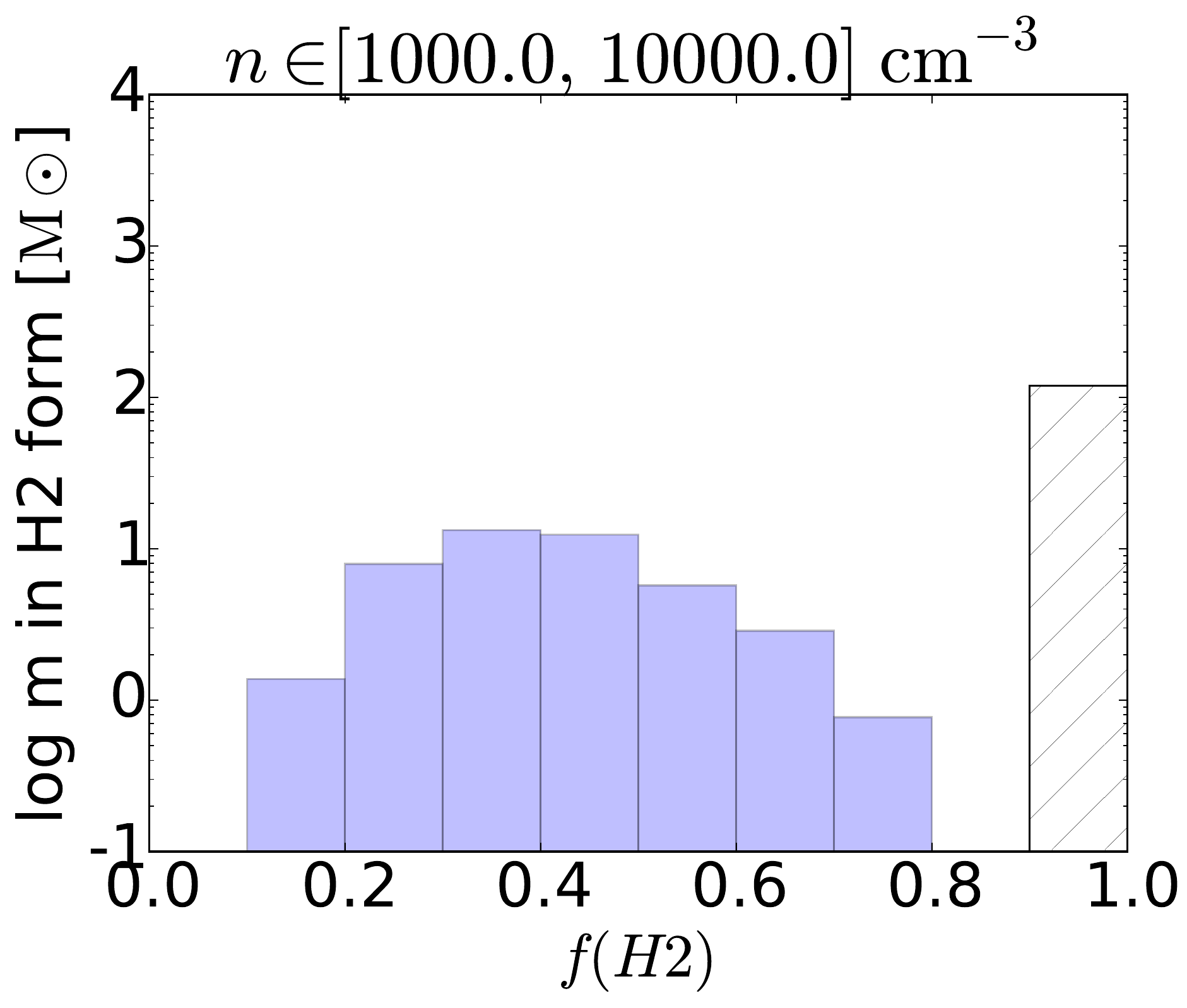} &
    \includegraphics[width=.23\textwidth]{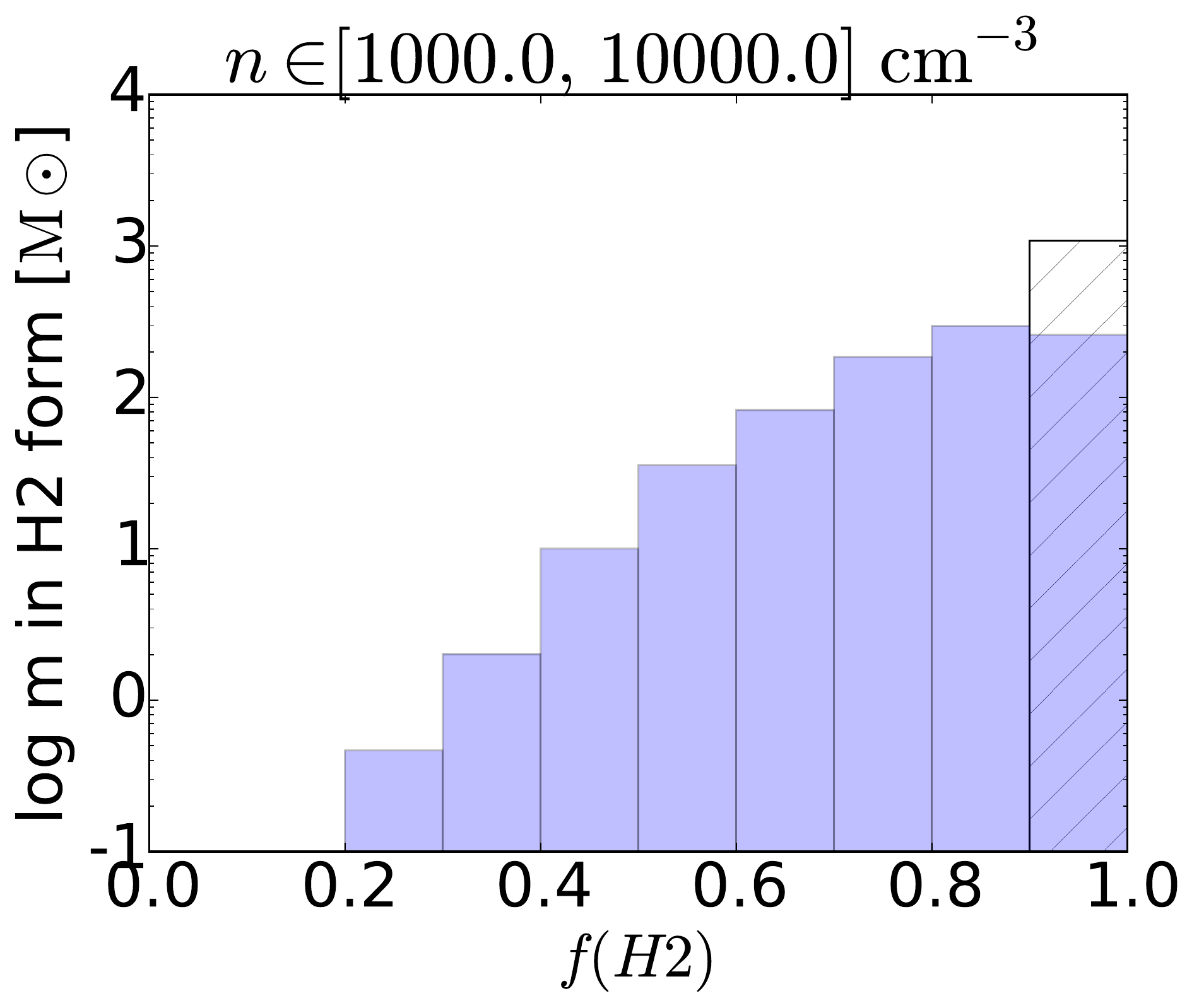} &
    \includegraphics[width=.23\textwidth]{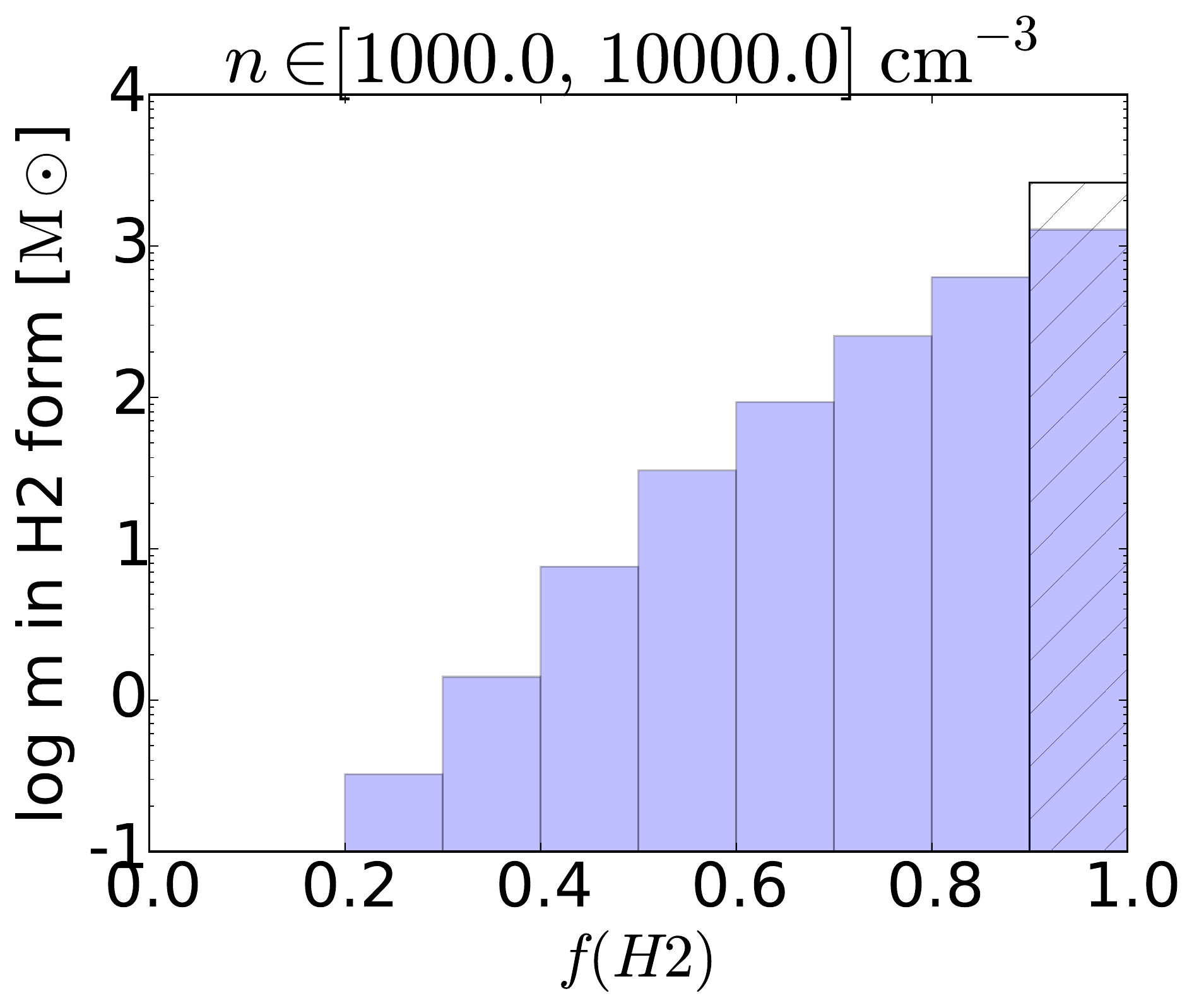} &
    \includegraphics[width=.23\textwidth]{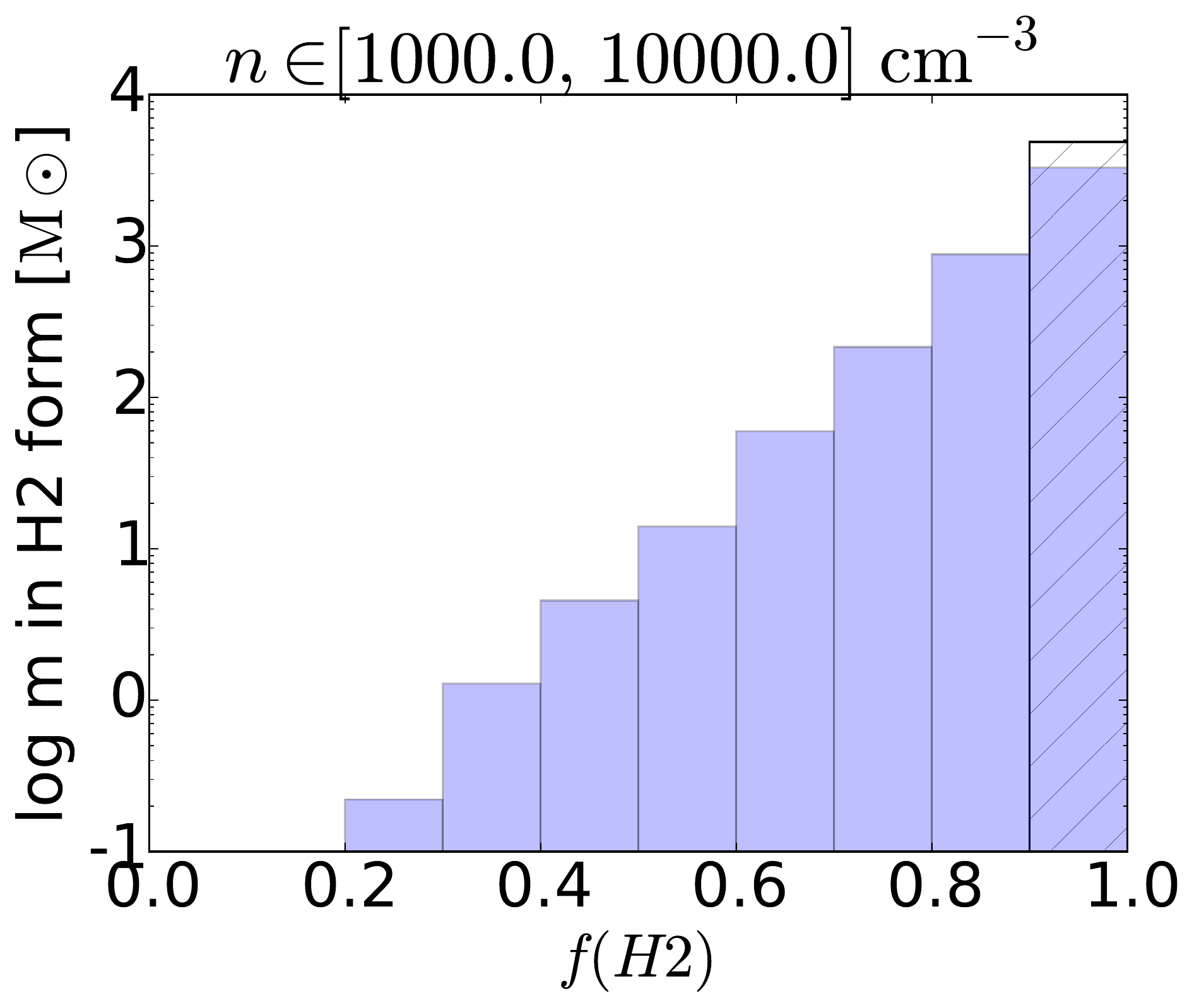}
  \end{tabular}
  \caption{Mass in H$_2$ form per density bin. Comparison between the simulation (blue) and the calculation at equilibrium (dashed) for different timesteps. From left to right: $t = 5, 10, 15,$ and $20~\mathrm{Myr}$. From top to bottom: density bins $n~\in~(0.1, 1), (1, 10), (10, 100), (100, 1000), (1000, 10000)~\mathrm{[cm^{-3}]}$. Note the change of vertical scale.}
\label{histo_n}
\end{figure*}

\begin{figure}%[htb]
\centering
   \includegraphics[width=7.5cm]{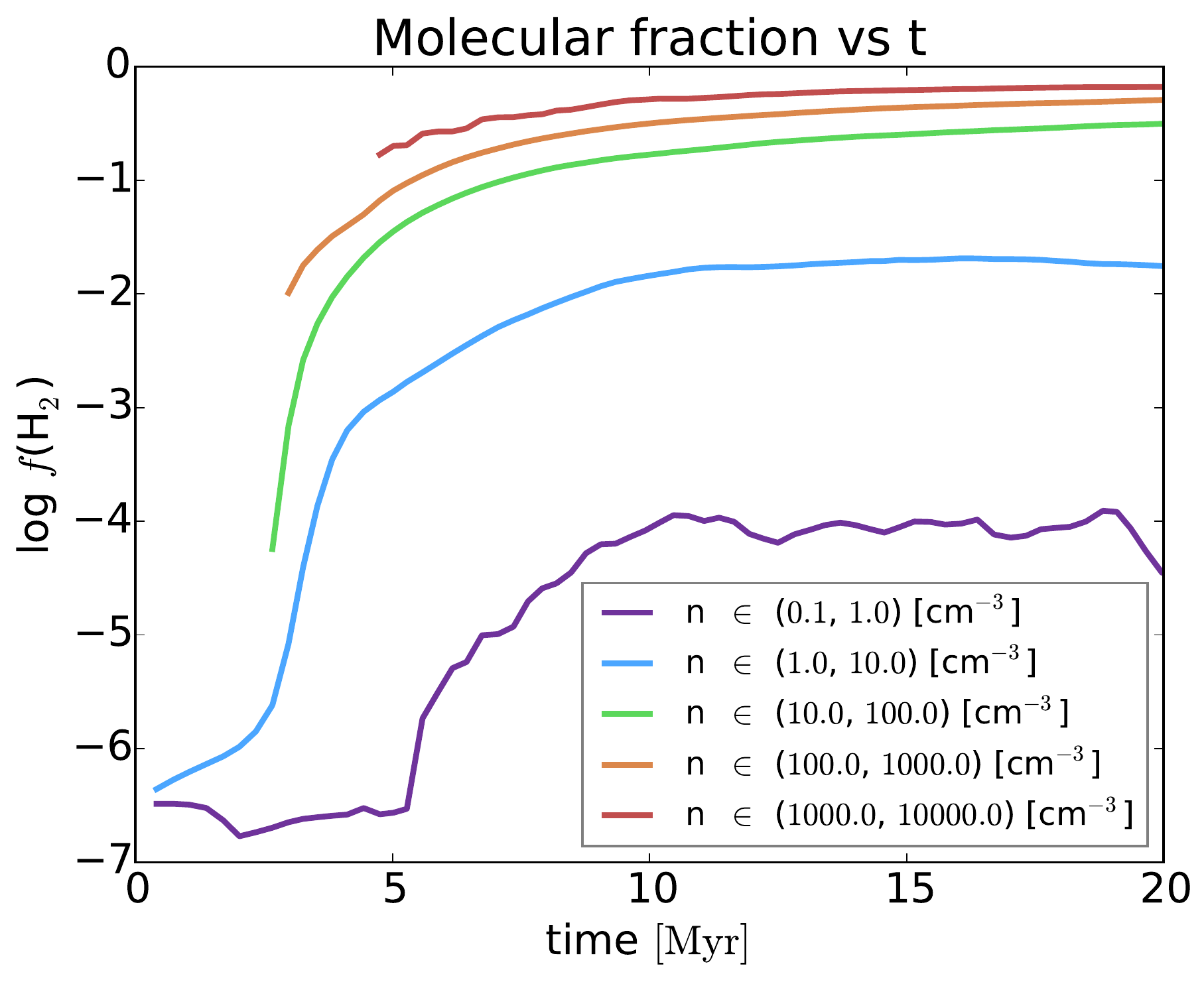} \\
   \includegraphics[width=7.5cm]{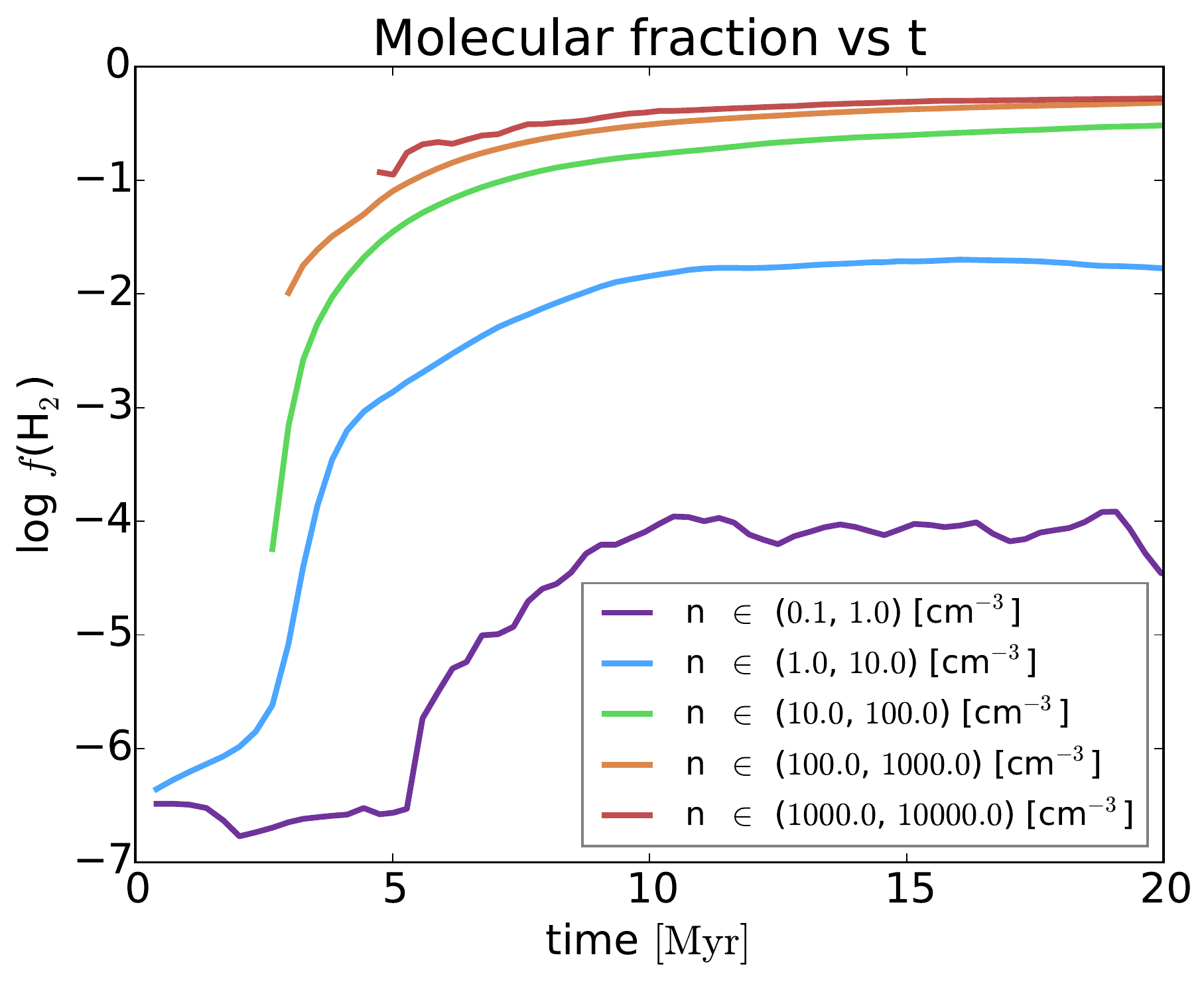} \\
   \includegraphics[width=7.5cm]{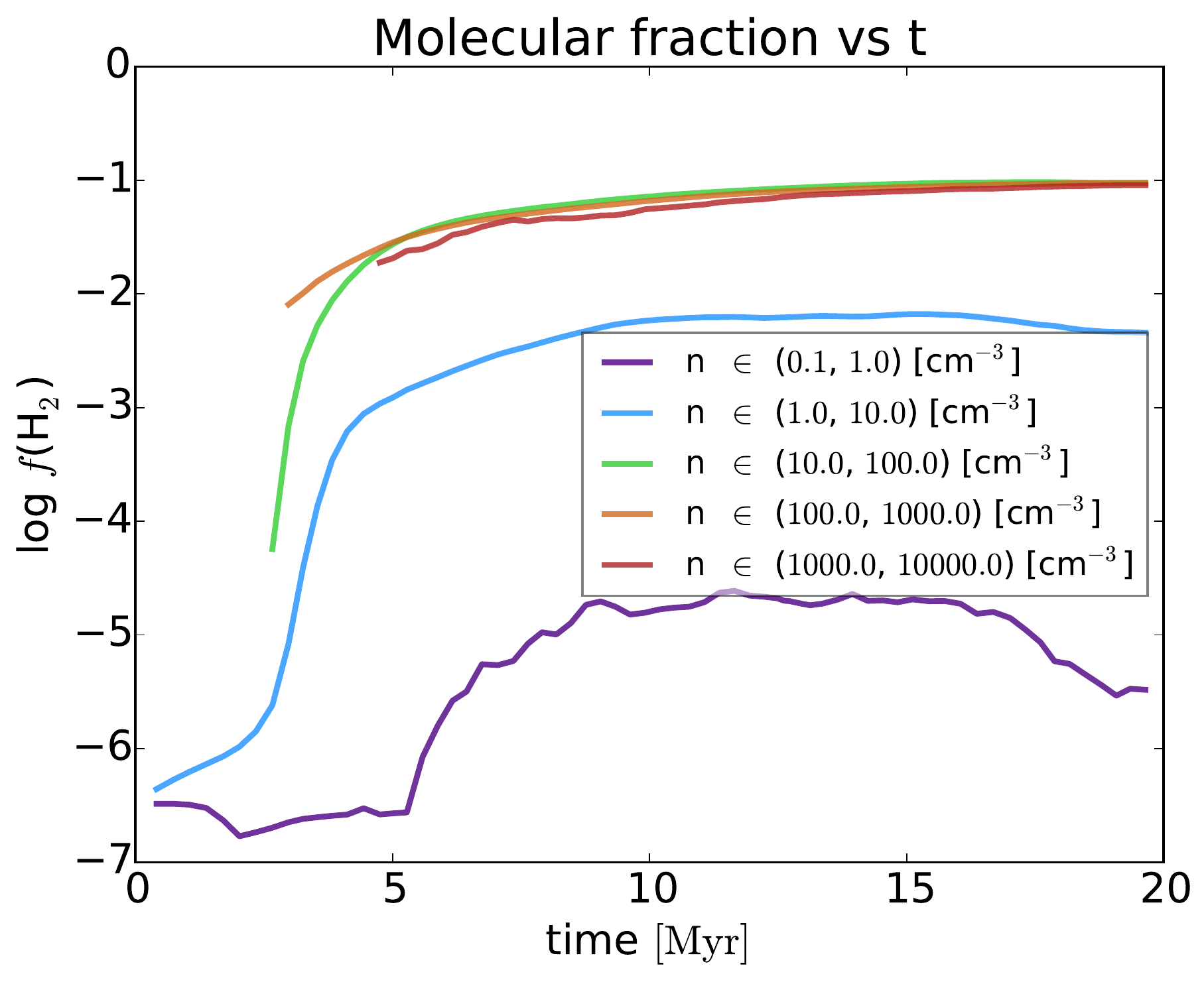} \\
\caption{Evolution of the molecular fraction, where the formation of H$_2$ has been suppressed for gas denser than a fixed 
threshold. The top panel corresponds to the standard case, with no suppression of H$_2$ formation. 
The middle panel shows the evolution for a density threshold of $n = 1000~\mathrm{[cm^{-3}]}$, while the panel at 
the bottom corresponds to a density threshold of $n = 100~\mathrm{[cm^{-3}]}$}
\label{suppr_h2}
\end{figure}

\begin{figure}
\centering
\includegraphics[width=7.8cm]{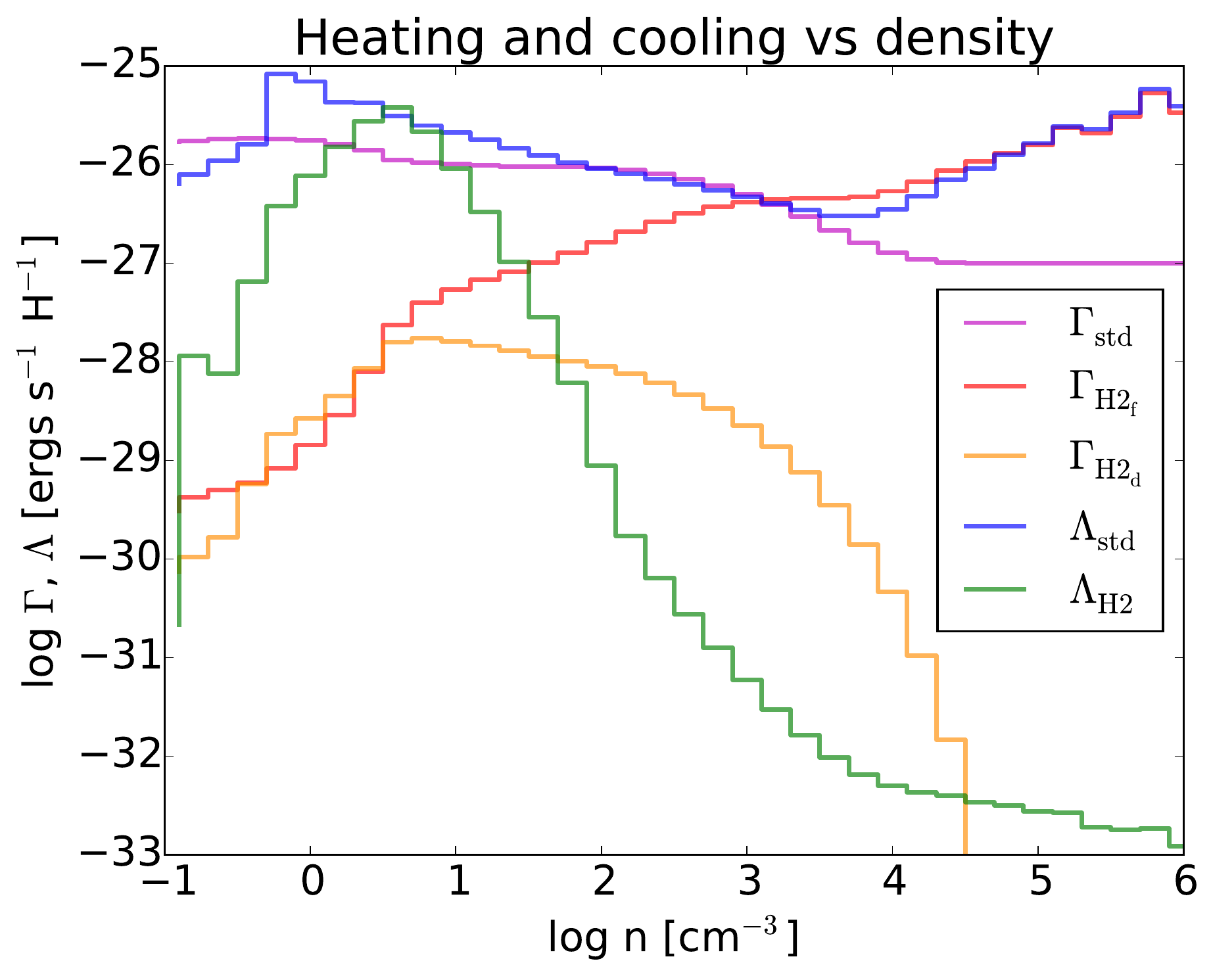}
\caption{Various cooling and heating contributions as a function of density. As can be seen, 
H$_2$ does not have a dominant influence except at densities of about 4 cm$^{-3}$ where 
H$_2$ cooling (green line) becomes comparable to the standard ISM cooling (blue line) and at high densities
where the heating by H$_2$ formation (red line) is significant (but without modifying the temperature substantially).}
\label{heat_cool}
\end{figure}

\section{Results}\label{results}
We now present the results of our calculations. We start by discussing the general 
properties of the clouds because it is essential to understand in which context 
the hydrogen molecules form. Then we discuss the abundance and 
distribution of H$_2$ itself. Finally, we perform various complementary calculations
aiming at better understanding in which conditions H$_2$ forms.

\subsection{Qualitative description of the cloud}

As many colliding flow calculations have been performed during the past decade, we do not attempt here 
to display all the steps that the flow experiences (see references above for accurate descriptions), so we quickly summarise the main steps. 
First, the WNM flows collide, which triggers the transition for a fraction of the gas into 
moderately dense gas \footnote{If the incoming flow is fast enough, a collision is not even necessary, as it is the case 
for example in the study presented by \citet{koyama_inu2002}.}, that is to say, densities of the order 
of 100-1000 cm$^{-3}$ as a consequence of the thermal instability and ram pressure. Second, when enough gas has been accumulated,  
gravity becomes important and triggers infall, first at small scales and then at larger ones. This leads 
to a continuous increase of the dense gas fraction.

The left panels of Figs.~\ref{NcolH2fH2} and \ref{shield_f} show the column 
density along the $z$-direction and a density cut in the $xy$ plane at time 20 Myr, respectively. 
As we show below, this corresponds to a phase where gravity already plays a significant role. 
The density cut shows that the flow is very fragmented. The dense gas is organised into dense clumps
that are embedded in diffuse and warm gas. The column density typically spans 
2-3 orders of magnitude from a few 10$^{21}$ cm$^{-2}$ to at least 10$^{24}$ cm$^{-2}$ in some 
very compact regions, the dense cores in which gravity triggers strong infall. 
The column density map is also clumpy, but less 
obviously structured than  the density cut.

\subsubsection{Density PDF and volume filling factor}
In many respects the density distribution is an essential cloud property that reveals the dynamical state of the cloud and strongly influences the formation of H$_2$. 

Figure~\ref{gen_PDF} shows the density probability distribution function (PDF) at time 5, 10, 15, and 20 Myr. 
The peak at low density ($n \simeq 1$ cm$^{-3}$) is simply a consequence of the initial conditions and boundary conditions (which inject WNM from the boundaries), the rest, at higher densities, represents the cold phase, 
whose formation has been triggered by the thermal instability. Because of the supersonic turbulence that develops
in the cold gas, the density PDF is broad and presents a lognormal shape \citep{kritsuk+2007,federrath+2008}. 
At later times, more cold gas accumulates and the PDF broadens. The high-density 
tail tends to become less steep. At intermediate densities ($\rho\sim$ 10$^3$-10$^4$ cm$^{-3}$)
the shape is compatible with a power law with an exponent between -1 and -2. This is typical 
of what has been found in super-sonic turbulent simulations that include gravity
\citep{kritsuk+2011}. At very high densities ($> 10^{4}- 10^{5}$ cm$^{-3}$), the density field 
flattens. This occurs because we did not use sink particles,
and the very dense gas therefore piles up and accumulates 
within a few grid cells. This feature is thus a numerical artefact and represents a limit of the simulations.

Figure~\ref{vol_fill} shows the density PDF for different computational box regions
at a time of 20 Myr.
The four lines show the results for four different box regions (as shown in the label). The black line 
shows the whole box, the red line is limited to the densest regions of the computational 
box.  The volume filling factor is clearly dominated by the warm and diffuse gas. Selecting 
gas of densities $n < 3$ cm$^{-3}$, we find that it 
typically occupies 70$\%$ for the region located between $x=27$ and $x=37$ (85$\%$ for the whole computational box). 
The dense gas, even in the densest region, occupies 
only a tiny fraction, for example we find that the gas denser than 100 cm$^{-3}$ has a filling factor 
of about 3$\%$ (respectively 1.5$\%$ for the whole box). The remaining 26$\%$ (13$\%$) are filled with 
gas of densities between 3 and 100 cm$^{-3}$. We therefore conclude that the interclump medium that occupies most of the 
volume, is itself made of two components: a warm gas that is similar to the standard WNM, but can be 
slightly denser, and a cold gas that is similar to the CNM, but contains, as we show below, a significant 
fraction of H$_2$. They typically occupy about 70$\%$ and 25$\%$ percent of the volume, respectively.

\subsubsection{Pressure of various phases}
Another important diagnostic for characterizing the dynamics of a medium are the different pressures. 
Figure \ref{rapport_pressions} shows the thermal, $P_{th}$, dynamical, $P_{ram}$, and magnetic, $P_{mag}$, pressures equal to
$P_{th}= nkT$, $P_{ram}=\rho V^2$ and $P_{mag}= B^2/8 \pi$, respectively.

The dynamical pressure clearly dominates the thermal pressure
by typically one to two orders of magnitude
in the dense gas ($n>10$ cm$^{-3}$), while they are more similar in the diffuse gas. The magnetic pressure
lies in between the two, with values a few times higher than that of the thermal pressure. $P_{ram}$ and $P_{mag}$ increase with density, and at densities 
of the order of $n \simeq 10^3$ cm$^{-3}$, they are about one order of magnitude higher than 
 their mean values at $n \simeq 10$ cm$^{-3}$. This means that
while the low-density gas that fills the 
volume provides some confining pressure, it has a limited influence on the clumps. 

Altogether, these results indicate that the molecular cloud produced in this calculation 
can be described by density fluctuations or clumps, induced by both ram pressure and gravity. The
clumps occupy a tiny fraction of the volume, which is filled by a mixture of warm diffuse and 
dense HI gas (occupying a fraction  >70$\%$ and >$20\%$ of the volume, respectively). This low-density 
material feeds the clumps, which grow in mass. This picture agrees
well with 
the measurement performed by \citet{williams+1995}, who found that the interclump medium 
typically has a density of a few particles per cm$^{-3}$ and a high velocity dispersion of several km s$^{-1}$.

\subsubsection{Clump statistics}
Because the cloud is organised into clumps, we further quantified the simulations by providing 
some statistics. To identify the clumps, we used a density threshold of 
1000 cm$^{-3}$ and a friends-of-friends algorithm. This structure is important for the chemistry 
evolution because very significant density and temperature gradients arise at the edge of clumps . 
Figure~\ref{fig_clump} shows the mass spectrum, the velocity dispersion, and the kinetic $\beta$ parameter
(that is to say, $P_{ram}/P_{mag}$) of the clumps.
The mass spectrum for masses above a few 0.1 $M_\odot$ presents a power law with 
an index of about $-1.7$, which is very similar to what has been shown in related works
\citep[e.g.][]{audithenne2010,heitsch_et_al_2008}. The velocity dispersion within the clumps
as a function of their size is about $\sigma  \simeq 1 ~{\rm km \, s^{-1}} (L / 1 {\rm pc})$,
which is close to the Larson relation \citep{larson1981,hennefalg2012}.
The third panel shows that the kinetic $\beta$ parameter 
is typically of the order of, or slightly below 1, showing that the magnetic 
field plays an important role within the clumps.

\subsection{Molecular hydrogen}
\label{molecular}
We now discuss the H$_2$ abundance.
The middle and right panels of Fig.~\ref{NcolH2fH2} 
show the H$_2$ column density and molecular abundance, $f(\mathrm{H}_2) = 2 n(\mathrm{H}_2)/n$. As expected, the high 
column density regions are dominated by the H$_2$ molecules, and values of $f({\rm H}_2)$ close to 1 are 
obtained there. The values of $f({\rm H}_2)$ are obviously significantly lower in the outer parts of the cloud. 

\subsubsection{UV screening factor}

The middle and right panels of Fig.~\ref{shield_f} show a cut of $n({\rm H}_2)$ and the 
value of the shielding parameter, $\chi _{\rm shield} = k_\mathrm{ph}/k_\mathrm{ph,0}$. This latter steeply decreases
when entering the cloud, where it takes values of the order of $10^{-3}$. 
A comparison between the density cut with  $k_\mathrm{ph}/k_\mathrm{ph,0}$ reveals that there are regions of diffuse 
material in which the shielding parameter is low. This is because these regions are surrounded 
by dense gas in which H$_2$ is abundant, which provides an efficient self-shielding.
As a consequence, there are low-density regions that contain a relatively high abundance of
H$_2$ (see left and middle panels of Fig.~\ref{shield_f}).

A more quantitative statement can be drawn from Fig.~\ref{UV_scatt}, which displays the distribution 
of the dust shielding, $\chi$ (top panel), and  of the total shielding
 $\chi _{\rm shield}$ as a function of density (bottom panel).  As expected, most low-density cells are associated with 
$\chi$ values that are close to 1, that is to say, they are weakly shielded. There is a fraction 
of them, however, that presents values of $\chi _{\rm shield}$ as low as 10$^{-5}$.
 At higher densities, 
the mean value of $\chi$ decreases with a steep drop between $n \simeq 10^3$ and  $10^4$ cm$^{-3}$ 
, which corresponds to the point where the dust significantly absorbs the UV external field.
For the total shielding, a steep drop is observed at about 10 cm$^{-3}$. 
There is a broad scatter, however, which increases from $n=1$ to $n=10^4$ cm$^{-3}$
for the dust shielding and tends to decrease for the total shielding $\chi _{\rm shield}$.
This indicates that the mean value of $\chi _{\rm shield}$ is not sufficient information to quantify 
the abundance of H$_2$ expected at a given density. 
This is a clear consequence of the complex cloud structure. Since 
$\chi _{\rm shield}$ plays a key role in the formation of H$_2$, this 
complex distribution certainly makes molecular hydrogen 
formation in a multiphase turbulent cloud a complicated problem.

\subsubsection{Global evolution of molecular hydrogen}

Figure~\ref{fH2_evo} shows $f({\rm H}_2)$, the mean molecular fraction within the whole cloud, 
as a function of time. During the first 5 Myr,  $f({\rm H}_2)$ exponentially increases
from nearly 0 to about $\simeq 0.1$. After this phase, $f({\rm H}_2)$ continues to increase 
in a more steady almost linear way. Near 15 Myr, about 40$\%$ of the gas 
is in H$_2$ molecules. 
Since, as discussed before, the cloud is rather inhomogeneous in density and temperature, 
we also plot the time evolution per bin of density and temperature. The corresponding 
curves are displayed in Fig.~\ref{evof_tbin}.
As expected, the total distribution closely follows the values of the densest bins, that is to say, it 
corresponds to densities higher than 100-1000 cm$^{-3}$. 
For these densities, the timescale for H$_2$ formation is thus of a few ($\simeq 6-8$) Myr.
As recalled previously, this typical timescale is about $10^9 / n$ yr. The 
timescale we observe in the simulation therefore agrees well with this scaling. 
Altogether, this agrees well with the results of \citet{glover2007a,glover2007b}, who 
found that H$_2$ could be formed quickly in molecular clouds because  it 
preferentially forms into clumps that are significantly denser than the rest 
of the cloud. In the present case, the mean density of our cloud at time 
5-10 Myr would be about 10-100 cm$^{-3}$ (depending on exactly which area is selected),
and this would lead to a timescale for H$_2$ formation of the order of 10 to 100 Myr.

The relatively short amount of time ($\simeq 10$ Myr) after which there is a significant amount of 
H$_2$ in low density gas ($n \simeq 1-10$ cm$^{-3}$) is somewhat
more surprising. 
The expected timescale for H$_2$ formation in this density range would be 
about 10$^8$ Myr.
Figure~\ref{evof_tbin} shows that the behaviour of $f({\rm H}_2)$ is similar, although 
not identical, to the behaviour at higher densities. There is a fast increase
followed by a slowly increasing phase. The slope 
changes at about 4 Myr however, and this is different from the higher density evolution. The steady evolution starts at about 8 Myr instead of 5-6 Myr for
the denser regions. 
This suggests that the  presence of H$_2$ at low density is triggered  by formation of 
H$_2$ in the dense gas. This can occur in various non-exclusive ways. First of all, 
some  dense gas can expand back and mix with diffuse gas. This may occur either through a sonic 
expansion, through evaporation, or through numerical diffusion.   
Second of all, some of the diffuse gas may be surrounded by 
dense gas in which $f({\rm H}_2)$ is large and therefore has a low $\chi _{\rm shield}$. 
Below we attempt to analyse these mechanisms in more
detail. 

Finally, we note that in the bottom panel of Fig.~\ref{evof_tbin} there is a small (1-10$\%$)
but nevertheless non-negligible fraction of H$_2$ 
within gas of relatively high temperature, that is, $T>300$ K. Since H$_2$ is the first 
molecule involved in producing other molecules and since 
some of them, such as  CH$^+$, are formed through reactions with high activation barriers \citep[of the order of $4300~\mathrm{K}$ for CH$^+$, see][]{agundez2010}, the presence of H$_2$ could have consequences
for the production of these species, as proposed in \citet{lesaf+2007}. 
In the same way, as discussed below, this warm H$_2$ contains molecules in excited high $J$
rotational  levels, which can therefore contribute to the gas cooling.

\subsubsection{Detailed distribution of molecular hydrogen}

To further quantify the distribution of H$_2$ in the cloud, we
display histograms of $f({\rm H}_2)$ per density bin (Fig.~\ref{histo_n}).
We also draw the distribution of $f({\rm H}_2)$ obtained at equilibrium
in the same panel. 
Knowing the density, $n$, the temperature, $T$ and the 
shielding factor, $\chi _{\rm shield}$, we solve Eq.~(\ref{eqH2}) at equilibrium.
We note that  this distribution is not fully self-consistent since 
the value of $\chi _{\rm shield}$ should in principle be recalculated to be consistent 
with this equilibrium distribution (this would imply performing several iterations).
In practice, the goal here is simply to compare with the time-dependent 
distribution to gain insight into the H$_2$ production mechanisms, and it is therefore easier to have exactly the 
same formation and destruction rates. 

In particular, the equilibrium distribution is expected to be different from the 
time-dependent distribution for at least two main reasons. First, since the H$_2$ formation 
timescale is somewhat long, if the time-dependent $f({\rm H}_2)$ lies below the 
equilibrium timescale, it will indicate that the H$_2$ abundance has been limited by the 
time to form the molecule. Second, if the time-dependent $f({\rm H}_2)$ is above 
the equilibrium value, it will indicate that $f({\rm H}_2)$ has increased because 
of an enrichment coming from denser gas. As discussed above, this could operate 
through the expansion or the evaporation of cold clumps,   a process that we call turbulent diffusion,
 or through numerical diffusion. The latter is quantified in the appendix by performing convergence studies
that suggest that it remains limited.
Another  possibility is that the H$_2$ excess has been produced during  a phase where the
fluid parcel was more shielded by the surrounding material and therefore the UV field 
was lower. The two effects probably act simultaneously and are difficult to separate.

At a time of 5 Myr (left panels), all density bins show an excess of the equilibrium distribution with respect
to the time-dependent distribution. This is clearly due to the long timescale needed to form H$_2$. 
The differences between the equilibrium and time-dependent distributions 
become eventually less important. For example, the two distributions are obviously 
closer at a time of 20 Myr (right panels) than at a time of 5 Myr. However, they are not 
identical. Since all distributions at a time of 15 Myr and 20 Myr are similar, 
the persistence of the differences between the equilibrium and 
time-dependent distributions indicates that this is most likely the result of a 
stationary situation. It most likely reflects the accretion process, that is to say, 
HI gas (possibly mixed with a fraction of H$_2$) is continuously being accreted 
within  denser clumps and therefore the dense gas does not become 
fully molecular. This interpretation is also consistent with the fact that 
the effect is more pronounced for the fourth density bin (line 4)
than for the fifth and highest one (line 5). 
Considering now the lowest density bin (first line), we find the reverse effect at a time of 
10 and 15 Myr: the time-dependent distribution dominates the equilibrium distribution. 
This is most likely an effect of the turbulent diffusion. Because
of the low mass contained 
in this low-density bin, this remains a limited effect, however. 

The second density bin (line 2), which corresponds to $n$ between 1 and 10 cm$^{-3}$ , is 
slightly puzzling. No significant differences are seen between the two 
distributions, which is surprising. One possibility is that various effects compensate for each other. 
That is to say, the time delay to form H$_2$ (clearly visible for the three denser 
density bins) may be compensated for by an enrichment from the denser gas. 
This  might possibly occur for the second density bin at time 20 Myr 
(fourth column, second line) where overall a small excess
of time-dependent H$_2$ is found, but for $f({\rm H}_2)>0.7$ the equilibrium distribution
dominates. 

\subsubsection{Spatial fluctuations}
As a complement to the time evolution of $f({\rm H}_2)$, it is also worthwhile to 
investigate the spatial fluctuations at a given time. In this respect, 
the clumps constitute natural entities for studying the spatial variations.  
The bottom panel of Fig.~\ref{fig_clump} shows $f({\rm H}_2)$ within the clumps. 
For the most massive clumps, the value of $f({\rm H}_2)$ is close to 1 and the dispersion remains weak. 
In contrast, the dispersion becomes very high for the less massive clumps,
and the value of $f({\rm H}_2)$ can in some circumstances be rather low. 
This shows that  clump histories, that is to say, their ages and the local UV flux
in which they grow, have a major influence on $f(\mathrm{H_2})$.

\subsubsection{Further analysis of  H$_2$ formation}
To better quantify the interdependence of the different density bins, we 
conducted three complementary low-resolution runs for which the formation of 
H$_2$ was suppressed ($k_\mathrm{form}$ was set to zero) above a given density threshold. 
Figure~\ref{suppr_h2} displays the result. In the top panel, no threshold is applied. 
In the middle and bottom panels a threshold of 1000 cm$^{-3}$  and 100 cm$^{-3}$, respectively, is 
applied. For the threshold 1000 cm$^{-3}$, the values of $f({\rm H}_2)$ are not 
significantly changed except for the highest density range (for $n>1000$ cm$^{-3}$).
On the other hand, with a threshold  100 cm$^{-3}$, the values of $f({\rm H}_2)$ decrease by 
a factor of about 3 for all density bins. This clearly shows that most of the 
H$_2$ molecules in the low-density gas form at a density of a few 100 cm$^{-3}$.
While the shielding provided by molecules in gas denser than this value could 
contribute to enhance the more diffuse gas, the filling factor of the dense gas 
is too small to strongly affect the diffuse gas, which has a much larger 
filling factor (see Fig.~\ref{vol_fill}). Therefore we conclude that the 
H$_2$ abundance within the low-density gas (<100 cm$^{-3}$) is very likely a consequence 
of turbulent mixing and gas exchange between diffuse and dense gas.

\subsection{Thermal balance}
To quantify the influence of the H$_2$ molecule on thermodynamics, 
 we present the contributions of the various heating and cooling terms to the 
 thermal balance as a function of density.
Figure~\ref{heat_cool} shows the standard ISM cooling (blue curve) and the cooling by 
H$_2$ (green curve). The latter clearly remains below the former throughout, except at a
density of about 3-5 cm$^{-3}$ , where they become similar.
At densities above $3\times 10^{4}~\mathrm{cm^{-3}}$ 
the "standard" heating curve (magenta) is dominated by the heating caused by cosmic rays and 
reaches the intermediate value for shielded gas proposed by \citet{goldsmith2001}.  
While the heating due to H$_2$ destruction remains negligible at any density (orange curve), 
the heating due to its formation (red curve) is the dominant heating 
mechanism at high densities. 
This does not affect the temperature very strongly, however.
It remains equal to about 10 K. 
Moreover, at higher densities (above $\sim 10^{5}~\mathrm{cm^{-3}}$ ) other processes that are not included in our simulations can take over, such as collisions with dust grains and the cooling by other molecular lines, such as CO. 

Finally, as can be seen for densities between $1$ to $\sim 50 ~\mathrm{cm^{-3}}$, the 
cooling dominates the heating by a factor of a few. This indicates that there is 
another source of heating equal to the difference, which is due to
the mechanical energy dissipation. This latter therefore appears to have a
contribution similar to the UV heating. This explains why warm gas 
is actually able to survive within molecular clouds. 
In the same way that density is much higher than in the rest of the ISM, 
kinetic energy is also higher and provides a significant heating \citep[e.g.][]{henneinut2006}.

\begin{figure}
\centering
\includegraphics[width=9cm]{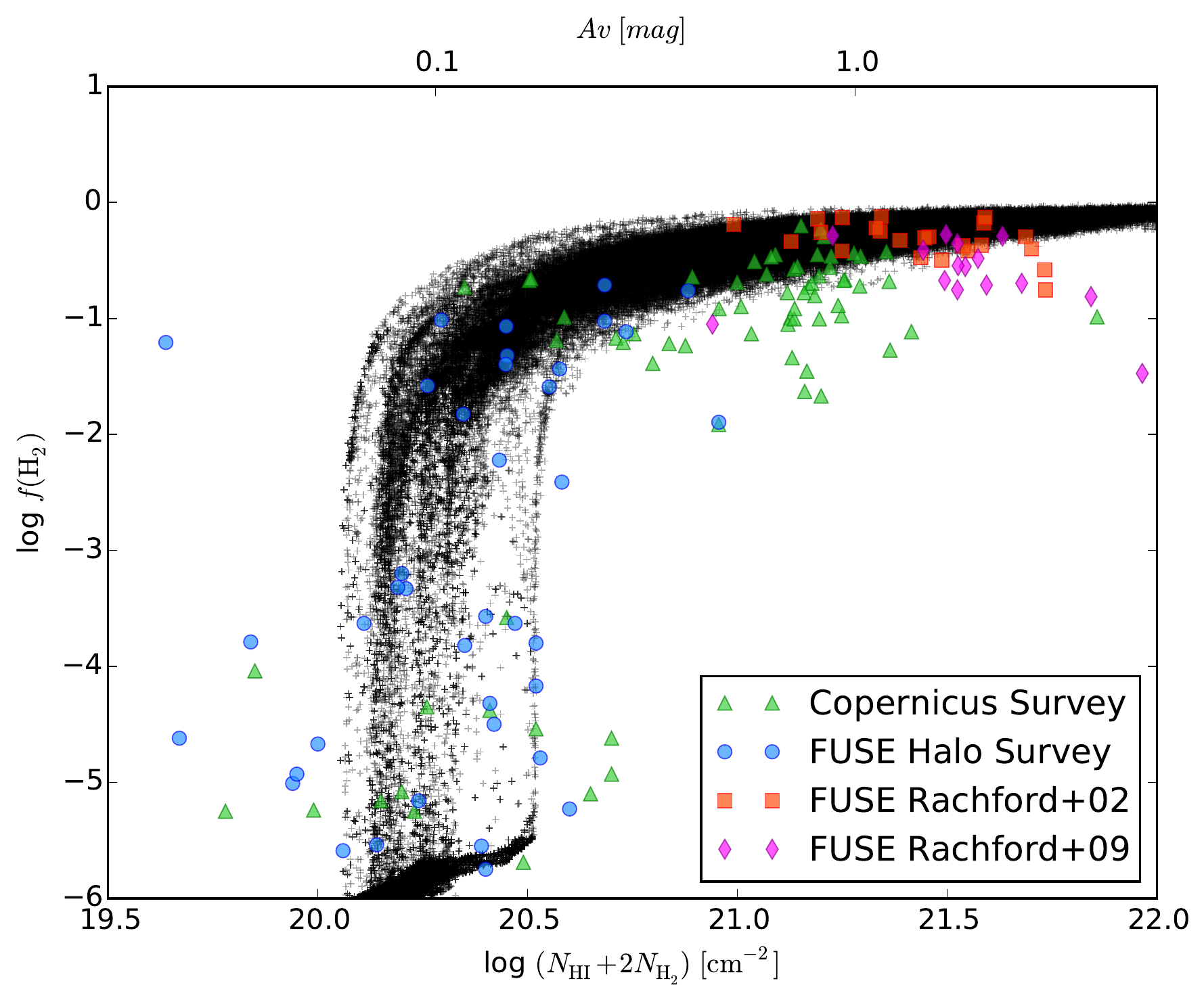}
\caption{Molecular fraction as a function of total hydrogen column density. Comparison with the \emph{Copernicus} survey \citep{savage1977} and \emph{FUSE} \citep{gillmon2006, rachford2002, rachford2009}.}
\label{col_H2}
\end{figure}

\section{Comparison to observations}
\label{observation}
We now compare our results with various observations that fall into two categories. 
The first observations have attempted to measure the molecular fraction, that is, the 
total column density of H$_2$ with respect to the total column density. The second observations have measured the excited levels of H$_2$, therefore  presumably tracing high-temperature gas. These two sets of observations are therefore complementary 
and very informative.

\subsection{Molecular fraction vs column density}
%The H$_2$ column density has been estimated along several lines of sight mainly 
%(though not exclusively, e.g. \citet{lee2012}) with the Copernicus 
%\citep{savage1977} and FUSE observatory \citep{gillmon2006, rachford2002, rachford2009}.
%These observations are probing lines of sight spanning a wide range of column densities
%and therefore constitute a good tests for our simulations although a difference to keep in 
%mind is that the lines of sight extracted from our simulations are all taken from a 
%50 pc regions and therefore are more homogeneous than the observed lines of sight.

The H$_2$ column density has been estimated along several lines of sight mainly 
(though not exclusively) with the \emph{Copernicus} satellite
\citep{savage1977} and the \emph{FUSE} observatory \citep{gillmon2006, rachford2002, rachford2009}.
These observations probe lines of sight spanning a wide range of column densities
and therefore constitute a good test for our simulations, although a difference to keep in 
mind is that the lines of sight extracted from our simulations are all taken from a 
$50~\mathrm{pc}$ region and therefore are more homogeneous than the observed lines of sight.

Figure~\ref{col_H2} shows the molecular fraction $f({\rm H}_2)$ as a function of total column density for 
all lines of sight of the simulations (taken along the $z$-direction) and 
the available lines of sight reported in \citet{gillmon2006}, and \citet{rachford2002, rachford2009}.
The simulation results and the observations agree rather well, 
many observational data directly fall into the same regions as simulated data. 
In particular, the two regimes, that is, the vertical transition branch at column densities of between $10^{20}$ and 
$3 \times 10^{20}~\mathrm{cm}^{-3}$ as well as the higher column density region,  
 are clearly seen both in observations and in the simulation.

This said, there are also data points that are not reproduced by any of the 
lines of sight from the simulations. This is particularly the case
for column densities higher than $3 \times 10^{21}~\mathrm{cm}^{-2}$ and for the \emph{Copernicus} survey
at column densities of around $\simeq 10^{21}~\mathrm{cm}^{-2}$.
The most likely explanation is that the UV flux is different from the standard value we assumed
in our study. We stress in particular that the measurements were made in absorption 
toward massive stars. Because these are strong emitters of UV radiation, it may not be too 
surprising that our measurements lead to higher values of $f({\rm H}_2)$. A more 
quantitative estimate should entail a detailed modelling of every line of sight, including
 the UV flux in the regions of interest and specific cloud parameters, such as column densities. 
This is beyond the scope of the present paper.

\begin{figure}
\centering
\includegraphics[width=8.cm]{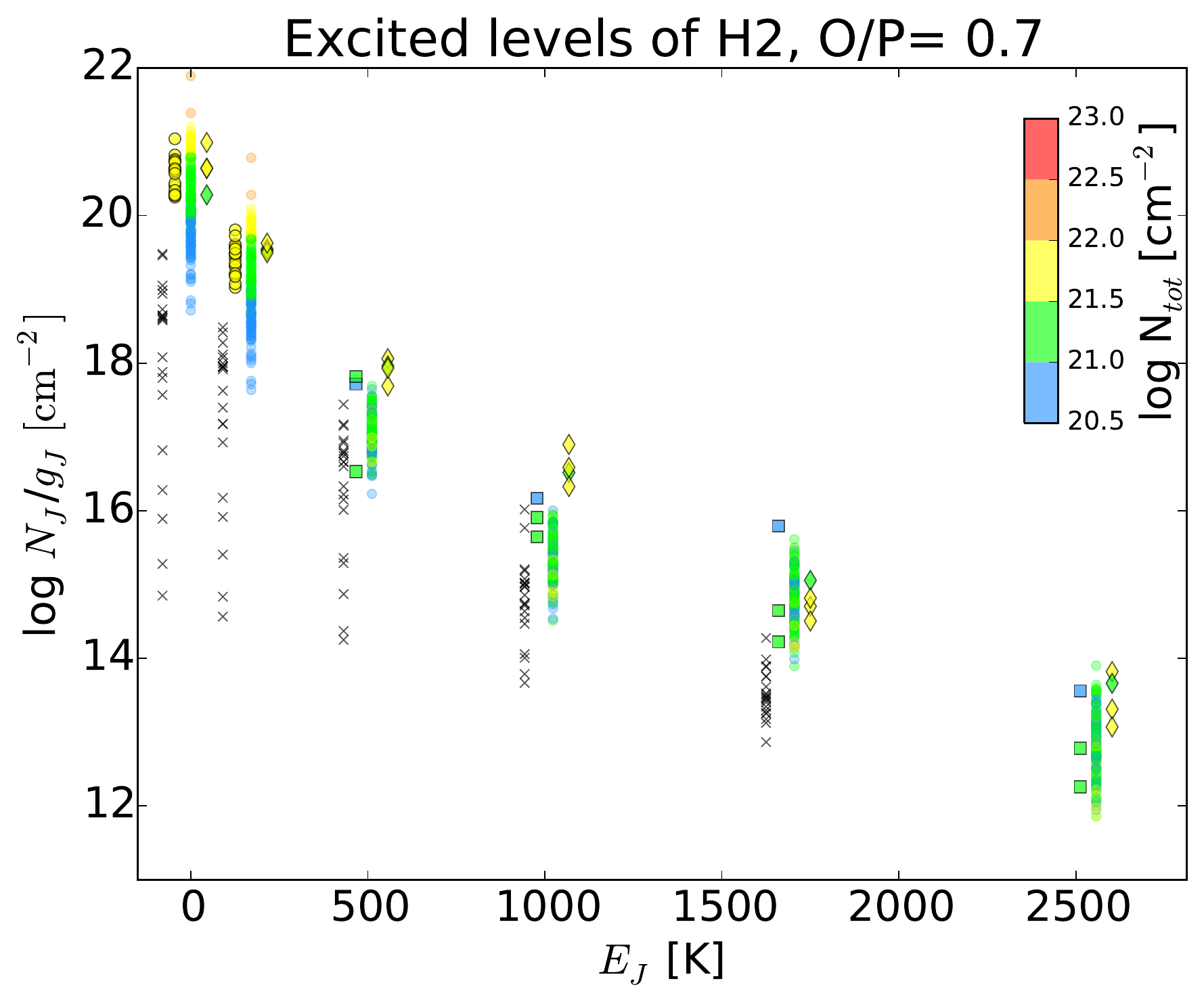}\\ \includegraphics[width=8.cm]{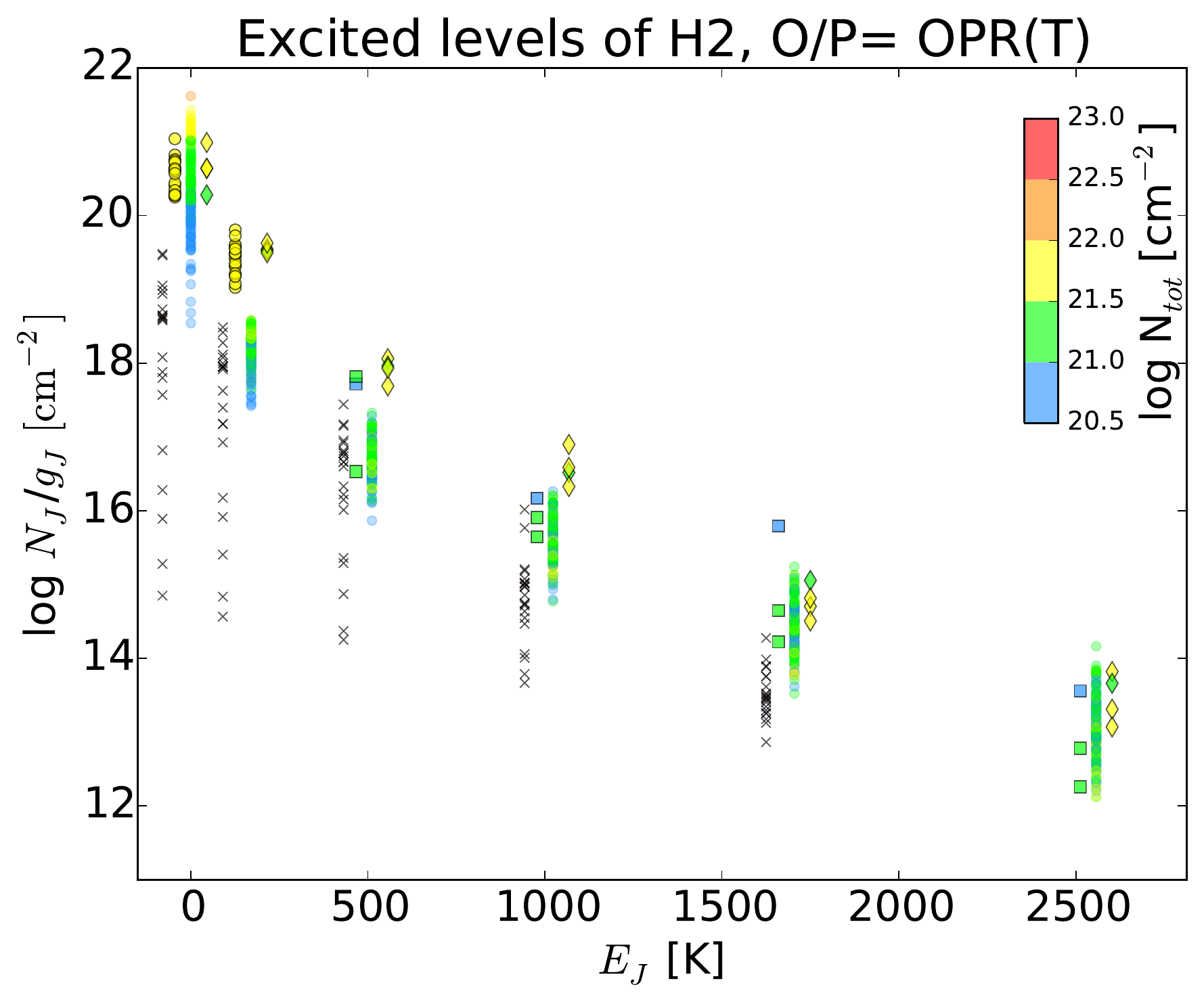}
\caption{H$_2$ population distribution along different lines of sight at a time $t=20~\mathrm{Myr}$, using an ortho-to-para ratio of $0.7$ (top), and at thermal equilibrium (bottom). The symbols  correspond to the observational data of 
\citet{wakker2006} (cross), \citet{rachford2002} (circle), \citet{gry2002} (square), and 
\citet{lacour2005} (diamond). Colours correspond to the total column density range.}
\label{Jpop_opr}
\end{figure}

\begin{figure}
\centering
\includegraphics[width=7.8cm]{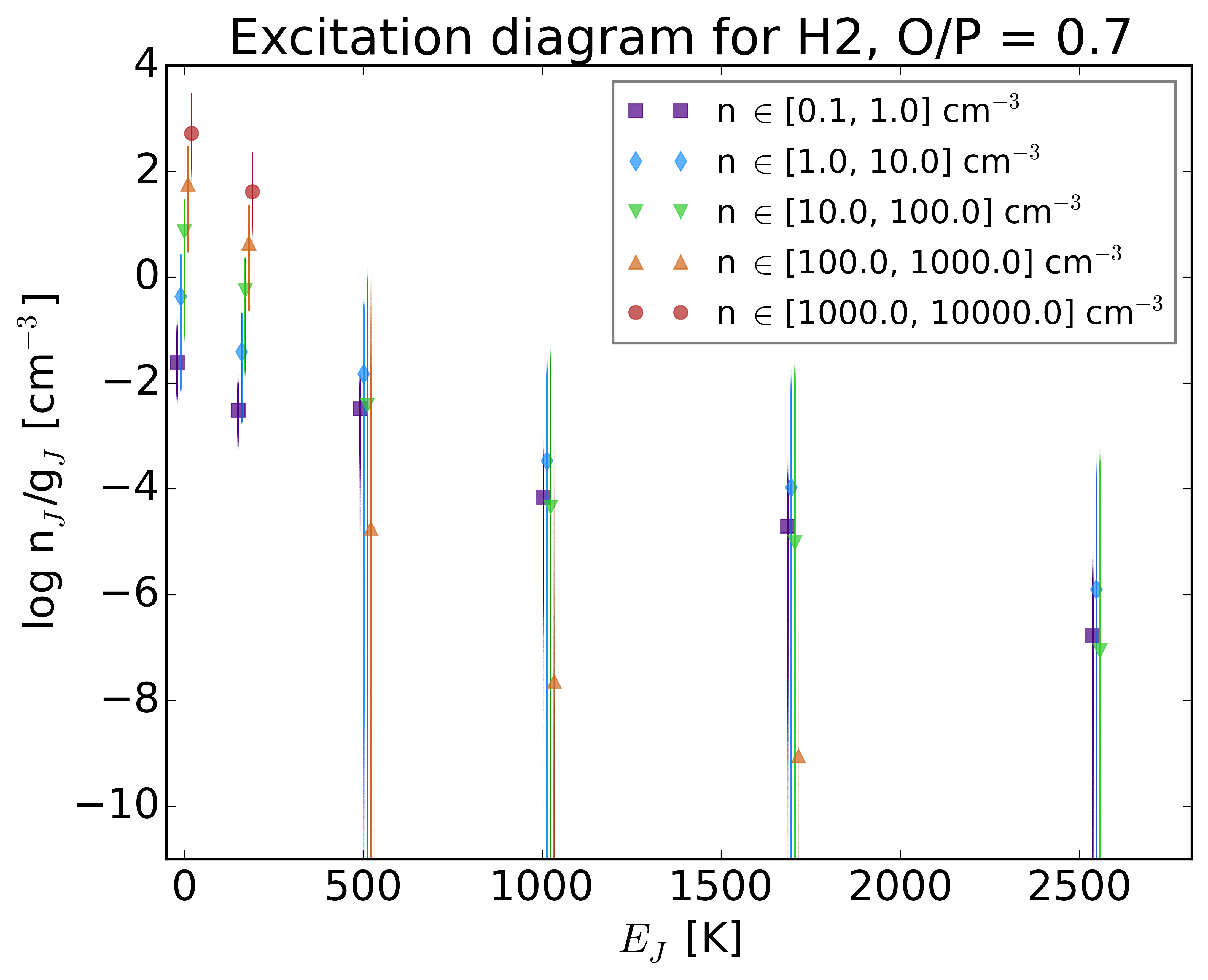}
\includegraphics[width=7.8cm]{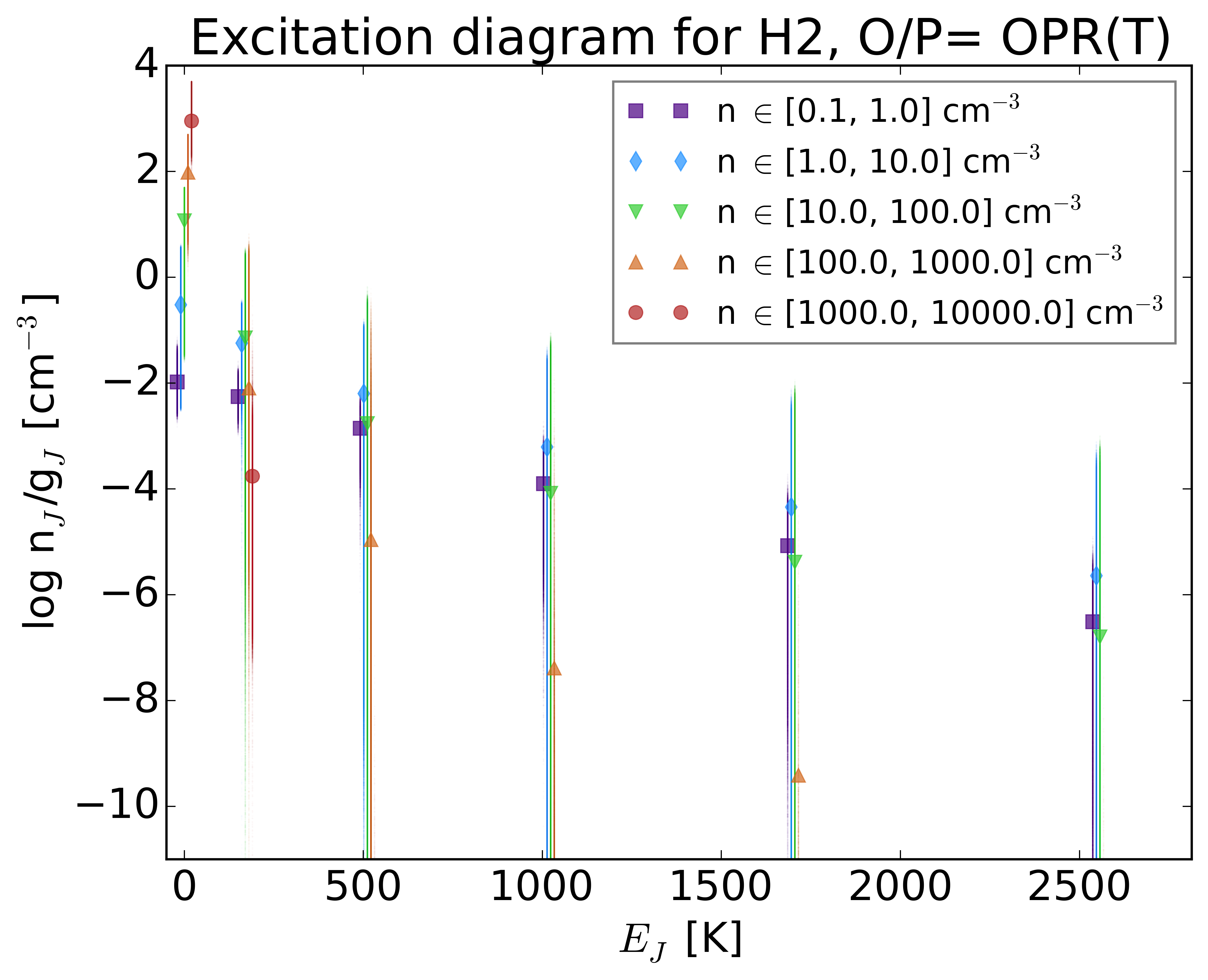}
\caption{H$_2$ level population distribution for individual cells in the simulation calculated at $20~\mathrm{Myr}$: using an ortho-to-para ratio OPR = $0.7$ (top) and at thermal equilibrium (bottom).  }
\label{Jpop_cell}
\end{figure}

\begin{figure}%[htb]
\centering
\includegraphics[width=7.8cm]{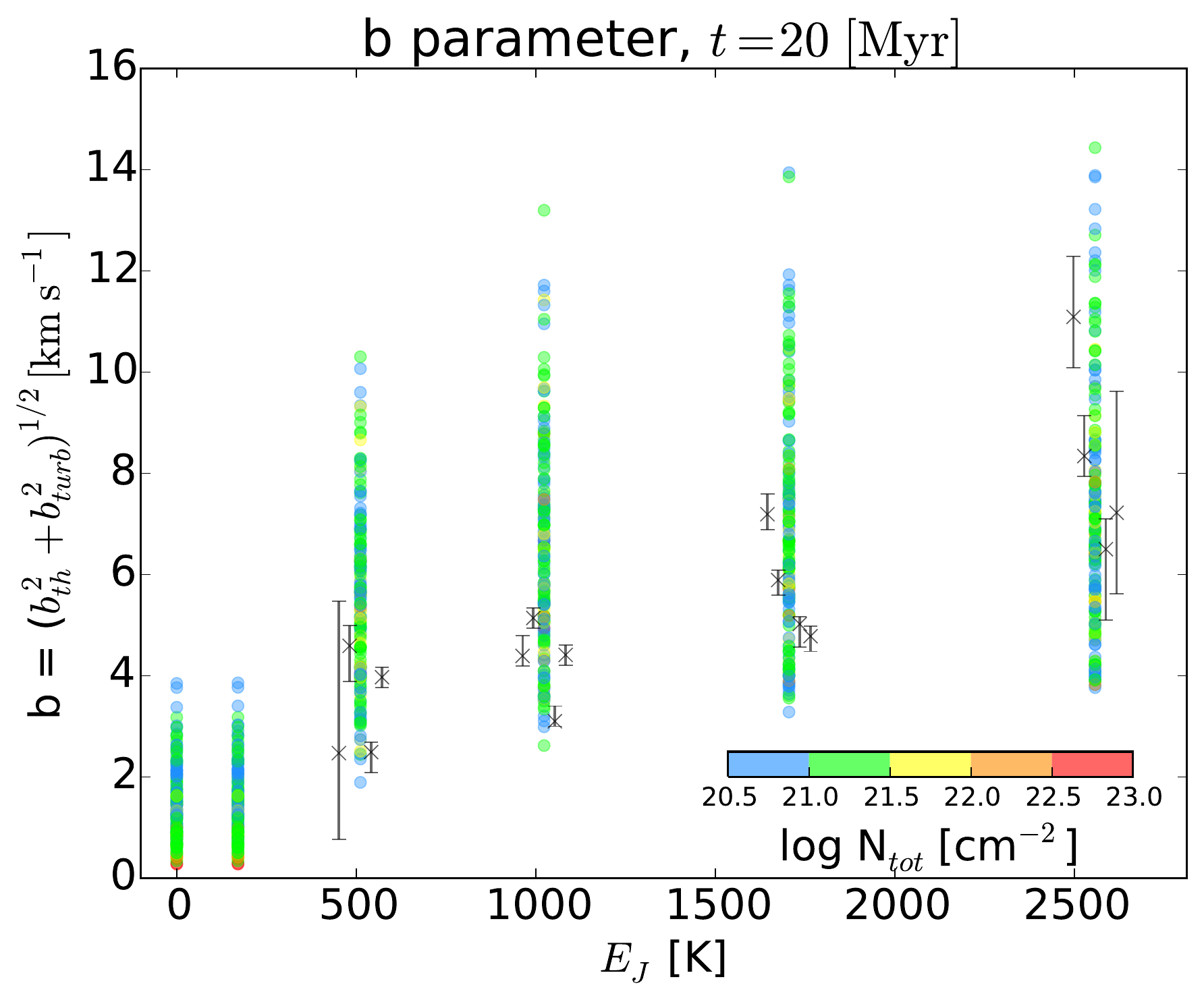} 
\includegraphics[width=7.8cm]{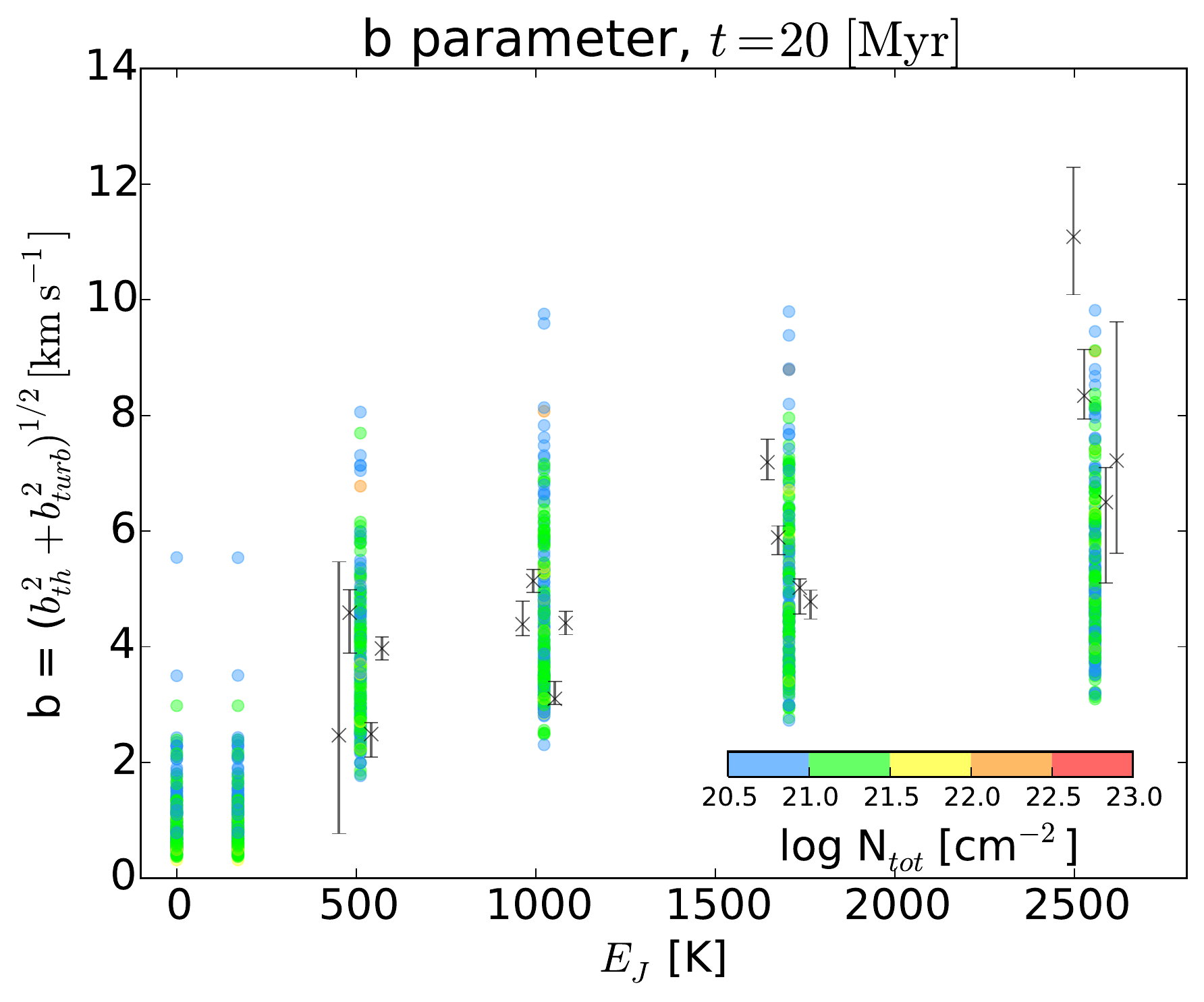}
\includegraphics[width=7.8cm]{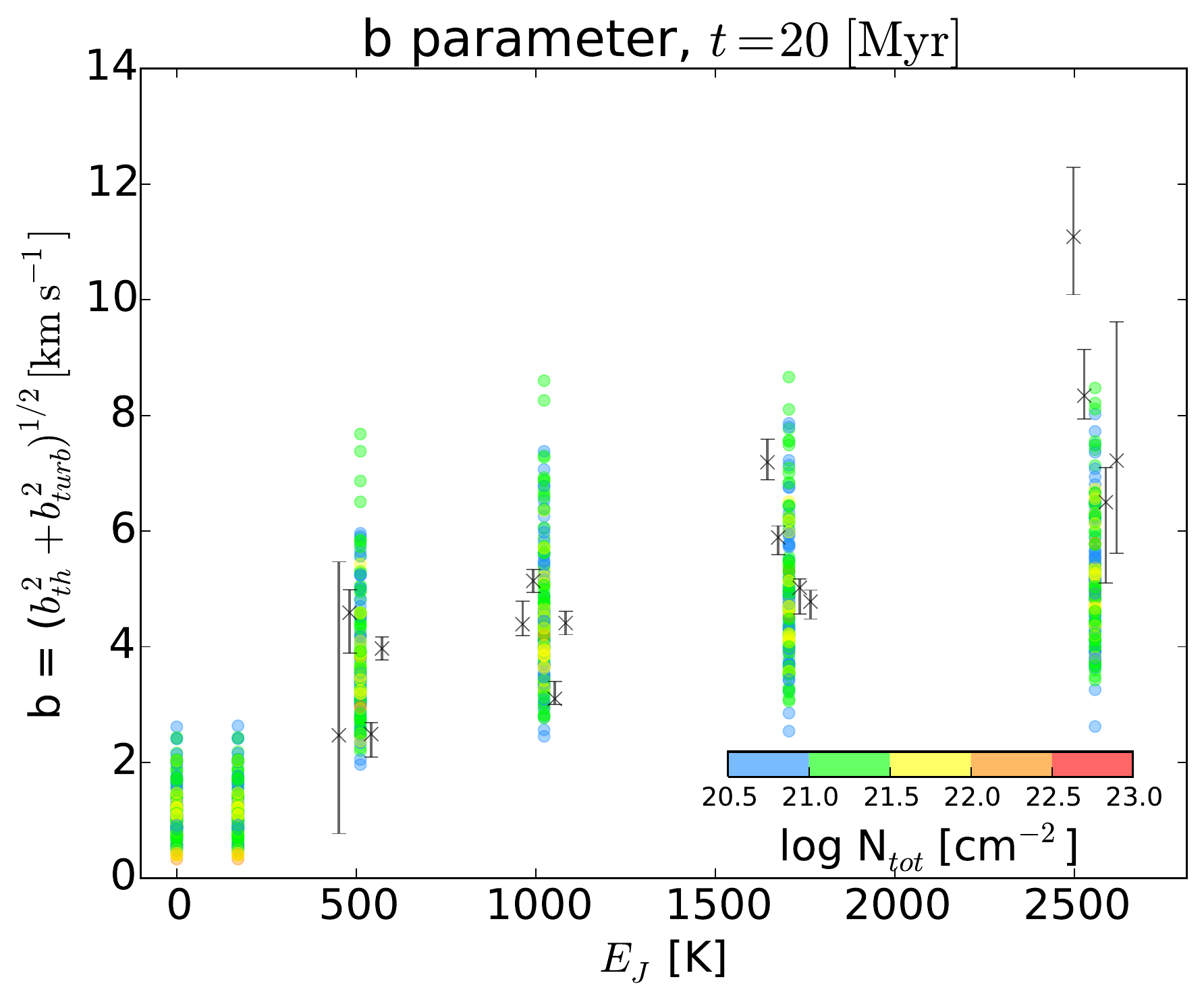}
   \caption{$b$-parameter along several lines of sight for excited levels of H$_2$ and comparison 
with the data from \citet{lacour2005}. The top panel shows lines of sight along the $x$-axis, the middle panel those along the 
$y$-axis, and the bottom panel the lines along the $z$-axis.}
\label{sigma_H2}
\end{figure}

\subsection{Excited levels of H$_2$ }
We now compare our results with the observations of rotational levels of the H$_2$ molecule,
which require high temperatures to be excited. 

As investigated by previous authors \citep{lacour2005,godard+2009}, these 
data cannot be explained by pure UV excitation. 
While \citet{lacour2005} concluded that to explain the observations, a warm layer 
associated with the cold gas should be present, \citet{godard+2009} performed 
detailed modelling that entailed shocks or vortices. The general idea is that in such dissipative small -scale structures the temperature reaches high values because of the intense 
mechanical heating. Interestingly, \citet{godard+2009} obtained a nice agreement between 
their model and the data. We stress that the small dissipative scales 
that are determinant in these models cannot be described in our simulations, which 
have a limited resolution of the order of $\simeq0.1~\mathrm{pc}$. On the other hand, as
our simulations contained large quantities of warm H$_2$, it is worthwhile to 
investigate whether the inferred populations of excited H$_2$ can be reproduced quantitatively.

\subsubsection{Calculation of population levels and column densities}
We calculated the population of the first six rotational levels ($J$) of $\mathrm{H_2}$ as in \citet{flower1986}, 
based on the approach of \citet{elitzur1978}. This calculation was made in post-processing for all grid cells. It needs the total 
hydrogen number density, the $\mathrm{H_2}$ number density, the $\mathrm{He}$ number density (we assumed $10 \%$ of the total $\mathrm{H}$), 
the temperature, and the ortho-to-para ratio (OPR). It assumes thermal equilibrium and includes spontaneous de-excitation 
and collisional excitation by $\mathrm{HI}$, by $\mathrm{He}$, and by other $\mathrm{H_2}$ molecules. 

The OPR is not very well known in the ISM
\citep[e.g.][]{lebourlot2000} and may need a specific time-dependent treatment, which is beyond 
the scope of this paper. Even though the OPR of newly formed H$_2$ molecules is thought to take values close to 3 \citep{takahashi2001}, observations differ. In high-density regions ($n(\mathrm{H_2}) \sim 10^5~\mathrm{cm}^{-3}$) observations seem to favour values as low as $0.25$ \citep{neufeld2006, pagani2011, dislaire2012}, while in translucent clouds intermediate values close to $0.7$ are preferred \citep{myers2015, gry2002, lacour2005, nehme2008, ingalls2011, rachford2009}. Because we investigate the role of low-density gas regions in the distribution of excited rotational levels of H$_2$, we assumed an OPR= $0.7$, suitable for translucent clouds such as those produced in our simulations. We also considered for comparison an OPR given by thermal equilibrium that reaches the statistical weight of 3 at high temperatures ($T \gtrsim 300~\mathrm{K}$).   
 
After we obtained the populations of the H$_2$ rotational levels
and column density, we integrated along several lines of sight in the $z$-direction. 

\subsubsection{Comparison with observed column densities}
Figure~\ref{Jpop_opr} shows the distribution of the column densities corresponding 
to the various rotational levels
in the simulation. The top panel shows results for an OPR equal to $0.7,$ while the bottom panel shows 
results for the thermal equilibrium assumption. 
The colour coding shows the total column density of the corresponding line of sight.
The symbols  correspond to the observational data of 
\citet{wakker2006} (cross), \citet{rachford2002} (circle), \citet{gry2002} (square), and 
\citet{lacour2005} (diamond). 

The simulation results are divided into two groups of points, very low column densities
 coming from the WNM that is located outside the molecular cloud (blue points)
and higher column densities (green and yellow points) that come from the gas
belonging to the molecular cloud. 
For the $J=0$ and 1 levels, the column densities are 
more or less proportional to the total column density. 
This is indeed expected since most of H$_2$ molecules are in one of these 
two levels. Their respective abundance significantly depends on the assumption used for the OPR.
While there is about a factor 10 between the first two rotational levels for $N_J/g_J$ when the OPR is equal to 0.7, it is almost a
factor thousand when the OPR is assumed to be at thermal equilibrium.

The situation is different for  the higher  levels. 
The highest column densities of the excited rotational levels do not 
correspond to the highest total hydrogen column densities, and
there is generally 
no obvious correlation between the two. This is because the 
high $J$ levels come from the warm gas (with temperatures of
between a few hundred
and a few thousand Kelvin), which itself has a low column density and is largely 
independent of the cold gas. 

The comparison with the observational data is very enlightening. The 
agreement with the simulation results is very good in general. The observational 
data points have column densities that are very similar to the simulation data. The best agreement is found for OPR$=0.7$ , for which the simulation data points are slightly above the 
observational ones, while for an OPR at thermal 
equilibrium simulation data points are below the observed points. This probably suggests that the actual OPR 
lies in between, close to $0.7$. This might be in conflict with some of the data of 
\citet{gry2002} and \citet{lacour2005} for $J=2$ by about a factor on the order
of 3-10. If confirmed, this would indicate that another mechanism, such as 
the one proposed by \citet{godard+2009}, could operate and contribute 
to excite  H$_2$.

To help understand the origin of the excited rotational levels
in the simulation, Fig.~\ref{Jpop_cell} displays
 the mean density of the H$_2$ excited levels for five density bins and the complete distribution for the two OPR used in this work.
Figure~\ref{Jpop_cell} shows that the choice of the OPR mainly
affects the dense, and thus colder gas, which is the main contributor to the populations of the two first levels ($J=0$ and $J=1$). The OPR at thermal equilibrium for very dense and cold gas is close to zero, which dramatically affects the ratio between the populations of these two levels. At lower densities in warmer gas, the OPR becomes closer to $3$, but the populations do not differ much from those calculated using OPR= $0.7$.
For the excited level the highest contributions are found in 
gas of total density between 1 and 10 cm$^{-3}$. This means that
 very diffuse gas with $n<1$ cm$^{-3}$ and moderately dense gas (with $n$
between 10 and 100 cm$^{-3}$) contribute in roughly equal proportions
to the excited levels. We recall that the peak of the 
cooling contribution of H$_2$ lies precisely in this density domain (see Fig.~\ref{heat_cool}).

\subsubsection{Velocity dispersion of individual rotational levels}
The conclusion we can draw so far is that the population of the excited 
levels of H$_2$ ($J \geq 2$) is dominated by the warm H$_2$ that 
is interspersed between the cold and dense molecular clumps. 
This "layer" of warm gas is directly associated with the molecular 
cloud and agrees well with the proposition made by \citet{lacour2005}.
These authors made another interesting observation. They 
found that the width of the lines associated with the 
excited levels increases with $J$. To investigate whether 
our simulation exhibits the same trends, we calculated the 
mean velocity dispersion along several (125 along each axis) lines of sight for 
each excited level. Figure~\ref{sigma_H2} shows the 
mean Doppler-broadening parameter, $b$, linked to the velocity dispersion 
$\sigma$  by $b^2=2 ^{1/2} \sigma^2$, and where
\begin{eqnarray}
\sigma^2 =  { \int \rho \left( (v_i - \langle v_i\rangle )^2 + C_s^2 \right) dx \over \int \rho  dx  },
\end{eqnarray}
\noindent where $C_s$ is the local sound speed, $v_i$ is the $i$ component of the local velocity, and $\langle v_i\rangle$ is the mean velocity along the $i$-direction. 
Figure~\ref{sigma_H2}  displays the results. 
Since the colliding flow configuration is highly non-isotropic, we calculated
these velocity dispersions
along the  $x$-axis (top panel), the $y$-axis (middle panel), and the $z$-axis (bottom panel).
While a more accurate treatment would consist of simulating the line profiles and then 
applying the same algorithm as the authors used, this simple estimate 
is already illustrative. From $\sigma^2$, we can infer the width of 
 the line and the $b$ parameter. 

The trends in the simulation and in the observations are 
similar. Higher $J$ levels tends to be associated with larger $b$.
The agreement is quantiatively satisfying for the $y-$ and $z$-directions.
Similar to the observations, the simulated lines of sight present 
a slightly higher velocity dispersion for higher $J$.
The agreement is poor for the lines of sight along the 
$x$-direction because they present a dispersion that is too high with 
respect to the observations. This is most likely an artefact 
of the colliding flow configuration.

The trend of larger $b$ for higher $J$ stems for the fact that 
higher levels need warmer gas to be excited. Therefore the fluid 
elements, which are enriched in high rotational levels, have higher 
temperatures and therefore higher velocity dispersions (since both are usually correlated). 

Altogether, these results suggest that the excited rotational level abundances 
reveal the complex structure of molecular clouds that is two-phase in nature and 
entails warm gas deeply interspersed with the cold gas due to the mixing induced by turbulence. 

\section{Conclusions} \label{conclusions}

We have performed high-resolution magneto-hydrodynamical simulations to describe the formation of molecular clouds
out of diffuse atomic hydrogen streams. We particularly focused on the formation of the 
H$_2$ molecule itself using a tree-based approach to evaluate UV shielding \citep[see][]{valdivia2014}. 

In accordance with previous works \citep{glover2007a}, we found that H$_2$ is able to form much faster 
than simple estimates, based on cloud mean density, would predict. The reason for this is that because the clouds are
supersonic and have a  two-phase structure, H$_2$ is produced in clumps that are much denser than the clouds on average.

As a result of a combination of phase exchanges and high UV screening deep inside the multiphase
molecular clouds (numerical convergence tests suggest that numerical diffusion is not responsible 
of this process), a significant fraction of H$_2$ develops even in the low-density and warm 
interclump medium.  This warm H$_2$ contributes to the thermal balance of the gas, and in the range of density $3-10~\mathrm{cm}^{-3}$ its cooling rate is similar to the standard cooling rate of the ISM.

Detailed comparisons with \emph{Copernicus} and \emph{FUSE} observations showed a good 
agreement overall. In particular, the fraction of H$_2$ varies with the total gas column 
density in a very similar way, showing  a steep increase between column densities 
$10^{20}$ and $3 \times10^{20}~\mathrm{cm}^{-3}$ and a slow increase at higher column 
densities. There is a trend for the high column density regions to 
present H$_2$ fractions significantly below the values inferred from the simulations, however. 
This is a possible consequence of the constant UV flux assumed in this work. 
Interestingly, the column densities of the excited rotational levels obtained at 
thermal equilibrium reproduce the observations fairly well. This is a direct 
consequence of the presence of H$_2$ molecules in the warm interclump medium
and suggests that the H$_2$ populations in excited levels reveal the two-phase structure of molecular 
clouds.

\begin{acknowledgements}
We thank the anonymous referee for the critical reading and valuable comments. We thank J. Le Bourlot and B. Godard for the insightful discussions.\\
V.V. acknowledges support from a CNRS-CONICYT scholarship. This research has been partially funded by CONICYT and CNRS, according to the December 11, 2007 agreement.\\
P.H. acknowledges the financial support of the Agence National pour la Recherche through the COSMIS project. This research has received funding from the European Research Council under the European Community’s Seventh Framework Program (FP7/2007-2013 Grant Agreement No. 306483).\\
M.G. and P.L. thank the French Program "Physique Chimie du Milieu Interstellaire" (PCMI)\\
This work was granted access to the HPC resources of MesoPSL financed
by the Region Ile de France and the project Equip@Meso (reference
ANR-10-EQPX-29-01) of the programme Investissements d’Avenir supervised
by the Agence Nationale pour la Recherche
\end{acknowledgements}

\bibliographystyle{bibtex/aa} % style aa.bst 
\bibliography{biblio_val} % your references Yourfile.bib

%%%%%%%%%%%%%%%%%%%%%%%%%%%%%%%%%%%

\appendix

\section{Numerical convergence: resolution study}\label{res_study}

\begin{figure}[h]
\centering
\includegraphics[width=7.5cm]{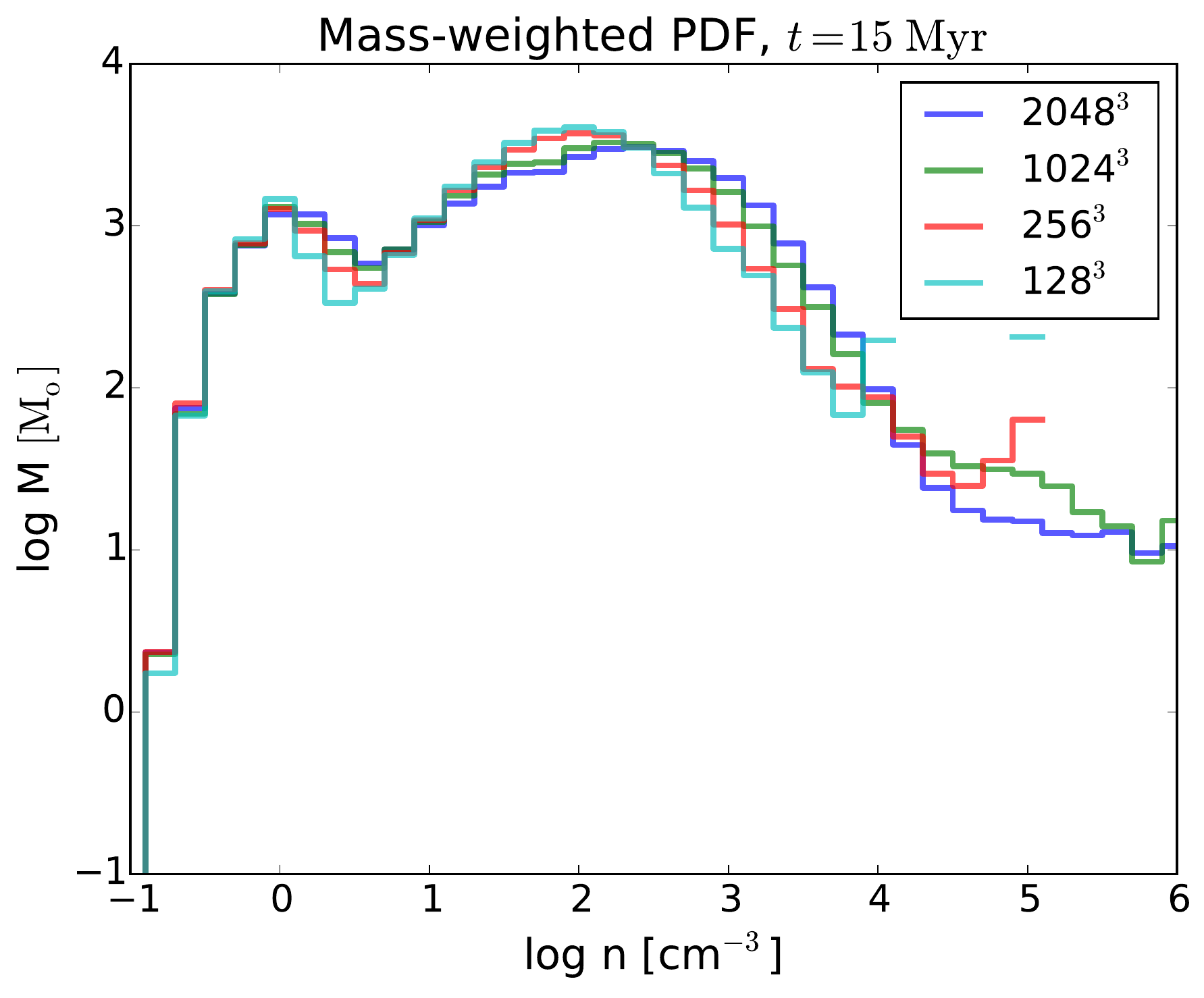}
\caption{Mass-weighted density PDF for a series of numerical resolutions. The discrepancy between low- and high-resolution runs is very significant. The difference 
between the resolution of the standard and high resolution runs is low, except for the 
highest density bins, which contain little mass.}
\label{reso_PDF}
\end{figure}

\begin{figure}[htb]
\centering
  \begin{tabular}{@{}cc@{}}
     $t = 10~\mathrm{Myr}$ & $t = 15~\mathrm{Myr}$ \\
    \includegraphics[width=.23\textwidth]{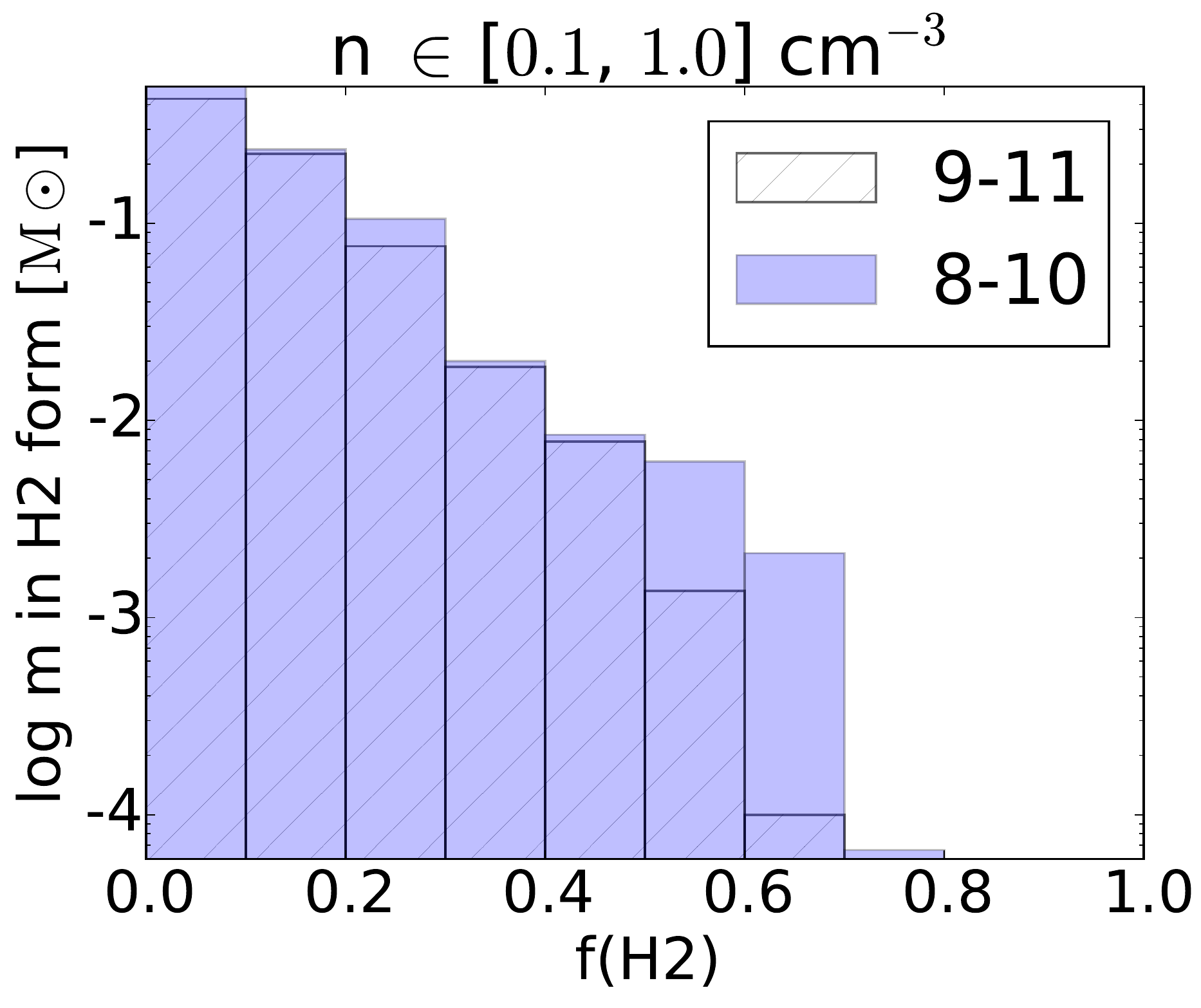}   &
    \includegraphics[width=.23\textwidth]{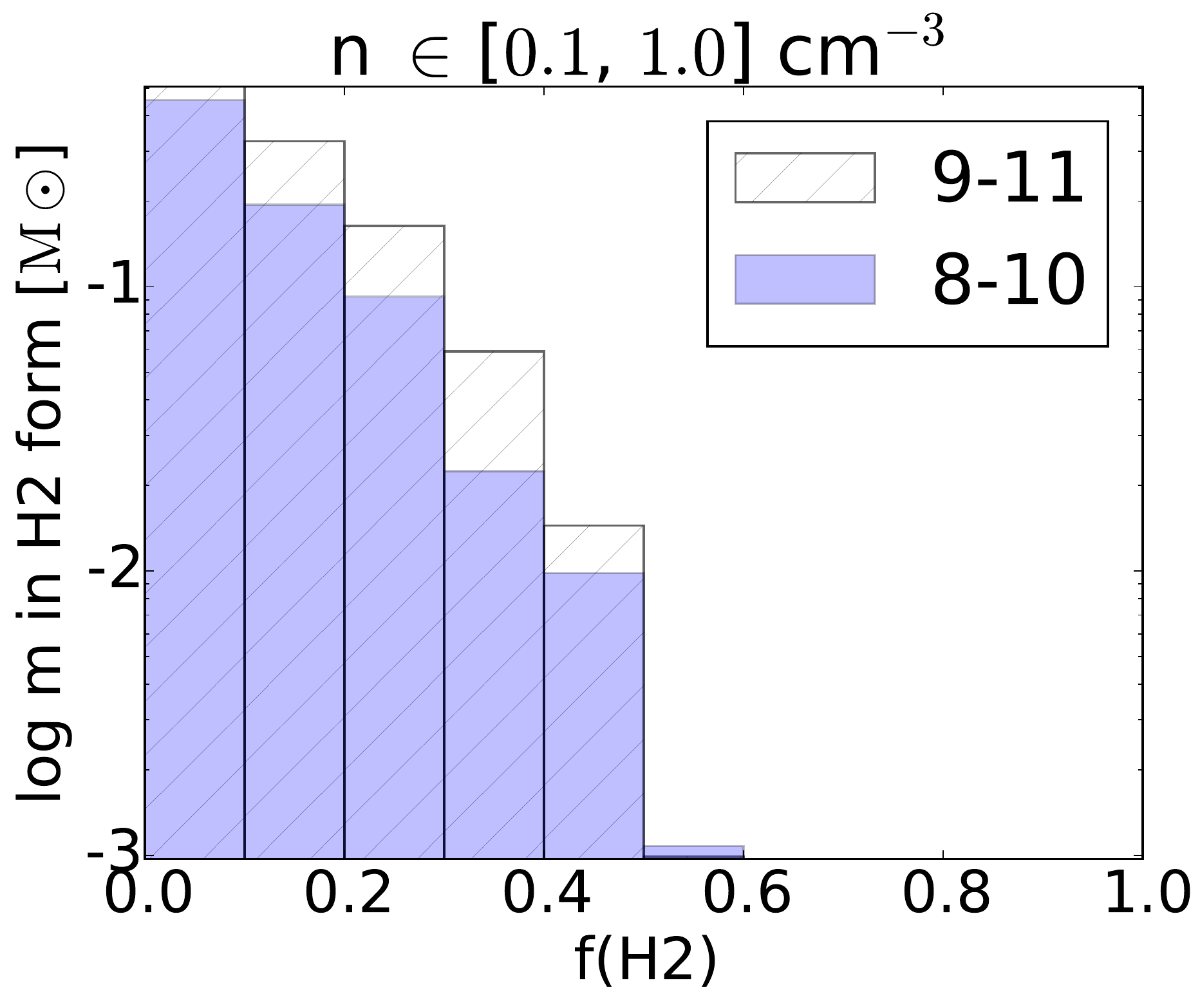}   \\
    \includegraphics[width=.23\textwidth]{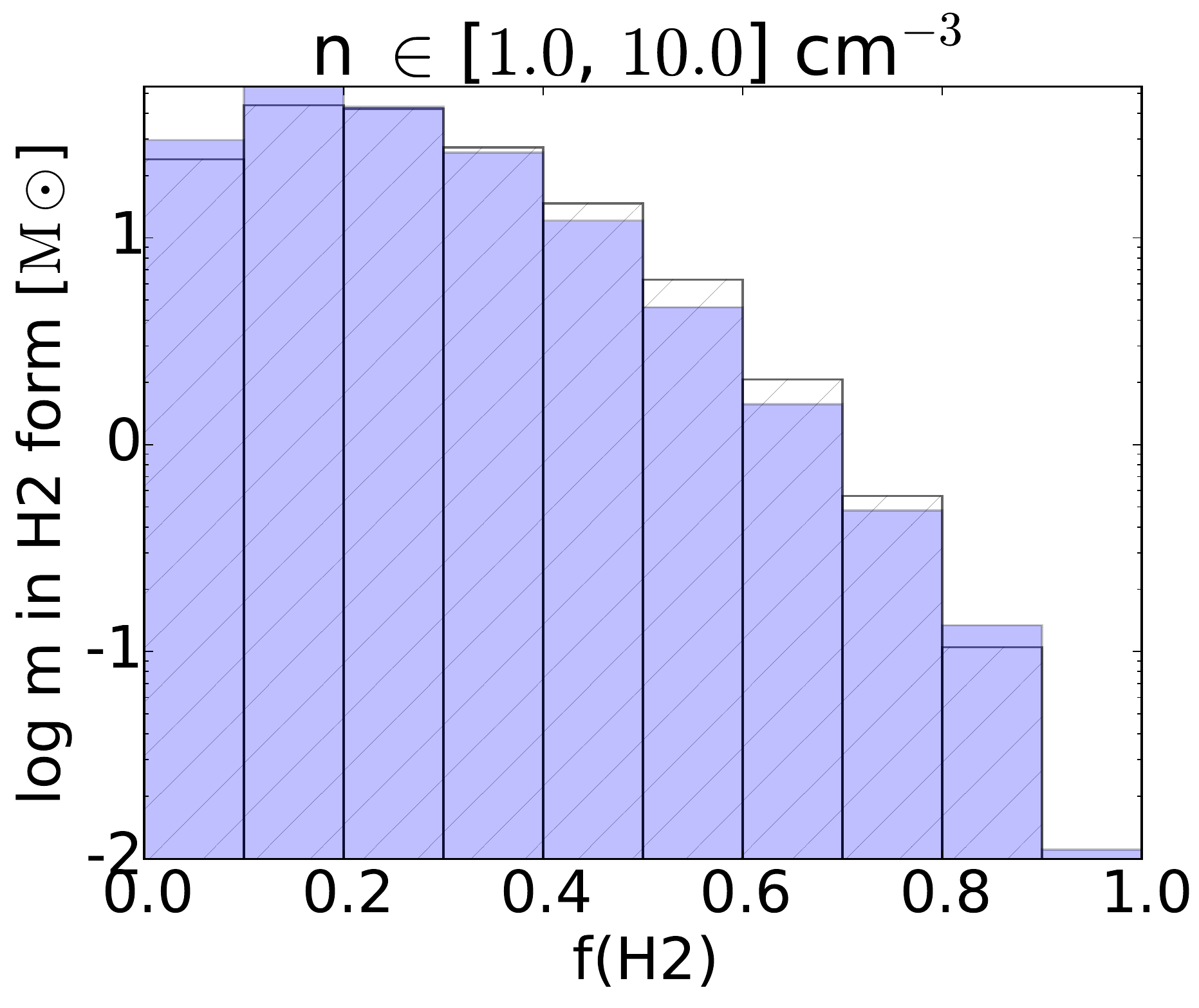}   &
    \includegraphics[width=.23\textwidth]{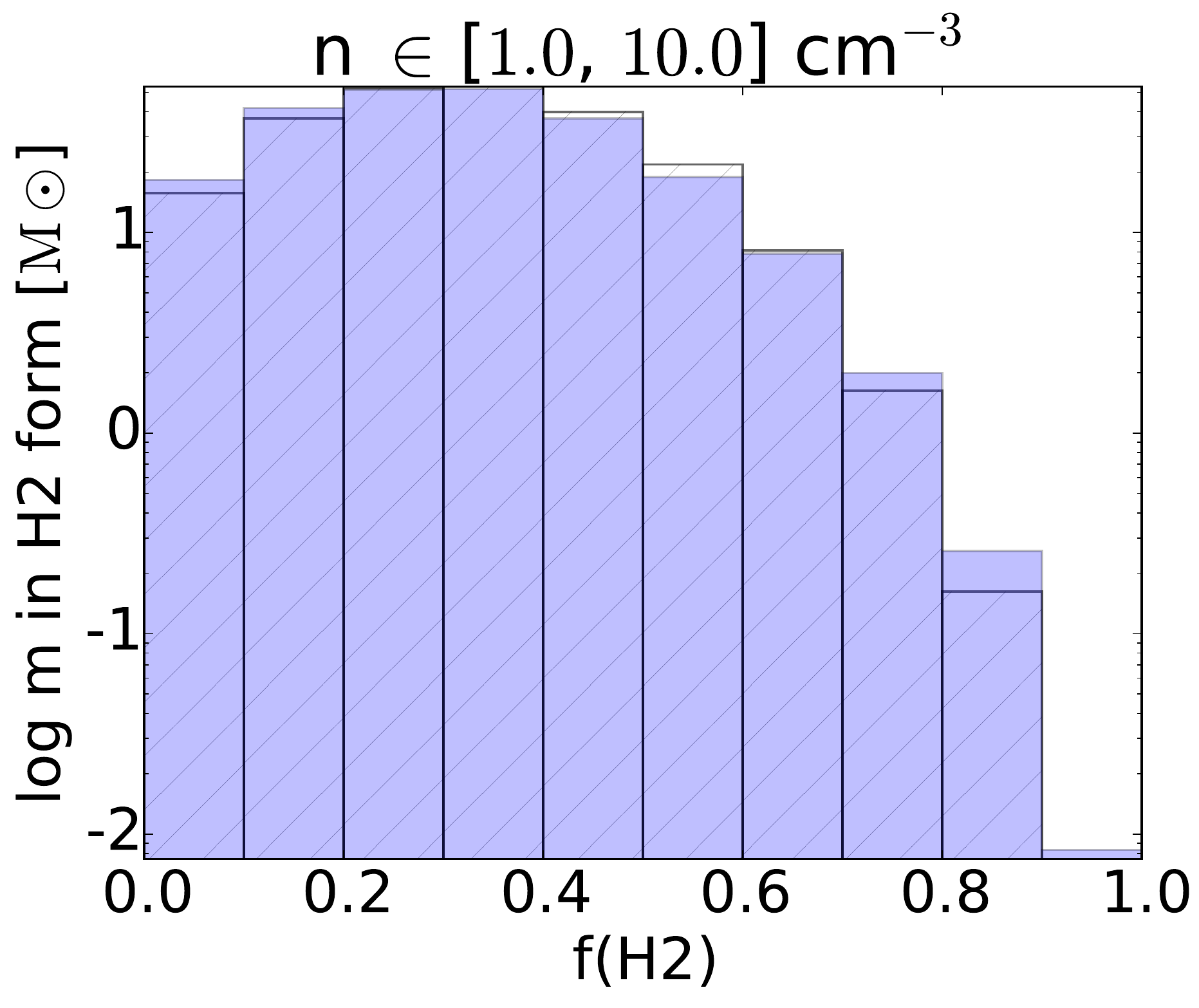}   \\
    \includegraphics[width=.23\textwidth]{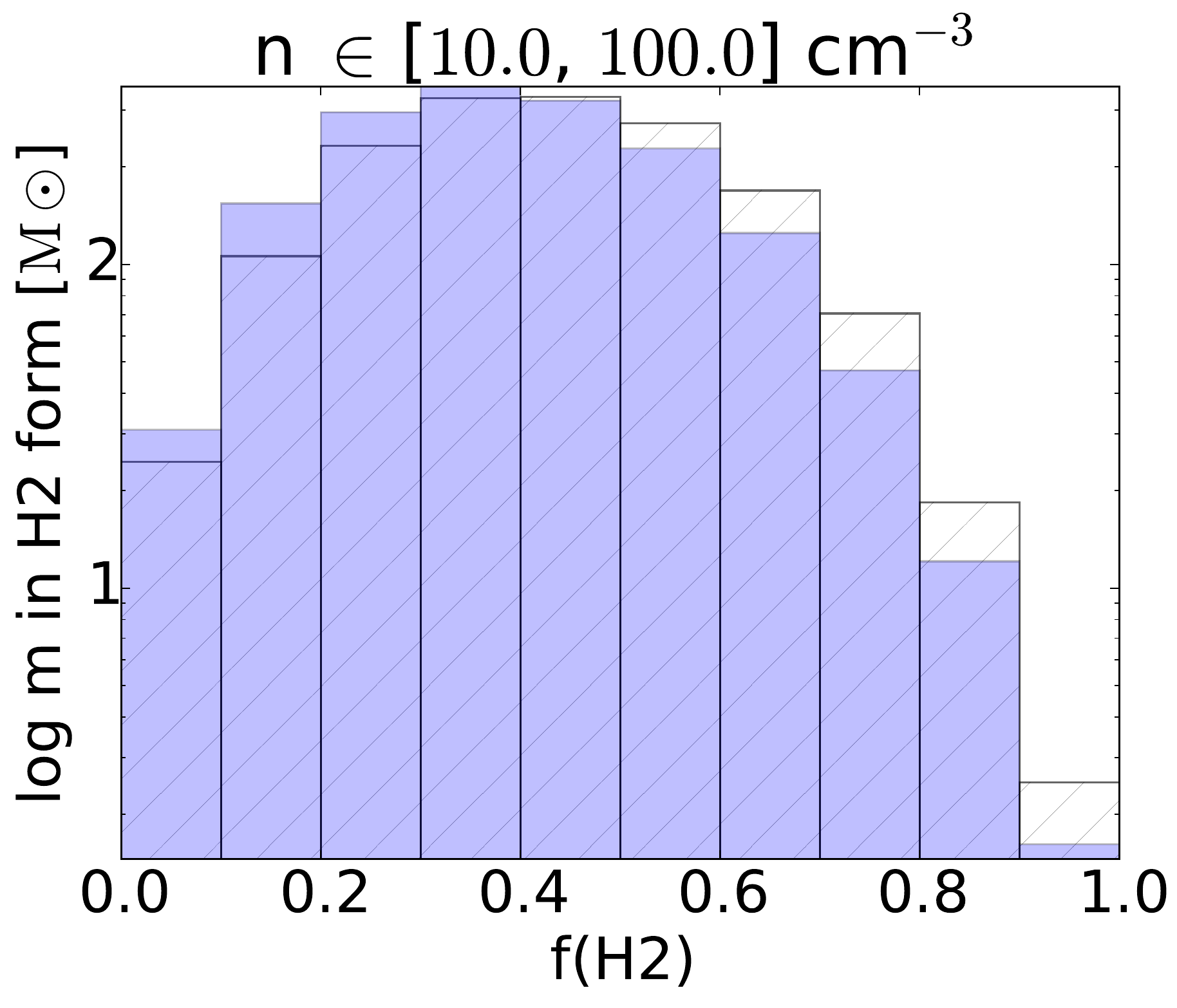}   &
    \includegraphics[width=.23\textwidth]{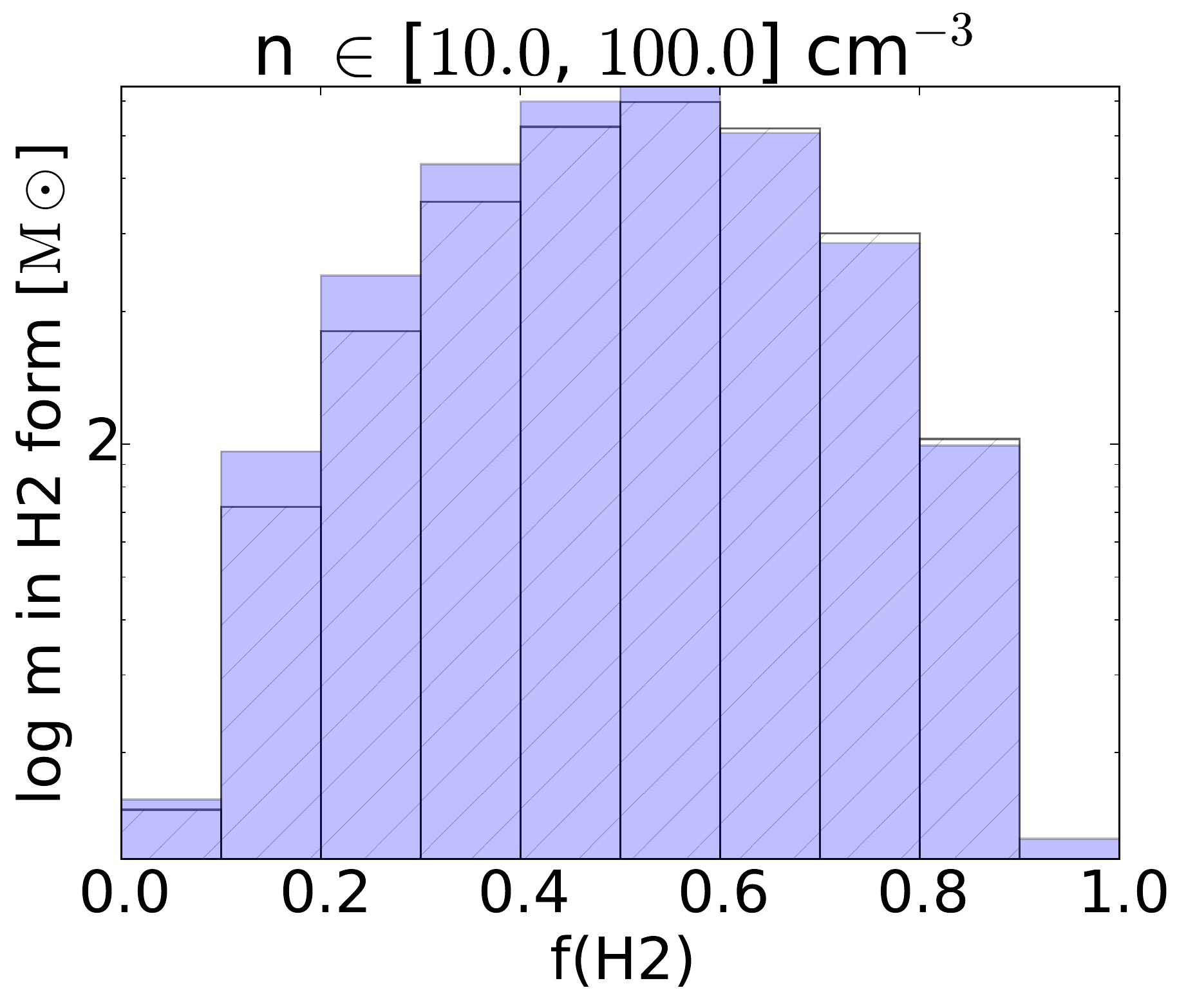}   \\
    \includegraphics[width=.23\textwidth]{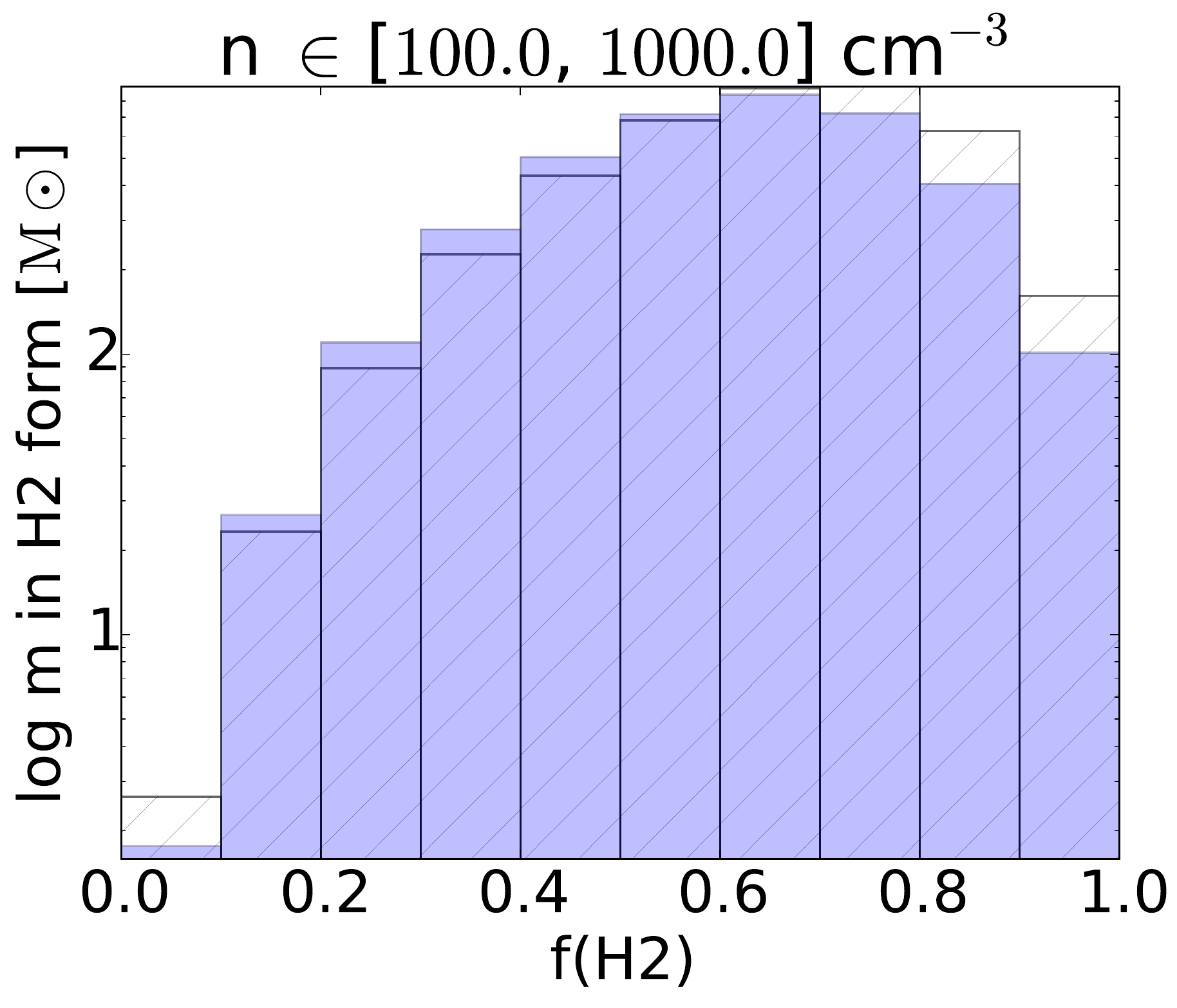}   &
    \includegraphics[width=.23\textwidth]{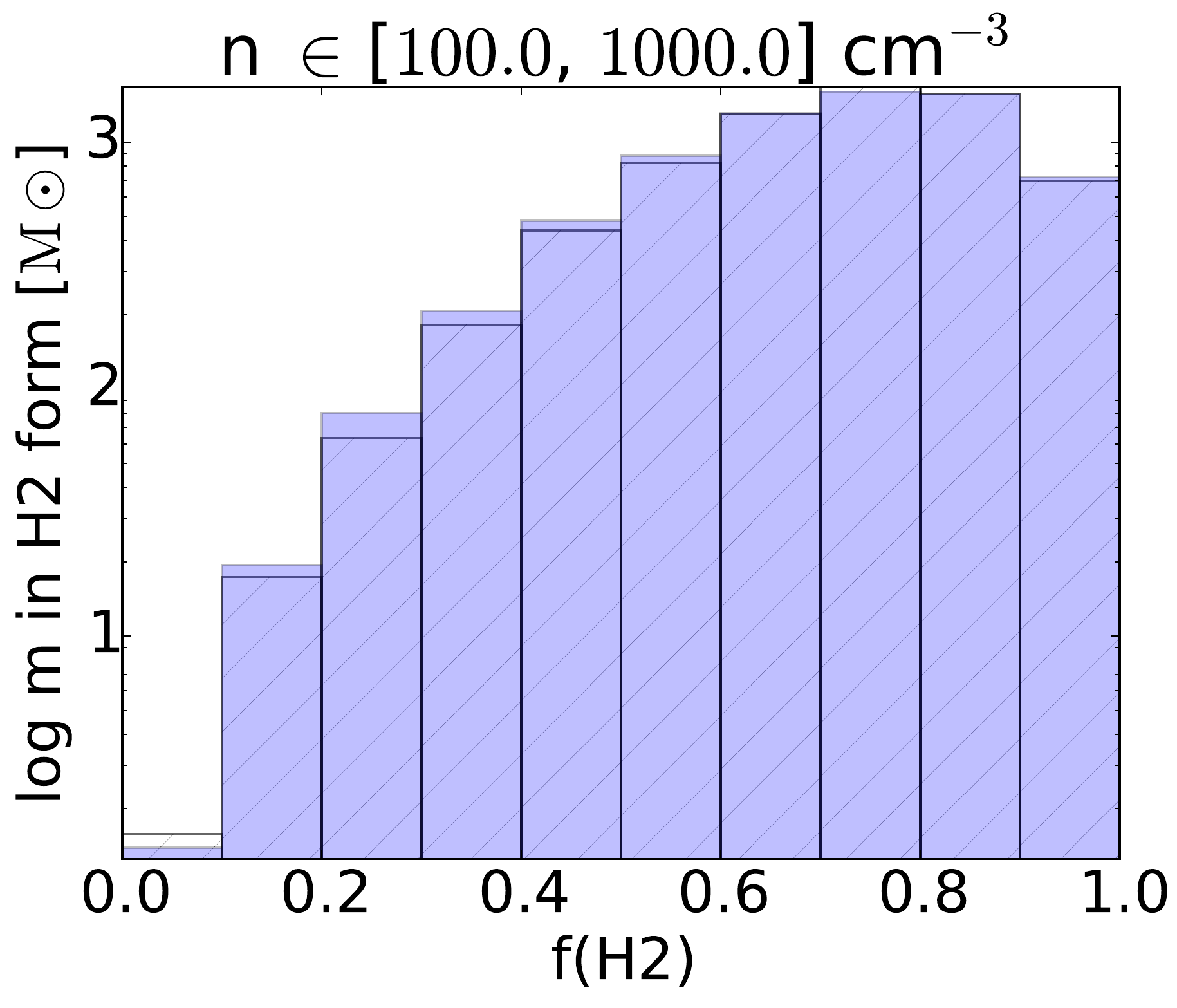}   \\
    \includegraphics[width=.23\textwidth]{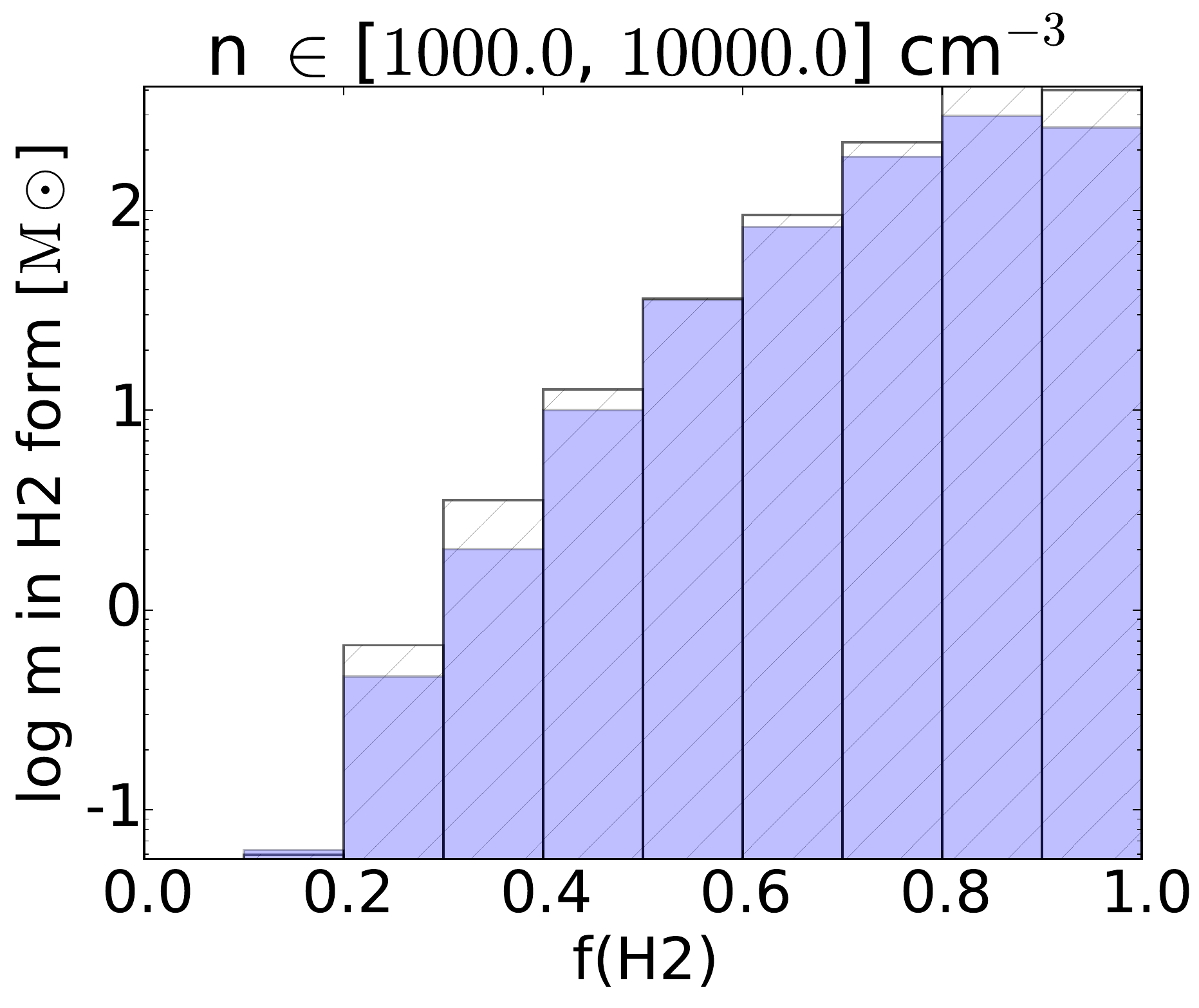}   &
    \includegraphics[width=.23\textwidth]{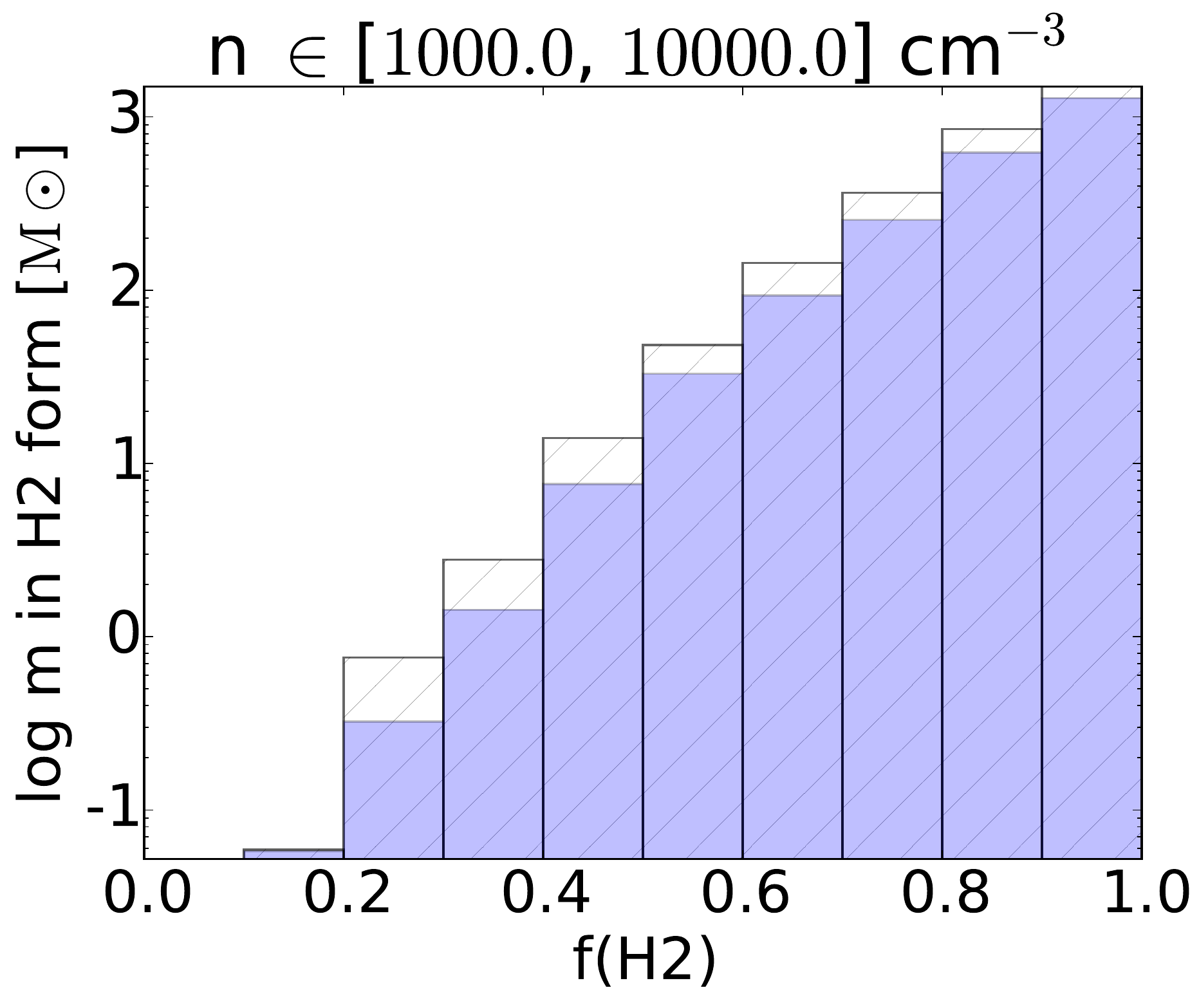}   
  \end{tabular}
  \caption{Mass in H$_2$  per density bin. Comparison between the standard resolution simulations (levels 8-10, in blue)
and the high-resolution simulations (levels 9-11, dashed) at
a time of 
  $10$ and $15~\mathrm{Myr}$. From top to bottom: density bins $n~\in~(0.1, 1),\ (1, 10),\ (10, 100),\ (100, 1000), \ (1000, 10000)~\mathrm{[cm^{-3}]}$.
The differences between the two remain fairly limited, showing that numerical convergence is not a great problem.}
\label{histo_n2}
\end{figure}

Numerical resolution is a crucial issue particularly in the context of 
chemistry. Here we compare runs of various 
resolutions. 

Figure~\ref{reso_PDF} shows the mass-weighted density PDF at a time of $15~\mathrm{Myr}$ for four different resolutions, $128^3$,  $256^3$, 
$1024^3$ , and $2048^3$ (for the two last and higher resolutions this represents the highest effective resolution
since AMR was used, as described above). The density PDF of low-resolution runs is quite 
different from the higher ones, which shows the importance of the numerical resolution. The two highest resolution 
runs present similar but not identical PDFs. This suggests that for the highest resolution runs, numerical 
convergence for the density PDF is nearly reached, although strictly speaking, it would be necessary to 
perform runs with an even higher resolution. This conclusion agrees well with the results of 
\citet{feder2013}, who found that the density PDF converges
for a resolution inbetween $1024^3$ and $2048^3$  for isothermal self-gravitating
simulations.

Figure~\ref{histo_n2} shows $f({\rm H}_2)$ distribution in $\text{five}$ density bins at times of
 $10$ and $15~\mathrm{Myr}$ for the standard and high-resolution runs 
(effective $1024^3$ and $2048^3$ resolution). Overall, the two simulations agree well. 
The differences are similar to the differences found in the density PDF.
In particular, this suggests that  numerical diffusion is not primarily responsible of the 
fraction of warm H$_2$ that we observe at low density (as discussed in Sect.~\ref{molecular}).

%%%%%%%%%%%%%%%%%%%%%%%%%%%%%%%%%%%
\section{Influence of the $b$ Doppler parameter}\label{bdop_infl}

\begin{figure}
\centering
\includegraphics[width=8cm]{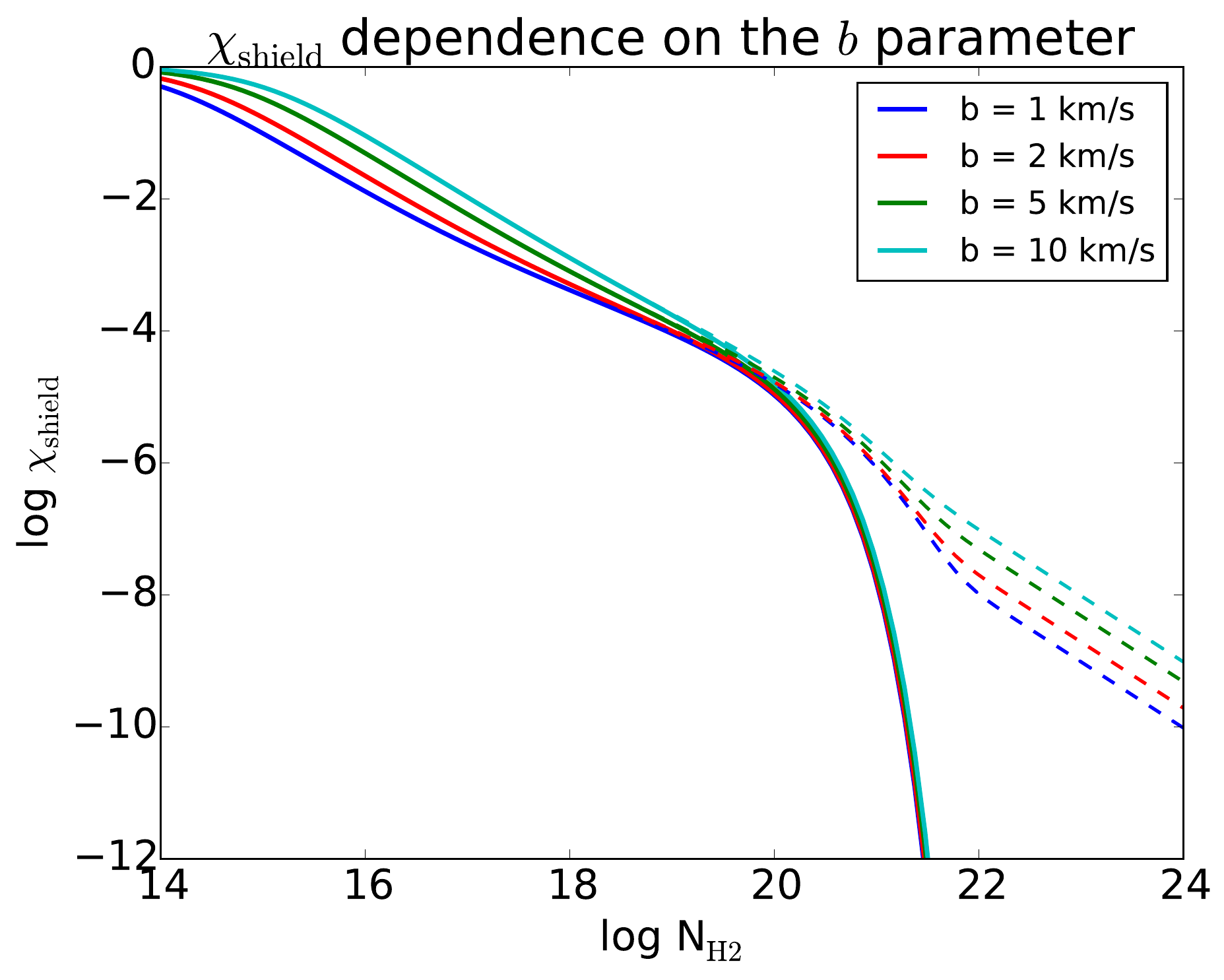}
\caption{Total shielding coefficient for H$_2$, $\chi_\mathrm{shield}$, as a function of total H$_2$ column density for several values of the $b$ Doppler parameter. The dashed lines show the expected value in absence of shielding by dust.}
\label{bdopdep_fig}
\end{figure}

Figure \ref{bdopdep_fig} shows the dependence of the total shielding coefficient for H$_2$ (shielding by H$_2$ and by dust) on the $b$ Doppler parameter as a function of total H$_2$ column density. The total column density was approximated as $\mathcal{N}_\mathrm{tot} \approx 2\times\mathcal{N}_\mathrm{H_2}$, which sets a lower limit to the influence of dust shielding. The $\chi_\mathrm{shield}$ parameter varies within less than one order of magnitude for different $b$ parameters ranging from $1$ to $10~\mathrm{km\ s^{-1}}$. In the H$_2$ column density range of interest ($\mathcal{N}_\mathrm{H_2} \sim 10^{17} - 10^{22}~\mathrm{cm^{-2}}$) all values are very similar. At higher column densities the shielding is dominated by dust.

\end{document}